\pdfoutput=1
\documentclass[usenatbib]{mn2e}
\usepackage{apjfonts}
\usepackage{amssymb}
\usepackage{amsmath}
\usepackage{ctable}
\usepackage{fixltx2e} %% forces figure order to remain fixed (otherwise figure*'s can move out-of-order)

\newcommand{\be}{\begin{equation}}
\newcommand{\ee}{\end{equation}}

\newcommand{\heff}{h_{\rm eff}}

\newcommand{\msun}{M_{\sun}}

\newcommand{\fgas}{f_{\rm gas}}

\newcommand{\scaleup}{}

\newcommand\plotone[1]
 {\centering \leavevmode \includegraphics[width={0.99\columnwidth}]{#1}}
\newcommand\plotonexa[1]
 {\centering \leavevmode \includegraphics[width={0.97\columnwidth}]{#1}}

\newcommand{\plotside}[1]
 {\centering \leavevmode \includegraphics[width={0.95\textwidth}]{#1}}

\newcommand{\plotsidesmallest}[1]
 {\centering \leavevmode \includegraphics[width={0.75\textwidth}]{#1}}
\newcommand{\acknowledgments}{\begin{small}\section*{Acknowledgments}\end{small}}
\newcommand\altaffilmark[1]{$^{#1}$}
\newcommand\altaffiltext[1]{$^{#1}$}
\title[Angular Momentum Transport and AGN Fueling]{An Analytic Model of Angular Momentum Transport by Gravitational Torques:  From Galaxies to Massive Black Holes}

\author[Hopkins and Quataert]{
\parbox[t]{\textwidth}{ 
Philip F. Hopkins\altaffilmark{1}\thanks{E-mail:phopkins@astro.berkeley.edu},
\&\ Eliot Quataert\altaffilmark{1}
} 
\vspace*{6pt} \\
\altaffiltext{1}{Department of Astronomy and Theoretical Astrophysics Center, University of California
  Berkeley, Berkeley, CA 94720 \voffset=-0.6in} \voffset=-0.6in
}

\date{Submitted to MNRAS, June, 2010 \voffset=-0.6in}
\begin{document}
\maketitle
\label{firstpage}

\begin{abstract}

We present analytic calculations of angular momentum transport and gas inflow in galaxies, from scales of $\sim$ kpc to deep inside the potential of a central massive black hole (BH).   We compare these analytic calculations to numerical simulations and use them to develop a sub-grid model of BH growth that can be incorporated into semi-analytic calculations or cosmological simulations.  Motivated by both analytic calculations and simulations of gas inflow in galactic nuclei, we argue that the strongest torque on gas arises when non-axisymmetric perturbations to the stellar gravitational potential produces orbit crossings and shocks in the gas.   This is true both at large radii $\sim 0.01-1$ kpc, where bar-like stellar modes dominate the non-axisymmetric potential, and at smaller radii $\lesssim 10$ pc, where a lopsided/eccentric stellar disk dominates.  The traditional orbit crossing criterion is not always adequate to predict the locations of, and inflow due to, shocks in gas+stellar disks with finite sound speeds. We derive a modified criterion that predicts the presence of shocks in stellar dominated systems even absent formal orbit crossing.    We then derive analytic expressions for the loss of angular momentum and the resulting gas inflow rates in the presence of such shocks.   We test our analytic predictions using hydrodynamic simulations at a range of galactic scales, and show that they successfully predict the mass inflow rates and quasi-steady gas surface densities with a small scatter $\simeq 0.3\,$dex.   We use our analytic results to construct a new estimate of the BH accretion rate given galaxy properties at larger radii, for use in galaxy and cosmological simulations and semi-analytic models.  While 
highly simplified, this accretion rate predictor captures the key scalings in the numerical simulations.   By contrast, alternate estimates such as the local viscous accretion rate or the spherical
Bondi rate fail systematically to reproduce the simulations and have significantly larger scatter.

\end{abstract}


\section{Introduction}
\label{sec:intro}

Gas inflows into galactic nuclei play a critical role in galaxy formation, 
initiating a wide range of phenomena including starbursts, the formation of bulges, spheroidal 
galaxies, and nuclear star clusters, 
and accretion onto supermassive black holes (BHs). 
The  initial triggers for these inflows are diverse and may include galaxy mergers  and tidal encounters; bar and spiral disk instabilities; secondary and tertiary 
instabilities (``bars within bars''); and large-scale triaxial or binary structures in galaxy 
halos. 

 Intense star formation in galactic nuclei also has a major effect on the structure of the resulting systems, with  dissipative processes dominating the formation of the inner $\sim$kpc of bulges at surface mass densities reaching $\sim 10^{11-12}\,\msun\,{\rm kpc^{-2}}$
\citep{ostriker80,carlberg:phase.space,gunn87,kormendy:dissipation,
  hernquist:phasespace,hopkins:cusps.ell,hopkins:cores}.   There is also strong indirect evidence that the most luminous starbursts, ultra-luminous infrared galaxies (ULIRGs), are accompanied or 
 followed by periods of luminous quasar activity \citep{sanders88:quasars,sanders88:warm.ulirgs,
  dasyra:pg.qso.dynamics}.

On galactic scales, numerical simulations have shown that in gas-rich mergers, tidal torques lead to rapid gas inflow into the central $\lesssim$ kpc \citep{hernquist.89,barnes.hernquist.91,
  barneshernquist96}.   Minor mergers and instabilities in 
  self-gravitating disks (bars and spiral waves) can drive similar inflows 
  \citep{hernquist.mihos:minor.mergers,bournaud:minor.mergers,
elichemoral:pseudo.from.minor,younger:minor.mergers}. 
And in high redshift, gas-rich disk galaxies, the sinking of massive star-forming clumps  may produce analogous strong gravitational torques \citep{bournaud:disk.clumps.to.bulge,genzel:highz.rapid.secular}. 

% At both low and high redshifts, many rapidly star-forming galaxies appear to have unusually high gas densities due to gravitational disturbances 
% \citep{sanders96:ulirgs.mergers, schweizer98,jogee:review,
 % dasyra:mass.ratio.conditions,woods:tidal.triggering,woods:minor.mergers,
 % veilleux:ir.bright.qso.hosts.merging,
 % dasyra:highz.ulirg.imaging.not.major,puech:minor.merger.at.z06,
 % hammer:hubble.sequence.vs.mergers,
  %tacconi:smg.mgr.lifetime.to.quiescent}.
 %Both observations and simulations of these systems show that the gas has large, 
 %super-sonic inflow velocities  that can approach the circular velocity itself 
% in the most extreme cases.  The inflows that produce these high star formation rates 
 %also have a major effect on the structure of the resulting systems, with 
 %dissipative processes dominating the formation of the inner $\sim$kpc of 
%bulges at surface mass densities reaching $\sim 10^{11-12}\,\msun\,{\rm kpc^{-2}}$
%\citep{ostriker80,carlberg:phase.space,gunn87,kormendy:dissipation,
%  hernquist:phasespace,hopkins:cusps.ell,hopkins:cores}.   There is also strong indirect evidence that the most luminous starbursts, ultra-luminous infrared galaxies (ULIRGs), are accompanied or 
 %followed by periods of luminous quasar activity \citep{sanders88:quasars,sanders88:warm.ulirgs,
  %dasyra:pg.qso.dynamics}.

How gas is transported from  $\sim 1$ kpc to the much smaller 
 scales relevant for massive BH growth remains uncertain (e.g.,
\citealt{goodman:qso.disk.selfgrav}). Once gas reaches sub-kpc scales, 
the torques from large-scale bars, and even from major mergers, become 
less efficient.   Local viscous stresses -- which are believed to dominate angular
momentum transport near the central BH (e.g.,
\citealt{balbus.hawley.review.1998}) -- are inefficient at radii
$\gtrsim 0.01-0.1$ pc (e.g.,
\citealt{shlosman:inefficient.viscosities,
  goodman:qso.disk.selfgrav,thompson:rad.pressure}).   
  And scattering of individual gas clumps or molecular clouds 
  cannot produce the high accretion rates needed to explain bright quasars 
  \citep{hopkins:seyferts,kawakatu:disk.bhar.model,
  nayakshin:forced.stochastic.accretion.model}.
  As a consequence,
most models implicitly or explicitly invoke some form of global gravitational torques (``bars
within bars''; \citealt{shlosman:bars.within.bars}) to continue
transport to smaller radii.   
Recently, numerical simulations using a variety of 
techniques have been able to follow the inflow of gas to smaller scales in galaxies 
\citep{escala:bh.mgr.idealized,escala:nuclear.gas.transport.to.msigma,
  mayer:bh.binary.sph.zoom.sim, wise2007:protogalaxy.collapse,
  levine2008:nuclear.zoom, hopkins:zoom.sims}.  
   This has allowed simulations to begin to address some of the long-standing
questions about gas inflow in galactic nuclei.

In a previous paper \citep{hopkins:zoom.sims}, we presented a large suite of $\sim100$ simulations 
that begin from initial conditions of either gas-rich major mergers or bar-unstable disks, and follow the inflow of gas from $>10\,$kpc to $<0.1\,$pc; at the smallest radii, the system begins to resemble a traditional, viscous accretion disk.  Our method uses a series of independent numerical simulations 
on successively smaller scales (rather than an AMR or particle-splitting approach), which 
allows us to systematically survey how a variety of initial conditions and galaxy properties affect gas dynamics in galactic nuclei.   These calculations show that  large-scale bars can indeed drive gas to $\lesssim$kpc; and if the disky material at smaller radii becomes self-gravitating, it triggers a cascade of secondary gravitational instabilities that transports gas deeper in the potential well. This qualitatively resembles 
the ``bars within bars'' scenario.  However, there are several critical differences.
First, the secondary instabilities on intermediate scales ($\sim 10-100$\,pc)
exhibit a diverse range of morphologies. A more accurate characterization would thus be 
``it's non-axisymmetric features all the way down'' (or ``stuff
within stuff'').    This is consistent with the fact that observed systems that do show nuclear asymmetries can show nuclear rings, barred rings, one or three-armed modes, and clumpy/irregular structures 
\citep{martini:seyfert.host.morph, peletier:seyfert.morph.imaging,knapen:seyfert.morphology,
laine:nested.bars.in.seyferts,kanpen:nuclear.region.in.bars.vs.host.prop,
greene:2009.sigma.gas.in.agn}.  
 Moreover, in our simulations, the inflow rates and gas morphology are time-variable, leading to modest duty cycles that may explain the generally weak correlation in observations between mode structure and inflow/outflow on different spatial scales.  
A second key difference between our numerical results and the canonical ``bars within bars'' model is that once gas reaches the radii where the gravity of the BH begins to dominate ($\sim 10$ pc), bar-like modes no longer survive.  However, a new instability appears, in the form of a slowly precessing lopsided/eccentric disk or one-armed spiral ($m=1$) mode, which dominates the angular momentum transport to $<0.1\,$pc. 
The resulting torques can drive gas inflows at rates comparable to those required to 
sustain the brightest observed 
quasars, $\sim 10\,\msun$ yr$^{-1}$. And 
the surviving stellar relics of these modes closely resemble the 
observed nuclear disks in M31 \citep{lauer93} and other 
galaxies \citep{hopkins:m31.disk}. 

Given the simplifying assumptions necessary in numerical simulations 
(e.g., sub-resolution star formation and feedback models), 
it is clearly important to understand analytically the physics of angular momentum transport.  
Moreover it is not possible to simulate the entire parameter space relevant to 
galactic inflows, so a way of generalizing the numerical results would be useful, both for gaining physical intuition and for developing 
subgrid models that can be used in calculations that lack the 
physics/resolution to directly capture gas inflow on sub-kpc scales.  

There is at present no analytic theory that accurately predicts the inflow rates of gas 
in mixed stellar-plus-gas systems, despite the rich history of analytic studies of gravitational instabilities in disks 
\citep[see e.g.][]{goldreich:1965.spiral.stability, goldreich:1965.sheared.spirals,lin.shu:spiral.wave.dispersion,
toomre:spiral.group.velocity,lynden-bell:1972.spiral.amplification,
goldreichtremaine:disk.satellite.interaction,binneytremaine,shu:gas.disk.bar.tscale,
toomre:spiral.structure.review,toomre:swing.instability,
lau:spiral.wave.dispersion.relations,
adams89:eccentric.instab.in.keplerian.disks,shu:gas.disk.bar.tscale,
ostriker:eccentric.waves.via.forcing}. 
Almost all previous calculations have simplified the problem 
dramatically by focusing on {\em strictly} stellar or gas disks
 (as opposed to gas+stellar disks) and by ignoring the 
presence of shocks and cooling/dissipation in the gas. 
In addition, the majority of the analytic 
development and intuition has focused on the case of weak perturbations.  
In these limits, the leading-order instantaneous gravitational torques 
associated with the asymmetric gravitational potential cancel exactly. 
Gravitational angular momentum exchange occurs in narrow resonances 
\citep{goldreichtremaine:spiral.resonances}, 
and the net rate at which the gas loses angular momentum is second-order 
in the fractional magnitude of the perturbation \citep[denoted $|a|$ 
throughout;][]{lynden-bell:1972.spiral.amplification}. 
It is typically further suppressed by strong powers in the radial 
and azimuthal wavenumbers of the perturbation
\citep[in both the global and local perturbation limits;][]{kalnajs:1971}. 
These results lead to the prediction of relatively weak torques, with sub-sonic mean gas inflow velocities 
$v_{\rm inflow} \lesssim 0.1\,|a|^{2}\,Q^{-1}\,c_{s}$ 
\citep[where $Q$ is the standard Toomre $Q$ parameter and $c_{s}\ll V_{c}$ is the 
gas sound speed;][]{goodman:qso.disk.selfgrav}. 
%In this regime, the net effect of gravitational torques can be parameterized as 
%an effective local viscosity, from turbulent velocities generated by gravitational motions
%\citep{gammie:2001.cooling.in.keplerian.disks}. 
%As a result, gravitational torques in these simplified systems 
%are not necessarily dominant over other ``purely hydrodynamic'' sources of angular momentum 
%transfer such as an effective viscosity associated with turbulence generated by stellar feedback, or cancellation of angular momentum in shocks between 
%stellar winds or merging gas-rich disks \citep{escala:nuclear.gas.transport.to.msigma,
%nayakshin:forced.stochastic.accretion.model,hobbs:turbulence.agn.feeding}.

In the numerical simulations described above of galaxy mergers, disk instabilities, and secondary and tertiary instabilities, none of the simplifications often utilized in the study of angular momentum transport are in fact realized. 
Most of the angular momentum exchange occurs in the non-linear regime 
in which the non-axisymmetric perturbations to the gravitational potential 
due to the stars are sufficiently large that they induce orbit crossing, shocks, and dissipation in the gas.    
The multi-component dissipative nature of the system is thus critical.  
In, for example, instabilities of galactic disks, the gas orbits respond to the 
asymmetric potential of the collisionless (stars plus dark matter) material; 
when the asymmetry is sufficiently large, the gas
is driven into shocks, which dissipates energy and allows the torques from 
the asymmetric potential to remove angular momentum efficiently. 
The ``pile up'' of material in these shocks manifests itself observationally 
as dust lanes and enhanced star formation in the leading edge of 
spiral arms \citep{roberts:gas.dynamics.in.bars,athanassoula:bar.dust.lanes,
wada:gas.orbits.in.weak.bars,bournaud:bar.lifetimes}. 
The same basic process also dominates the angular momentum loss 
in major galaxy mergers, with the asymmetry in the stellar 
potential being excited by the companion \citep{hernquist.89,barnes.hernquist.91,
  barneshernquist96,hopkins:disk.survival}.
The resulting mean inflow velocities are typically 
super-sonic, with inflow rates up to $\sim100-1000$ times
larger than the predictions based on linear torques in single-component systems! 
The simulations also explicitly demonstrate that ``hydrodynamic'' pressure forces do not contribute significantly to the torques and inflow \citep[see][]{barneshernquist96,
berentzen:gas.bar.interaction,hopkins:disk.survival,hopkins:zoom.sims}: 
the process driving angular momentum loss is overwhelmingly {gravitational}. 

The basic processes summarized above -- gas responding to gravitational torques predominantly 
from stars, ``piling up'' in shocks, and exhibiting supersonic mean radial inflow/outflow velocities as a result  -- have been directly observed in a variety of nearby galaxies, 
where the role of gravitational torques can be mapped 
\citep{haan:nuga.gas.dynamics.maps,durbala:obs.isolated.gal.torque.maps,
block:obs.bar.torque.maps,sakamoto:bar.driven.mol.gas.transport,
quillen:inflow.along.bar.ngc7479}. 
These observations also extend to galactic nuclei where secondary instabilities ("bars within bars") may become important \citep{laine:nested.bars.in.seyferts,schinnerer:nuclear.bar.starburst,
kanpen:nuclear.region.in.bars.vs.host.prop}, 
and to the gas surrounding bright AGN \citep{combes:liner.nuclear.ring,
garcia.burillo:torques.in.agn.nuclei.obs.maps.no.inflow}. 
The torques inferred in many of these systems correspond to  the gas
losing a large fraction of its angular momentum in a few dynamical 
times.  This is much stronger than can be accounted for by torques in linear, single-component systems and highlights the need for a better physical understanding of gravitational torques and inflow under realistic galactic conditions.

In this paper, we develop an analytic model that attempts to account for the realistic complications 
of angular momentum redistribution in systems with both collisional (gas) and 
collisionless (stellar/dark matter) components, in the presence of shocks and dissipation. 
This is an analytic companion to the numerical simulations in \citet{hopkins:zoom.sims}, 
which in turn builds on the large history of analytic and numerical studies of angular momentum transport in galaxies described above. The remainder of this paper is organized as follows.   In \S~\ref{sec:response}, we calculate the response of gas to general asymmetric perturbations, deriving the conditions for shocks and the resulting angular momentum transport; we consider the linear regime of no shocks (\S~\ref{sec:deriv.mom.weak}), strong orbit crossings that induce shocks (\S~\ref{sec:deriv.mom.strong}), and marginal orbit crossings (\S~\ref{sec:deriv.mom.inflow:marginal}). 
We then briefly describe the balance between inflow and star formation in gaseous disks, and the resulting quasi-steady gas density distribution (\S~\ref{sec:deriv.gasden}).   In \S~\ref{sec:sims.main}, we describe our suite of high-resolution hydrodynamic simulations of mergers and unstable disks, with multi-scale re-simulations of secondary instabilities, which we use to test our analytic scalings as a function of galaxy properties and physical scale. In \S~\ref{sec:unified}, we use the results of \S~\ref{sec:response} to develop an analytic model that predicts the inflow rate to the BH given the conditions at large radii in a galaxy; we show that this predictor is much more accurate than the spherical accretion prediction or a local viscous model.   Finally, we summarize and discuss our results in \S~\ref{sec:discussion}.   

\vspace{-0.5cm}
\section{Gas Response in Asymmetric Systems}
\label{sec:response}

\subsection{Definitions}
\label{sec:perturb}

We adopt a cylindrical coordinate frame, 
($R$, $\phi$, $z$), and consider an initially axisymmetric disk with an 
arbitrary spherical (BH+bulge+halo) component.   We use the test particle approximation to calculate the response of the disk to a specified non-axisymmetric perturbation to the potential, largely neglecting pressure forces.    Since we are interested in the behavior of cold, rotationally supported,
gas, the gas of interest is always in a thin ($h/R \ll1$) disk, aligned with the stellar disk.

The initial potential in the disk plane can be 
written $\Phi_{0} = \Phi_{0}(R)$, and other properties 
are defined using standard notation: 
\begin{align}
\Phi_{0} &= \Phi_{0}(R) \\
V_{c}^{2} & = R\, \frac{\partial \Phi }{\partial R} 
\approx \frac{G\,M_{\rm enc}(<R)}{R} \\ 
\Omega & \equiv  t_{\rm dyn}^{-1} = V_{c}/R \\ 
\kappa^{2} &\equiv  R\,\frac{{\rm d} \Omega^{2}}{{\rm d} R} + 4\,\Omega^{2} = 
\frac{\partial^{2}\Phi}{\partial R^{2}}+3\,\Omega^{2}
%\Omega^{2}\,{\Bigl (}1 + \frac{{\rm d}\ln{M_{\rm enc}}}{{\rm d}\ln{R}}  {\Bigr)}
\end{align}
where $V_{c}$ is the circular velocity, $\Omega$ the angular 
velocity, and  $\kappa$ the epicyclic frequency.   We now consider a global perturbation in the disk plane and define the perturbed potential 
\begin{equation}
\Phi \rightarrow \Phi_{0}(R) + \Phi_{1}(R,\,\phi)
\end{equation}
In what follows, the subscript ``1'' will denote all perturbed quantities.    We will consider a frame rotating with the perturbation pattern 
speed $\Omega_{p}$, so that in these coordinates the perturbation 
is stationary.  An unperturbed orbit in the rotating frame is at fixed $R$ with angular speed $\dot{\phi_0}=\Omega-\Omega_{p}$.   We can represent the perturbed 
potential as a sum over modes: 
\begin{equation}
\Phi_{1} \equiv a(R)\,\Phi_{0}\,f(\phi)
= \Phi_{0}(R)\,\sum_{m=1}^{\infty}\, a_{m}(R)\,\cos{(m\,\phi)} \ .
\end{equation}  
For all perturbed quantities $X$, we define 
$X_{1}(R,\,\phi) = X_{a}(R)\,\cos{(m\,\phi)}$. Thus $\Phi_{a}=|a|\,\Phi_{0}$, etc.
Note that for a given particle, the perturbed and unperturbed quantities depend 
implicitly on time via $R(t)$ and $\phi(t)$.

\vspace{-0.3cm} 
\subsection{Torques and Inflow: Linear Regime}
\label{sec:deriv.mom.weak}

When the perturbation is weak ($|a|\ll 1$), it is  
straightforward to expand the equations of motion for a test particle in the perturbed potential using 
the epicyclic approximation. With the 
substitutions 
\begin{align}
\nonumber R & \rightarrow R_{0} + R_{1}(R,\,t) = R_{0}+R_{a}\,\cos{[m\,(\Omega-\Omega_{p})\,t]} \\
\phi & \rightarrow \phi_{0}(t) + \phi_{1}(R,\,t) = \phi_{0}(t) + \phi_{a}\,\cos{[m\,(\Omega-\Omega_{p})\,t]}
\end{align}
the linearized equations of motion become:
\begin{align}
R_{1}(R,\,t) = & {\Bigl (}\frac{f_{0}}{\Delta}{\Bigr )}\,{\cos{[m\,(\Omega-\Omega_{p})\,t]}} \nonumber \\ 
\dot{\phi}_{1}(R,\,t) = & 
-2\,\Omega\,\frac{R_{1}(t)}{R_{0}} - 
\frac{\Phi_{a}}{R^{2}\,(\Omega-\Omega_{p})}\,
\cos{[m\,(\Omega-\Omega_{p})\,t]} 
\label{eqn:eom.epicyclic}
\end{align}
where 
\begin{align}
\label{eqn:f0} f_{0} & \equiv -{\Bigl [}\frac{{\rm d}\Phi_{a}}{{\rm d}R} 
+ \frac{2\,\Omega\,\Phi_{a}}{R\,(\Omega-\Omega_{p})}
{\Bigr ]}_{R} \\ 
\label{eqn:Delta} \Delta & \equiv {\kappa^{2} - m^{2}\,(\Omega-\Omega_{p})^{2}}\ .
\end{align}

When the orbits of particles do not cross, i.e., there are no shocks, the net torque integrated over an orbit is identically zero to first order in $\Phi_1$. 
To see this, note that the instantaneous torque per unit mass $\tau$ is 
\begin{equation}
\tau \equiv -({\bf R}\times \nabla{\Phi})_{z} = -\frac{\partial \Phi}{\partial \phi}
\end{equation}
Integrated over an orbit, the net angular momentum loss is therefore  
\begin{align}
\nonumber \Delta\,J & = \int \tau\,{\rm d}t \\
\nonumber & = -\int_{0}^{2\pi}\,\frac{\partial \Phi}{\partial \phi}\,(\dot{\phi})^{-1}\,{\rm d}\phi\ \\ 
& = -\int_{0}^{2\pi}\,\Phi_{a}(R)\,\frac{\partial f(\phi)}{\partial \phi}\,(\dot{\phi})^{-1}\,{\rm d}\phi\ .  \label{J1}
\end{align}
To first order, we must use $\dot \phi = \dot \phi_0 = \Omega - \Omega_p$, so that 
$\Delta J \approx 
-\Phi_{a}(R)\,(\Omega-\Omega_{p})^{-1} \int_{0}^{2\pi} (\partial f(\phi)/\partial \phi) {\rm d}\phi$, 
which trivially evaluates to zero because $f(\phi)$ is periodic. 

Expanding the orbits to second order does not change this
conclusion -- the net torque is again zero.  \citet{goldreichtremaine:spiral.excitement,
goldreichtremaine:spiral.resonances} and \citet{athanassoula:bar.orbit.morphology,
athanassoula:bar.slowdown} derive this result rigorously.  They show that there can only be a non-zero second-order torque at the resonance points -- where the epicyclic expansion breaks down -- 
or if the perturbation magnitude is growing in time at a rate 
that is large compared to $\Omega-\Omega_{p}$.

The above derivation applies for collisionless particles.   However, gas is collisional, and thus obeys a slightly different dispersion relation.  For systems consisting of both gas and stars -- which are the focus of this paper -- the gas mode is offset from the stellar mode by a small angle $\delta$ due to dissipation.   This gives rise to a torque on the gas by the stars \citep[typically $\delta\sim10-15\, \deg$ in 
simulations;][]{noguchi:merger.induced.bars.gas.forcing,
barneshernquist96,barnes:review,
berentzen:gas.bar.interaction}. 
Following \citet{wada:gas.orbits.in.weak.bars}, the phase shift $\delta$ and the resulting orbit-averaged torque on the gas can be derived in the limit of  weak modes that do not induce orbit crossing or shocks in the gas:
\begin{equation}
\tau = \frac{dJ}{dt} = 
-\frac{m}{2}\,f_{\ast}\, V_c^2\,\left|\frac{\Phi_{a}(R)}{V_{c}^2}\right|^2\,\sin{(\delta)}\,\tilde{B}(R) 
\label{eqn:tau.weak.torques} 
\end{equation}
where $f_{\ast}$ is the stellar mass fraction in the disk,
\begin{align}
\delta & = \tan^{-1}{{\Bigl(} \frac{m\,\kappa\,(\Omega_{p}-\Omega)\,\Lambda}{\Delta} {\Bigr)}}, \\ 
\tilde{B} &= \frac{(2\,[1-\Omega_{p}/\Omega]^{-1} + \partial \ln \Phi_{a}/\partial \ln R)^{2}}
{\Omega^{-2}\,\sqrt{
\Delta^{2} + 
m^{2}\,\Lambda^{2}\,\kappa^{2}\,(\Omega-\Omega_{p})^{2}}},
\end{align}
and $\Lambda\sim0.1$ is a dimensionless constant that parameterizes the dissipation in the gas (formally, $\Lambda$ is the dimensionless constant in a phenomenological `friction' term in the momentum equation that is proportional to the radial velocity; the value $\sim0.1$ is motivated by comparison with hydrodynamic 
simulations). Note that the net torque 
is second-order in the perturbation amplitude $|a| \sim \Phi_a/V_c^2$.  In addition, for a 
bar ($m=2$) mode inside co-rotation, $\delta>0$, 
and the gas mode leads the stellar mode (implying net angular momentum loss by the gas); this reverses outside co-rotation. 
%This can be an important means 
%of re-arranging the stellar angular momentum, but in simulations is usually 
%sub-dominant. 

Since the gas follows nearly circular orbits ($J\approx\Omega\,R^{2}$), 
the orbit-averaged gas inflow rate follows trivially from the torque $\tau = dJ/dt$: 
\begin{equation}
\dot{M}  = 2\pi\,\Sigma_{\rm gas}\,R^{2}\,\Omega\,
\frac{\tau}{V_{c}^{2}\,(1 + {\partial}\ln V_{c} / {\partial}\ln R)}\ . 
\label{eqn:mdot.basic.eqn}
\end{equation}
Note the sign convention: positive $\dot{M}$ is outflow, negative is inflow.
For the simple case of a constant-$V_{c}$ disk with $\Phi_{a}\sim|a|\,V_{c}^{2}$ 
and typical values for $\delta \sim 10^\circ$ calibrated from simulations, equation~\ref{eqn:tau.weak.torques} implies 
\begin{equation}
\dot{M}  \approx - | a |^{2}\,\frac{\alpha\,m\,\Sigma_{\rm gas}\,R^{2}\,\Omega\,} 
{(1-\Omega_{p}/\Omega)^{2}\,|2-m^{2}\,(1-\Omega_{p}/\Omega)^{2}|} 
\label{eqn:mdot.weak.torques}
\end{equation}
where $\alpha\sim2-3$. Far from resonance, this reduces to 
$\dot{M}\sim |a|^{2} \, M_{\rm gas}\,\Omega$.   

In addition to the torque between the gas and stars, non-axisymmetric waves can also directly transport angular momentum (with the coupling at resonances); this yields a qualitatively 
similar torque that is also second-order in $|a|$ 
\citep{lynden-bell:1972.spiral.amplification}. 
\citet{kalnajs:1971} derives an exact solution for the logarithmic spiral, for which the inflow rate can be written 
\be
\dot{M}_{\rm K71} = -|a|^{2}\,\left(\frac{c_{s}}{V_{c}} \right)\,
\frac{\partial |K|^{2}}{\partial |kR|}\,\frac{\pi^{2}\sqrt{3}\,m}{2\sqrt{2}\,Q}\,
\Sigma_{\rm gas}\,R^{2}\,\Omega
\label{eqn:mdot.kalnajs}
\ee
where $K\equiv\Gamma[(m+i\,kR+1/2)/2]/\Gamma[(m+i\,kR+3/2)/2]$ and $k$ is the radial 
wavenumber of the mode.    The resulting torque scales as $|a|^2$ but is smaller than the torque between gas and stars (eq.~\ref{eqn:tau.weak.torques}) 
by two factors:   (1) a term $\sim (c_s/V_c)\,Q^{-1} \sim \Sigma_{\rm gas}/\Sigma_{\rm tot}$, and (2) a pre-factor $\partial |K|^{2}/\partial |kR|$ that depends on the structure of the mode, 
which scales as $\approx |kR|/(1+m^{3})$ 
for $|kR|\ll 1$ or $\sim |kR|^{-2}$ for $|kR|\gg1$.
As a result, the predicted torques become small in both the global and local mode limits. 

\vspace{-0.3cm}
\subsection{Torques \& Inflow: Non-Linear Regime (Orbit Crossing)}
\label{sec:deriv.mom.strong}

\subsubsection{General Criteria}
\label{sec:deriv.mom.strong:overhead}

In the epicyclic approximation of the previous subsection, the perturbation to the orbit of a particle is periodic in $\phi$.   For a gaseous disk, however, this assumption breaks down if the perturbed orbits of particles cross each other, because orbit crossing in gas leads to dissipation (generally shocks) and inflow.  
Such dissipation breaks the periodicity in $\phi$ that leads to the cancellation of the first order (in $\Phi_a \propto a$) contribution to the torque in equation (\ref{J1}).  As a result, in the limit of strong orbit crossing an order of magnitude estimate of the inflow rate is
\be
\dot{M}\sim \frac{-|a|\,m\,\Sigma_{\rm gas}\,R^{2}\,\Omega}{1+{\partial\ln V_{c}}/{\partial \ln R}} \sim -|a|\,\Sigma_{\rm gas}\,R^{2}\,\Omega.
\label{eqn:mdot.strongtorque.simple}
\ee
Equation~\ref{eqn:mdot.strongtorque.simple} will derived more rigorously in what follows. 

\citet{papaloizou:gas.orbit.crossing} derived a useful criterion for when orbits cross.  Consider two quasi-circular orbits at the same $\phi$, with infinitesimal initial radial separation 
$\delta r_i$.  Using the linear solution for each particle (eq. \ref{eqn:eom.epicyclic}), the separation between the particles as a function of time is given by
$\delta r(t) = 
 \delta r_i\ [1 + (\partial  R_{a}  / \partial R_{0})\,\cos{(m\,\phi)} ]$.  Thus if
\begin{equation}
|\zeta| \equiv |\partial  R_{a}  / \partial R_{0} |= 
{\Bigl |}\frac{1}{\Delta}\,{\Bigl (}
\frac{\partial f_{0}}{\partial R} - \frac{f_{0}}{\Delta}\,\frac{\partial \Delta}{\partial R}
  {\Bigr)} {\Bigr |} \gtrsim  1, 
\label{eqn:strong.torques.crit.complex}
\end{equation}
then the orbits cross, i.e., $\delta r < 0$,  at some point (phase) in the orbit.
We caution that this is a sufficient, but not necessary criterion for orbit crossing. 
One can readily derive a similar criterion for when the particles cross in azimuth:  $|\partial \phi_{a}/\partial \phi | > 1$; and, clearly,  if $|R_{a}|>R$ there will be orbit crossing.   Near-resonances, one can have orbit-trapping (where the gas librates about 
a specific radius), which will lead to shocks, for 
$|\partial R_{a}/\partial R|$ as small as $\sim0.05-0.1$ 
\citep[see][]{binneytremaine}. 
In general, even for linear modes, there is no purely local criterion for orbit crossing -- 
one of the above conditions must be satisfied {\em somewhere} for 
orbit crossings to occur, but if so, it can in principle occur anywhere.   A full census requires calculating the mode structure everywhere and the resulting particle orbits.   Nevertheless we can make considerable analytic progress in several limits.

\vspace{-0.4cm}
\subsubsection{Strong Orbit-Crossing Limit}
\label{sec:deriv.mom.inflow:strong}

We now derive the net torque and inflow rate in the limit $|\zeta| \gg 1$, in which the
gas experiences strong orbit crossings and shocks.  Prior to equation~\ref{eqn:strong.torques.crit.complex}, we derived the pericentric separation of two
perturbed particles at a given $\phi$.   Generalizing this result, the separation after moving from 
some arbitrary $\phi_{i}$ to a final $\phi_{f}$ will be given by 
$\delta r(t)=  \delta r_i [1+(\partial R_{a}/\partial R_{0})\,(f(m\,\phi_{f})-f(m\,\phi_{i}))]$, 
where $R_{1}=R_{a}\,f(m\,\phi)$ represents the epicyclic expansion 
solution, i.e., $f(m\,\phi) = \cos(m\,\phi)$.    Because $\zeta \gg 1$, there is orbit crossing at a range of $\phi$.   We are thus not in a regime in which the epicyclic approximation 
is valid in a global sense; however, we can consider it locally valid for a small amount of time ("between shocks").   
%This is because gas, being collisional, cannot support strongly 
%non-circular orbits, but will instead move on quasi-circular orbits -- that we describe using the epicyclic %approximation -- with nearly continuous shocking.   
Globally, the gas orbits may be quite non-circular but we can approximate this motion with a series of small epicyclic `steps' separated by shocks that dissipate angular momentum and energy, allowing the gas to flow in.  

Given an initial separation of $\delta r_i$ at $\phi_i$, the orbits will cross at a final $\phi_f$
such that $f(m\,\phi_{f}) -f(m\,\phi_{i}) = -1/\zeta$.   We assume that  dissipation is efficient when the gas streams cross and shock because the cooling time of the gas is always short under the conditions of interest.   As a result, the colliding streams will dissipate the kinetic energy of their relative bulk motion.
This effectively ``resets'' the orbit to a circular orbit with the appropriate total angular momentum per unit mass of the streams at that time.   We use this approximation to estimate the angular momentum loss due to the non-axisymmetric torques; the same results could be derived using energy considerations.

The angular momentum per unit mass in the rest-frame is $j=R^{2}\,(\dot{\phi}+\Omega_{p}) = (R_{0}+R_{1})^{2}\,(\dot{\phi}_{0}+\dot{\phi}_{1}+\Omega_{p})$ (recall that $\phi$ is defined in the frame rotating at $\Omega_{p}$).   Expanding to first order, the change in angular momentum of the streams from $\phi_{i}$ to $\phi_{f}$ is 
$\Delta j \equiv j_{f}-j_{i} \approx R_{0}\,
(2\,R_{a}\,\Omega +R_{0}\,\dot{\phi}_{a})\,(f[m\,\phi_{f}]-f[m\,\phi_{i}])$ where we have used $R_1 = R_a f(m\,\phi)$ and $\dot \phi_1 = \dot \phi_a f(m\,\phi)$.
Using equation~\ref{eqn:eom.epicyclic} and the condition for when the streams shock ($f[m\,\phi_{f}]-f[m\,\phi_{i}]=-1/\zeta$), we find
\begin{align}
\Delta j &= -\frac{R\,[2\,R_{a}\,(\dot{\phi}_{0}+\Omega_{p})+
R_{0}\,\dot{\phi}_{a}]}{\zeta} \\ 
& = \frac{\Phi_{a}}{\Omega-\Omega_{p}}\,\frac{1}{\zeta}
\label{eqn:dj}
\end{align}
The time required to move in azimuth by $\Delta \phi = \phi_f - \phi_i$ is given by
\begin{equation}
\Delta t = \frac{\Delta\,\phi}{|\Omega - \Omega_{p}|} = 
\frac{f^{-1}(f(m\,\phi_i)-\zeta^{-1}) - m\,\phi_{i}}{m\,|\Omega-\Omega_{p}|} 
\label{eqn:delta.t.collision}
\end{equation}
Note that because the $f(m\,\phi)$ term ranges from $-1$ to $+1$, 
the sign of $\zeta = \partial R_{a}/\partial R_{0}$ in the above simply amounts 
to a phase offset in the shock location, so 
the above can be generalized by the replacement $\zeta \rightarrow |\zeta|$.

The average rate of angular momentum loss is simply given by $\Delta j/\Delta t$. Using the above solutions for $\Delta t(\phi_{i})$ and $\Delta j$, this can be written as a sum over the ``number of shocks'' encountered in a circuit of $\Delta\phi=2\pi$.   In the $|\zeta|\gg1$ limit, we write this as continuous average over $\phi$, with $\Delta j/\Delta t \rightarrow \langle dj/dt \rangle = \tau$ and
\be
\langle \tau \rangle = \frac{1}{2 \pi}\int{\frac{\Delta j}{\Delta t[\phi_{i}]}\,d\phi_{i}}
\ee
where $\langle \tau \rangle$ is the azimuthally averaged torque.
The general expression for $\tau$ cannot be written in closed form, but the leading-order term in $\zeta^{-1}$ can be used to derive the following expressions for the torque $\tau$, inflow rate $\dot M$, and radial velocity $V_R$:
\begin{align} 
\langle \tau \rangle & = \frac{m}{2\pi}\,\Phi_{a}\,F(\zeta)\,{\rm sign}(\Omega-\Omega_{p}) \label{tau} \\ 
\dot{M} & = (\Sigma_{\rm gas}\,R^{2}\,\Omega)\,{\Bigl [}\frac{\Phi_{a}}{V_{c}^{2}} {\Bigr]}\,
\label{eqn:mdot.strong} 
\frac{m\,{\rm sign}(\Omega-\Omega_{p})}{(1+{\partial\ln V_{c}}/{\partial \ln R})}\,F(\zeta) \\ 
{\Bigl \langle}V_R {\Bigr \rangle}& 
= V_{c}\,{\Bigl [}\frac{\Phi_{a}}{V_{c}^{2}} {\Bigr]} \,
\frac{m\,{\rm sign}(\Omega-\Omega_{p})}{(1+{\partial\ln V_{c}}/{\partial \ln R})}\,F(\zeta) \\ 
F(\zeta) &\approx \left[2 - \frac{1}{2\,|\zeta|} + 
\mathcal{O}(|\zeta|^{-2})\, \right]\,{\rm sign}({\zeta})\ \ \ \ (|\zeta|\gg1) \label{defF}
\end{align}
Note that the net torque and inflow rate are independent of $\zeta$ for $\zeta \gg 1$.   They depend primarily on the magnitude of the non-axisymmetric potential $\Phi_a$.    This is because although the change in angular momentum per shock scales as $\zeta^{-1}$ (eq. \ref{eqn:dj}), the time between shocks itself  also scales as $\zeta^{-1}$ (this follows from Taylor expanding eq. \ref{eqn:delta.t.collision} and averaging over $\phi_i$).   Another way to say this is that there are $N \sim \zeta$ shocks each of which resets the gas to a circular orbit with $\Delta j\propto1/\zeta$, so that the net torque is independent of $\zeta$.  
%If, instead, the shocks all occurred in a single $\phi$ (arbitrarily thin region), one can show that  $F(\zeta)\sim1$, which happens to agree with equation~\ref{defF} as $\zeta\rightarrow1$. 
%It is also possible to solve the above 
%for $|\zeta|=1$ exactly (the limit of a razor-thin shock region), in which case $F(\zeta)=1$.
%(in agreement with the result if we extend the series to higher-order in $|\zeta|$).  
%{\em include only one of these, but this also suggests we could write 
%$F(\zeta)$ for this regime as 
%\be 
%F(\zeta)=2\,{\rm sign}(\zeta)\,\left[ 1 - \frac{1}{4\,(|\zeta|-1/2)} \right ]
%\ee
%motivated by a series sum that returns the appropriate $F(\zeta)=1$ for $\zeta=1$; of course, 
%in that regime, there is a lot more than a razor-thin orbit crossing because 
%of the points discussed below for marginal orbit crossing, so we dont necessarily need 
%to mention any of this.}

Dimensionally, given $\Phi_{a}\sim|a|\,V_{c}^{2}$, equation \ref{eqn:mdot.strong}  reduces to 
$\dot{M}\sim -|a|\,\Sigma_{\rm gas}\,R^2\,\Omega$, in agreement with our order of magnitude estimate in equation~\ref{eqn:mdot.strongtorque.simple}.    Thus in the strong-torque regime 
the inflow rates are {\em linear} in $|a|$:  when orbit crossings are important, gas inflow can easily be enhanced by an order of magnitude or more relative to the expectations from the weak-torque theory of \S \ref{sec:deriv.mom.weak}.

The sign() terms in equations \ref{tau}-\ref{defF} are important and arise from terms that in the full solution are continuous; i.e., $\zeta/\sqrt{1+\zeta^{2}}\rightarrow {\rm sign}({\zeta})$ for $|\zeta| \gg 1$.   Recall that with our sign convention, $\Phi_{a}$ is negative where $\Sigma_{a}$ is positive. For typical parameter values inside of co-rotation ($\Omega>\Omega_{p}$), 
$\zeta$ is negative and therefore $\dot{M}$ is also negative (inflow), whereas outside of 
corotation it is positive (outflow). This dependence on corotation is essentially a question of whether the shocks are in prograde or retrograde motion in the pattern frame, which changes at the corotation resonance.  The dependence on ${\rm sign}(\zeta)$ is due to the phase between the shocks and the surface density perturbation: when $\zeta<0$, shocks occur where $f(m\,\phi)-f(m\,\phi_{i})$ is maximized -- in phase with the positive maximum in $\Sigma_{1}$ -- and hence where (for $\Omega>\Omega_{p}$) $\Delta j$ is positive.  This generically leads to outflow. But when $\zeta<0$, the shocks occur leading the mode, where $\Delta j$ is negative and minimized with respect to the unperturbed orbit; this leads to inflow.

If instead of using real variables, we generalize to complex $\Phi_{a}$ and $\zeta$, a nearly identical derivation to that presented here leads to $\Phi_{a}\,{\rm sign}(\zeta) \rightarrow |\Phi_{a}|\,\cos{({\rm arg}[\Phi_{a}\,\zeta^{\ast} ] )}$, i.e., $|\Phi_{a}|\,\cos{\tilde{\phi}}$, where $\tilde{\phi}$ is the phase angle between $\zeta$ and $\Phi_{a}$; this can have 
an in-phase inflow and/or out-of-phase outflow term. Likewise, the ${\rm sign}(\Omega-\Omega_{p})$ term can be generalized  for complex $\omega=m\,\Omega_{p}+i\,\gamma$, by taking ${\rm sign}(\Omega-\Omega_{p})\rightarrow {\rm Re}[m\,|\Omega-\Omega_{p}|/(m\,\Omega-\omega)]$.

\vspace{-0.4cm}
\subsubsection{Marginal Orbit-Crossing Limit}
\label{sec:deriv.mom.inflow:marginal}

The derivation in \S \ref{sec:deriv.mom.inflow:strong} is only valid when 
$|\zeta| \gg 1$.   Indeed, when $|\zeta|\lesssim1$, the quantities used in these derivation are undefined, 
even though it is still very much possible that there are shocks and even orbit crossings (as discussed 
in \S~\ref{sec:deriv.mom.strong:overhead}).    Moreover, the requirement of formal 
orbit crossings for strong torques and shocks is a consequence of the fact that we have thus far (implicitly) considered only infinitely cold disks.   There can, in fact, be shocks when $|\zeta|\lesssim1$, even without orbit crossings, if the non-axisymmetric potential is dominated by the stars rather than the gas.   We now estimate the resulting torques and inflow rate in this limit.  

If the disk is not gas-dominated, then the gas orbits are given roughly by the standard 
equations of motion in response to the stellar potential.  Given radial variations in some property of the gas, the length-scale $h_{\rm eff}$ at which pressure forces become important is given by $h_{\rm eff}=2\,c_{\rm eff}^{2}/G\,\Sigma$ \citep[e.g.,][]{toomre:Q} where $c_{\rm eff}$ is the effective dispersion of the gas, including both the true thermal sound speed and any non-thermal turbulent motions:  $c_{\rm eff}^{2}=c_{s}^{2}+v_{t}^{2}$.  Note that in a spherical potential, $\heff$ is different from the 
vertical scale height $h$ by $\sim 2\, c_{\rm eff}\,\Omega/G\,\Sigma\sim Q$, the Toomre Q parameter (defined using $c_{\rm eff}$ not $c_s$).    The stellar motions do not explicitly appear in the effective sound speed here, because the key criteria is whether {\it  gas} shocks; $\Sigma$ does, however, represent the mass density of gas plus disk stars since both participate in the mode self-gravity.
%For a gaseous disk of scale height $h$, pressure forces will begin to
%dominate gravitational forces on length scales $\lesssim h_{\rm eff} \sim h$ 
%where $h_{\rm eff}$ characterizes the importance of pressure 
%forces for motions in the orbital plane (while the standard scale-height $h$ is the vertical thickness).   

If stellar streamlines approach to within a distance $\lesssim h_{\rm eff}$, 
they remain collisionless and conserve energy.   But if gas that is initially spatially separated is forced into a {\em supersonic} `near-collision' within a distance $\lesssim \heff$ by the potential of the stars, it will shock and radiate the difference in energy between the streamlines (conserving only their 
total momentum).    The key point is that when the non-axisymmetric 
potential is stellar dominated, gas ``collisions'' can happen not just 
at strict orbit crossings, but at a finite separation $\sim \heff$ between streamlines.

If two gas streamlines approach one another at a relative velocity that is much smaller than the sound speed $c_s$,  pressure effects can readjust the orbits without much dissipation.   
Thus significant dissipation probably requires relative stream velocities $>c_{s}$. 
What are the relative velocities?   For radial motions,
the maximum magnitude of the relative velocity over an orbit, 
from the epicyclic solution, is 
$\Delta v = |\partial v_{1,R}/\partial R|\,\delta r_i = [h_{\rm eff}/(1-|\zeta|)]\,|\partial \dot{R}_{1}/\partial R|$, where in the second expression we have used the maximum initial separation streamlines can have such that they are separated by $\heff$ at some point in the orbit, $\delta r_i = \heff/(1-|\zeta|)$.
%For radial perturbations in a disk, the length-scale $h_{\rm eff}$ at which pressure 
%forces become important is given by $h_{\rm eff}=2\,c_{\rm eff}^{2}/G\,\Sigma$ \citep[e.g.,][]{toomre:Q} 
%where $c_{\rm eff}$ is the effective dispersion of the gas, including both the true thermal sound speed and 
%any non-thermal turbulent motions:  $c_{\rm eff}^{2}=c_{s}^{2}+v_{t}^{2}$.  
%Note that in a spherical potential, $\heff$ is different from the 
%vertical scale height $h$ by $\sim 2\, c_{\rm eff}\,\Omega/G\,\Sigma\sim Q$, the 
%Toomre Q parameter (defined using $c_{\rm eff}$ not $c_s$).
The resulting criterion for gas streams to approach within a distance $\sim \heff$ with supersonic relative velocities $\Delta v > c_s$ is given by
\be
{\left|\frac{1}{\kappa}\, \frac{\partial v_{1,\,R}}{\partial R}  \right|} \,\left(\frac{c_{\rm eff}}{c_{s}}\right)
\gtrsim \frac{{1-|\zeta|}}{2\pi\,Q}.
\label{eqn:shock.criterion}
\ee
Since  $v_{1,R}$ is linear in the mode amplitude $|a|$, we can define 
$|{\partial v_{1,\,R}}/{\partial R}  |_{a=1}$ as the magnitude of 
${\partial v_{1,\,R}}/{\partial R}$ for $|a|=1$, and write equation~\ref{eqn:shock.criterion} as a condition on the mode amplitude
\be
|a| \gtrsim \frac{(1-|\zeta|)}{2\pi\,Q}\,\left(\frac{c_{s}}{c_{\rm eff}}\right) \, 
\left[ \kappa^{-2}\,\left( \left| \frac{\partial v_{1,\,R}}{\partial R} \right|_{a=1}^{2} +   
\left| \frac{\partial v_{1,\,\phi}}{\partial R} \right|_{a=1}^{2}
\right) \right]^{-1/2}
\label{eqn:shock.criterion.alt}
\ee
where we have also included the relative azimuthal velocity $v_{1,\,\phi}$ for completeness.

In equations \ref{eqn:shock.criterion} and \ref{eqn:shock.criterion.alt} we have allowed for the possibility that the disk has non-thermal random velocities via the effective sound speed $c_{\rm eff}$, which is also in our definition of $Q$.   One subtlety is whether the condition for effectively dissipating the non-axisymmetric motions is a relative velocity $\Delta v > c_{\rm eff}$ or $\Delta v > c_s$.  We suspect that the latter is appropriate.   When $c_{\rm eff} > \Delta v > c_{s}$, the energy per unit mass dissipated via non-axisymmetric motions is small compared to the overall rate at which supersonic turbulence is dissipating 
in the ISM (the latter being due to, e.g., star formation).
%\footnote{{\bf  Note that in sufficiently gas-dominated systems, if $c_{\rm eff}$ is a turbulent velocity and 
%$c_{\rm eff}\gtrsim \Delta v$, then those turbulent motions will lead to a cascade of 
%mode power into a wide range of modenumbers and the angular 
%momentum transport will be dominated by 
%small-scale shocks (the ``gravoturbulent regime''; 
%see e.g.\ \citealt{gammie:2001.cooling.in.keplerian.disks} 
%and \citealt{wada:starburst.torus.model,tasker:2008.gas.turb.vs.gal.prop,
%agertz:disk.clumps.drive.turbulence} for examples in galactic systems). However, 
%by definition this regime is outside of our regime in which the 
%modes induce large-scale shocks (as it should be), so our previously derived scalings 
%should not be applied. Otherwise, $c_{\rm eff}\gtrsim c_{s}$ is allowed, so 
%long as the gaseous system can still meaningfully be described as ``collisional.''
%}}  
Nonetheless, the ordered velocities induced by the non-axisymmetric potential can be efficiently dissipated if $\Delta v > c_s$; {we confirm this below in simulations that have resolved 
turbulent $c_{\rm eff} > c_{s}$.}

Equations~\ref{eqn:shock.criterion} and \ref{eqn:shock.criterion.alt} 
cannot be satisfied for $|a| < 1$ in a sufficiently cold disk with $Q\ll1$ and $|\zeta|\lesssim1$. 
In this limit, the condition for shocks and orbit crossing 
reduces to our previous criterion, $|\zeta|>1$.   
However, for larger $Q$, the effective
``thickness'' of a given gas stream increases and so for the same non-axisymmetric perturbation, 
it is easier to generate collisions of streamlines.   Of course, for sufficiently 
large $Q$ our assumption that the gas orbits are well described 
by test particles responding to the stellar potential will break down.   
But for the particularly relevant case of $Q \gtrsim 1$ and $c_{\rm eff} > c_s$, 
equations~\ref{eqn:shock.criterion} and \ref{eqn:shock.criterion.alt} show 
that shocks induced in the gas by the stellar potential are possible even 
though $|\zeta| \lesssim 1$.   For typical modes (say, $m=1$ modes in Keplerian 
potentials or bars in constant-$V_{c}$ disks), 
at the order of magnitude level, 
this condition on the mode amplitude becomes $|a| \gtrsim 0.2 \, Q^{-1} \, (c_s/c_{\rm eff})$.    
We reiterate that this is only true when the gravity of the perturbation is dominated 
by stars.   For gas dominated systems, the mode structure would already 
include pressure effects, invalidating our analysis.   The exact gas fraction at which this transition occurs 
is sensitive to the gas equation of state and phase structure; it is thus difficult to 
estimate precisely, beyond the order of magnitude estimate $M_{\rm gas}\gtrsim M_{\ast}$.

We now generalize the derivation in \S \ref{sec:deriv.mom.inflow:strong} to determine the inflow rate and torques on the gas in the $|\zeta| \lesssim 1$ limit.  We assume that both $\zeta$ and $\heff/R$ are small, so that we can Taylor expand in both quantities.  The separation between two streamlines is given by
$\delta r(\phi)=\delta r_i\,(1+\zeta\,[f(m\,\phi)-f(m\phi_i)])$ 
Gas that is initially within a maximum separation $\delta r_i=\delta R_{\rm max}\equiv h_{\rm eff}/(1-|\zeta|) \simeq \heff(1 + |\zeta|)$ will collide at some point in the orbit.  However, streamlines with initial separations less than a minimum value $\delta R_{\rm min}$ are never differentially accelerated by the perturbed potential to $\Delta v \sim c_s$ and thus never experience shocks.   The exact value of $\delta R_{\rm min}$ depends on $|a|$, $\zeta$, $|{\partial v_{1,\,R}}/{\partial R}  |_{a=1}$, etc.   As a rough estimate, we take $\delta R_{\rm min} \sim h_{\rm eff}$ because gas initially separated by $\delta R_{\rm min}\lesssim \heff$ is strongly affected by internal pressure forces and is unlikely to be accelerated to relative velocities $\gtrsim c_s$.      For a given $\delta r_i$ satisfying $\delta R_{\rm min}<\delta r_i < \delta R_{\rm max}$, the streamlines reach $\delta r < \heff$ (i.e., `collide') when $f(m\phi) = f(m\phi_i) + \zeta^{-1}(\heff/\delta r_i - 1)$.  For $\delta r_i = R_{\rm min}$, this corresponds to $f(m\phi) = f(m\phi_i)$ and $\Delta \phi = 0$, while for $\delta r_i = R_{\rm max}$, the shocks occur at $f(m\phi) = f(m\phi_i) \pm 1$ and $\Delta \phi = \pm (\pi/2m)$.   When the gas reaches the $\phi$ such that $\delta r(\phi)<h_{\rm eff}$, it dissipates its relative orbital energy but conserves momentum.    From \S \ref{sec:deriv.mom.inflow:strong}, the change in specific angular momentum $\Delta j$ for streamlines satisfying $\delta R_{\rm min}<\delta r_i < \delta R_{\rm max}$ is given by 
\be
\Delta j = -\frac{\Phi_a}{\Omega - \Omega_p}\frac{1}{\zeta}\left(\frac{h_{\rm eff}}{\delta r_i} - 1 \right).
\label{dj}
\ee
The average change in angular momentum over all initial streamlines separations is given by
\be
\langle \Delta j \rangle \simeq \frac{1}{\heff |\zeta|} \, \int_{\heff}^{\heff(1+|\zeta|)}{\Delta j} \ d(\delta r_i) \simeq \frac{1}{2}\,\frac{\Phi_a}{\Omega - \Omega_p} \, {\rm sign(\zeta)}
\label{djavg}
\ee
where we have used $R_{\rm min} = \heff$ and $R_{\rm max} = \heff(1 + |\zeta|)$ and in the second equality have Taylor expanded for $\zeta \ll 1$.   Note that $\heff$ has cancelled out in our result for $\langle \Delta j \rangle$ -- the value of the stream thickness does not influence how much angular momentum is lost. 

The integral over $\delta r_i$ in equation~\ref{djavg} is equivalent to an integral over the $\phi$ of the shocks.   Thus the average over $\delta r_i$ is equivalent to an average over $\phi$ and/or time. 
In particular, the range $\delta r_i = R_{\rm min}$ to $R_{\rm max}$ corresponds to $f(m\phi) = 0$ to $\pm1$.   This is equivalent to an integral over time for a stream circuit of 
$\phi=2\pi/m$; which takes a time $\Delta t = 2\pi/m\,|\Omega-\Omega_{p}|$. 
To leading order, we therefore obtain 
\begin{align}
\langle \tau \rangle &= \frac{m}{2\pi}\,\Phi_{a}\,F(\zeta)\,{\rm sign}(\Omega-\Omega_{p})\, \label{torque.low.zeta}\\ 
\dot{M} & = (\Sigma_{\rm gas}\,R^{2}\,\Omega)\,{\Bigl [}\frac{\Phi_{a}}{V_{c}^{2}} {\Bigr]}\,
\frac{m\,{\rm sign}(\Omega-\Omega_{p})}{(1+{\partial\ln V_{c}}/{\partial \ln R})}\,F(\zeta) \\ 
%F(\zeta) &\approx {\rm sign}(\zeta)\,\left[\frac{1}{2} - \frac{7}{6}|\zeta|+\frac{17}{12}\,\zeta^{2} \right]
%F(\zeta) &= {\rm sign}(\zeta)\,
%\zeta^{-2}\,\left[ 
%|\zeta| + (1-|\zeta|)\,\ln{[1-|\zeta|]} \right] \label{F.low.zeta}\\
F(\zeta) &\approx {\rm sign}(\zeta) \left(\frac{1}{2} + \mathcal{O}(|\zeta|^{-1}) \, \right)  \label{F.low.zeta}
%&\nonumber \approx {\rm sign}(\zeta)\,\left[\frac{1}{2} + \frac{|\zeta|}{6} + \frac{\zeta^{2}}{12}\right]
\end{align}
%(where the higher-order terms in $\zeta$ come from the analytic solution 
%keeping all terms in our previous equation). 
Note that equations~\ref{torque.low.zeta}-\ref{F.low.zeta} in the low $\zeta$ limit are comparable to equations~\ref{tau}-\ref{defF} derived in the $\zeta \gg 1$ limit (to a factor of $\sim 2$), subject to the additional restriction that equation~\ref{eqn:shock.criterion.alt} for the mode amplitude $|a|$ (and thus $\zeta$) must be satisfied in order for there to be shocks in the low $\zeta$ limit.  

\vspace{-0.3cm}
\subsubsection{Combined Scaling}
\label{sec:deriv.mom.inflow:all}

For $|\zeta|\approx1$, we can reasonably interpolate between the two regimes considered in the previous two subsections ($|\zeta|\ll 1$ and $|\zeta|\gg 1$) to find:
\begin{align}
\label{eqn:mdot.full}
\dot{M} & = (\Sigma_{\rm gas}\,R^{2}\,\Omega)\,{\Bigl |}\frac{\Phi_{a}}{V_{c}^{2}} {\Bigr |}\,
\frac{m\,S(\omega,\,\Phi_{1})\,F(\zeta)}{1+{\partial\ln V_{c}}/{\partial \ln R}} \\ 
S&(\omega,\,\Phi_{1}) = {\rm Re}\left[\frac{m\,|\Omega-\Omega_{p}|}{m\,\Omega-\omega} \right]\,
\cos{({\rm arg}[\Phi_{a}\,\zeta^{\ast}])} \label{phase} \\ 
F&(\zeta) \approx 
%\frac{1}{2} + \frac{|\zeta|}{6+2\,|\zeta|} + \frac{|\zeta|^{3}}{3+2\,|\zeta|^{4}}
\frac{1}{2} + \frac{|\zeta|}{6+2\,|\zeta|/3} 
+ \frac{|\zeta|^{2}}{1+(|\zeta|+9\,|\zeta|^{2}+|\zeta|^{3})/13}\ .
\end{align}
Recall again that in the low $\zeta$ limit, these expressions require that equation~\ref{eqn:shock.criterion.alt} for the mode amplitude $|a|$ be satisfied. 

Note some critical features of these results. The inflow rate $\dot M$ is linear in $|a|$ and thus, as advertised, shocks induced in the gas by the non-axisymmetric stellar potential (when present) will dominate the total angular momentum loss relative to the direct transport by the gaseous mode itself.
We collect the phase information in the term $S$ in equation~\ref{phase}
($S<0$ giving inflow, $S>0$ outflow).   If the modes are purely real and global
$S \rightarrow {\rm sign}(\Omega-\Omega_{p})\,{\rm sign}(\Phi_{a})\,{\rm sign}(\zeta)$.   The generalization of $S$ in equation~\ref{phase} allows for modes with 
radially varying phases, i.e.\ $\Phi_{a}\propto \exp{(i\,\int k\,dr)}$, as in \S \ref{sec:deriv.mom.inflow:strong}.   It is easy to show that  for a mode in a galaxy with typical structural properties, there is generally inflow 
induced inside of co-rotation, and outflow outside 
(for ${\rm sign}(\Phi_{a})<0$, then ${\rm sign}(\zeta)>0$ and 
${\rm sign}(\Omega-\Omega_{p})$ changes from positive to negative 
from inside to outside corotation); this is consistent with the shock-free, weak-torque results in \S \ref{sec:deriv.mom.weak}.   Also for typical structural parameters, the phase
changes at the Lindblad resonances, producing outflow inside of the ILR and inflow outside of the OLR.

\begin{footnotesize}
  \ctable[caption={{\normalsize Overview of the Simulations}
  \label{tbl:sim.summary}},center,star ]{lcccl}{
    \tnote[ ]{List of SPH simulations used to test our analytic predictions.  From left to right, columns summarize:
    {\bf (1)} Simulation ``class,'' each of which includes $\sim100$ individual 
    simulations, with different  structural properties (mass profiles, 
    kinematics) of the initial disks, bulge-to-disk ratios, 
    BH mass, gas fraction, baryonic mass, orbital parameters (in mergers), 
    and efficiency of stellar feedback and star formation.  Particle numbers range from $\sim 10^5-10^7$.
    {\bf (2)} Approximate range of spatial scales spanned by each simulation class.
    {\bf (3)} Corresponding total range in baryonic mass.   The particle mass is $\sim 10^{-6}$ times 
    smaller.  {\bf (4)} Typical spatial resolution/softening length 
    (often varied with galaxy scale length). 
    {\bf (5)} Reference(s) for the details of each simulation class. 
    Particular tables and figures in the original references that summarize the simulation properties are noted. % \\
    } 
%\tnote[a]{stuff}
    }{ \hline\hline 
    \multicolumn{1}{c}{Simulations} &
    \multicolumn{1}{c}{Spatial Scale} &
    \multicolumn{1}{c}{Mass Scale [$\msun$]} &
    \multicolumn{1}{c}{$\epsilon$ [pc]} & 
    \multicolumn{1}{c}{References} \\
    \hline
%\multicolumn{5}{c}{stuff} \\ 
%\hline
{\bf Major Mergers} & $0.1-100\,{\rm kpc}$ & $10^{9}-10^{12}$ & $10-100$ 
& \citet{cox:kinematics}, Tables~1-2; \citet{robertson:fp}, Table~1 \\ 
{\  } & \  & \  & \  
& \citet{hopkins:disk.survival}, Table~1 \&\ Fig.~16 \\ 
{\bf Minor Mergers} & $0.1-100\,{\rm kpc}$ & $10^{9}-10^{12}$ & $10-100$ 
& {\citet{younger:minor.mergers}, Tables~1-2} \\ 
{\bf Isolated Disks} & $0.1-10\,{\rm kpc}$ & $10^{9}-10^{12}$ & $\sim30$ 
& {\citet{younger:minor.mergers}, Tables~3-4; 
\citet{cox:massratio.starbursts}} \\ 
{\bf Intermediate-scale Re-simulations} & $10-1000\,{\rm pc}$ & $10^{7}-10^{10}$ & $\sim1$ 
& {\citet{hopkins:zoom.sims}, Tables~1-2} \\ 
{\bf Nuclear-scale Re-simulations} & $0.1-100\,{\rm pc}$ & $10^{6}-10^{8}$ & $\sim0.1$ 
& {{\citet{hopkins:zoom.sims}, Table~3}} \\ 
\hline\hline\\
}
\end{footnotesize}

\vspace{-0.5cm}
\section{Equilibrium Gas Densities}
\label{sec:deriv.gasden}

  \begin{figure*}
    \centering
    \scaleup
    \plotside{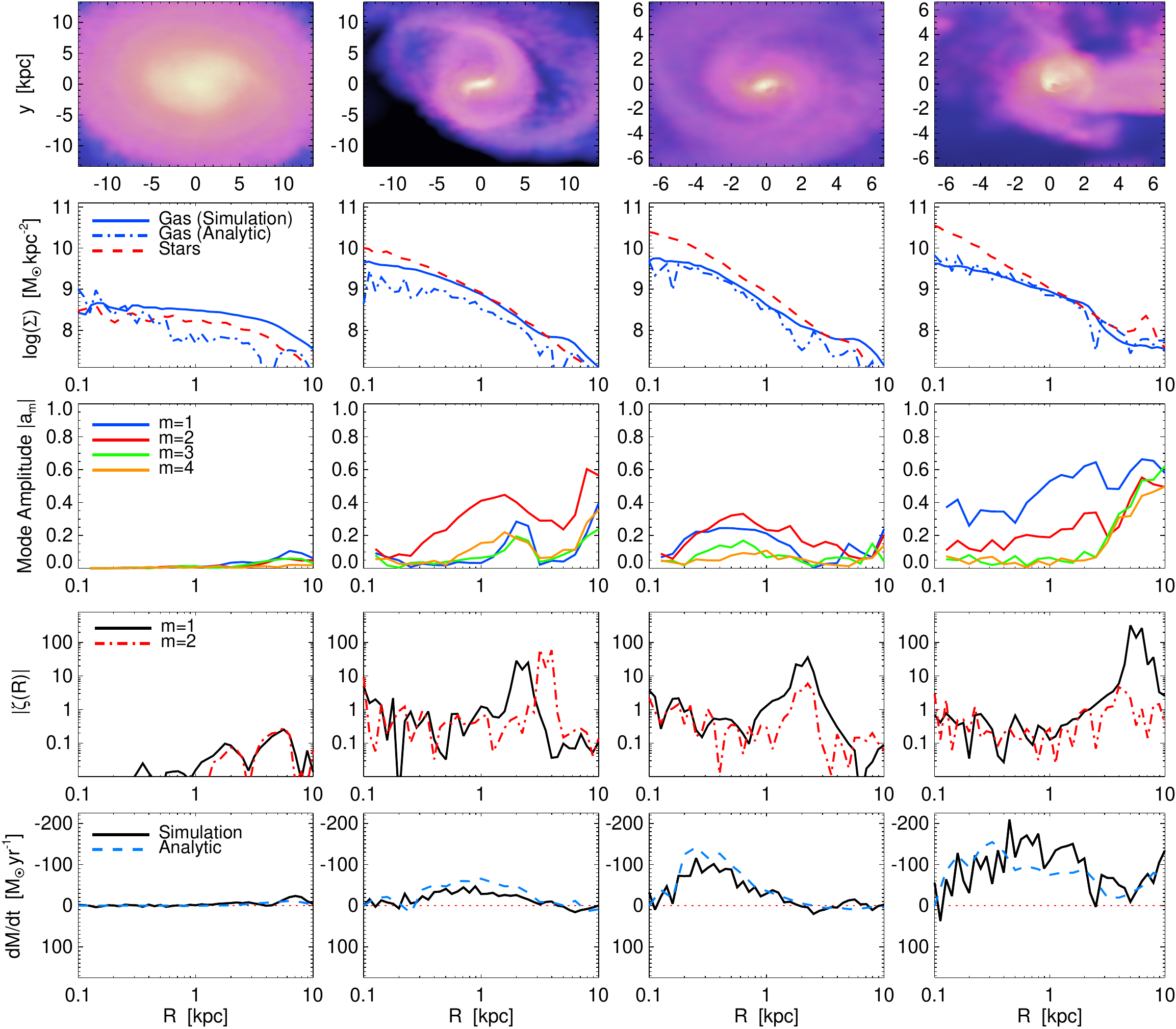}
    %\plotone{f1.pdf}
    \caption{Inflow properties in a typical gas-rich ($f_{\rm gas}\approx0.4$) merger 
    of two equal-mass $\sim L_{\ast}$ ($M_{\ast}=10^{11}\,\msun$) galaxies (run {b3ex} in \citealt{hopkins:zoom.sims}), 
    at different times: 
    before any interaction, when the system is unperturbed ({\em left}); 
    just after an early first passage ({\em center left}),  which induces a bar response 
    in the primary; somewhat later, just before second passage,  as the induced bar 
    feature has relaxed ({\em center right}); just 
    after final coalescence ({\em right}).   {\em Top:} Images of the gas density 
    at each time. Intensity reflects the gas surface density. Color reflects the 
    specific star formation rate (from blue being low, with 
    gas consumption timescale $\gg$Gyr to yellow being 
    high, with consumption timescale $\sim10^{7}-10^{8}\,$yr). Horizontal axis is cartesian ($x$, in kpc). 
    {\em Second from Top:}  Gas and stellar surface density profiles as a function of cylindrical radius $R$.
    We also show the analytic gas density profile from eq.~\ref{eqn:sgas.a}, 
    which assumes that gas inflow approximately balances star formation 
    at each radius (see \S~\ref{sec:deriv.gasden}).    
    {\em Middle:} Amplitude of non-axisymmetric 
    modes versus radius.
    {\em Second from Bottom:} Orbit-crossing measure $|\zeta(R)|\equiv |\partial R_{1}/\partial R|$ 
    (eq~\ref{eqn:strong.torques.crit.complex}); $|\zeta|>1$ implies strict orbit crossing, but 
    shocks are still present at many radii with $|\zeta|\lesssim1$. 
    {\em Bottom:} Gas inflow (negative) or outflow (positive) rate as a function of radius. 
    Dotted red line shows $\dot{M}_{\rm gas}=0$.  We compare the simulation 
    results to our analytic prediction based on strong shocks induced in the 
    gas by the non-axisymmetric stellar potential
        (\S~\ref{sec:deriv.mom.inflow:all}). 
        The predictions from the no-shock, weak-torque limit (eq~\ref{eqn:mdot.weak.torques}) or the direct transport  by the mode in the gas (eq~\ref{eqn:mdot.kalnajs}) are smaller by a factor $>10$, and 
        are would be indistinguishable from the $\dot{M}=0$ line. 
    \label{fig:illustrate.merger}}
\end{figure*}

In \S \ref{sec:response}, we described the physics that sets the inflow rate of gas due to gravitational torques.   Statistically, the {\em net} inflow rate through a given radius in a galaxy is set by a competition between these torques and star formation.    This competition determines the net inflow rate at very small radii (e.g., onto a central BH) and the resulting gas surface density profile.

According to the observed Schmidt-Kennicutt law, the average star formation rate surface density is given by
\begin{align}
\dot{\Sigma}_{\ast} = - \dot{\Sigma}_{\rm gas,\,\ast} = \frac{\Sigma_{K}}{t_{K}}\,
{\Bigl (} \frac{\Sigma}{\Sigma_{K}} {\Bigr)}^{\eta_{K}}\ .
\label{eqn:ks.law}
\end{align}
where the constants $\eta_K \sim 3/2-7/4$, $\Sigma_K$ and $t_K$ are discussed below.   In the limit of a very small supply of gas, the non-axisymmetries in the potential will in general be modest and the inflow of gas will not be able to compete with star formation, which will consume most of the gas.  In the opposite limit of a large supply of gas, however, the inflow rate approaches a linear function of $\Sigma_{\rm gas}$, while the star formation rate is a nonlinear function.   Thus the gas will be consumed rapidly until the inflow and star formation reach an approximate, time-averaged  equilibrium.  We will consider this in more detail in \S \ref{sec:unified:new} but for now note that inflow and star formation balance each other at a quasi-steady gas density $\langle \Sigma_{\rm gas} \rangle$ given by
\begin{equation}
\langle \Sigma_{\rm gas} \rangle \sim \Sigma_{K}\,
{\Bigl (} \frac{|\dot{M}|\,t_{K}}{\pi\,R^{2}\,\Sigma_{K}} {\Bigr )}^{1/\eta_{K}}\ .
\label{eqn:sgas.mdot}
\end{equation}
Moreover, because $\dot{M}$ itself depends on $\Sigma_{\rm gas}$, equation (\ref{eqn:sgas.mdot}) implies an equilibrium gas surface density {\em profile}.    For the simple dimensional scaling 
$|\dot{M}| \sim |a|\,\Sigma_{\rm gas}\,R^{2}\,\Omega$, we find
\begin{equation}
\Sigma_{\rm gas}(R) \sim \Sigma_{K}\,{\Bigl (} \frac{|a|\,
t_{K}}{\pi}\,\Omega(R) {\Bigr )}^{1/(\eta_{K}-1)}\ .
\label{eqn:sgas.a}
\end{equation}
Since $\Omega \sim \sqrt{G\,M_{\rm enc}(<R)/R^{3}}$, this implies that the scaling of the {\em total} surface density sets the equilibrium gas density profile, if we are in the strong torque regime.   

{We stress that equation \ref{eqn:sgas.a} and the associated balance between star formation and inflow are only approximately valid, and only then in a time-averaged, order-of-magnitude sense. 
There are regimes (e.g.\ the early stages of coalescence in a major merger) where 
gas can flow in on of order a single dynamical time
\citep{barnes.hernquist.91,mayer:2009.direct.collapse.bh}, much faster than star formation 
\citep[observationally supported by observations of e.g.\, ULIRGs with very high nuclear 
gas fractions; see][]{solomon.downes:ulirg.ism,bryant.scoville:ulirgs.co,
tacconi:ngc6240.gasdynamics,hibbard.yun:excess.light}.  In the opposite extreme, if gas densities are sufficiently low, the star formation and/or inflow timescales can 
be much larger than the dynamical time.   We show below 
that simulation gas densities typically scatter by an order of magnitude about the 
result implied by equation \ref{eqn:sgas.a}.   Nonetheless, this is a useful estimate of the gas densities achieved, particularly for analytic purposes.
%However, as an average statement, allowing for that scatter, it  is generally true in those simulations (even in 
%major mergers, despite the initial out-of-equilibrium inflow, 
%the timescale to equilibriate following said inflows is just a few dynamical times). 
%And observations that have attempted to quantify inflow rates infer 
%broadly similar behavior, albeit with large uncertainties 
%\citep{sakamoto:bar.driven.mol.gas.transport,davies:sfr.properties.in.torus,soto:ssp.grad.in.ulirgs}. 
%Indeed this must be the case, because of the 
%super-linear nature of star formation noted above. 
It is also important to note that the equilibrium between star formation and inflow still allows for an arbitrary range in observed gas fractions, as new stars are continuously produced.}

To use equation \ref{eqn:sgas.a} for specific predictions, we require values for $\Sigma_{K}$, $t_{K}$, 
and $\eta_{K}$. From the fit in \citet{kennicutt98}, 
$t_{K} \approx  0.63\times10^{9}\,{\rm yr}$ and $\eta_{K}\approx1.4-1.5$, for 
$\Sigma_{K}   =   10^{8}\,M_{\sun}\,{\rm kpc}^{-2}$. 
However, there is considerable uncertainty. For example, 
\citet{bouche:z2.kennicutt} 
suggest 
$t_{K} \approx  0.41\times10^{9}\,{\rm yr}$ and $\eta_{K}\approx1.71\pm0.05$, for 
$\Sigma_{K}   =   10^{8}\,M_{\sun}\,{\rm kpc}^{-2}$, 
from observations of high-redshift systems and more extreme starbursting 
galaxies. 
\citet{davies:sfr.properties.in.torus} and \citet{hicks:obs.torus.properties}
measure star formation in nuclear regions around AGN ($\sim1-100\,$pc), 
perhaps closest to the regions of interest here, and find results consistent
with the scalings in \citet{bouche:z2.kennicutt}.   When necessary, we use the latter throughout this paper.

\vspace{-0.4cm}
\section{Comparison to Hydrodynamic Simulations}
\label{sec:sims.main}

\begin{figure*}
    \centering
    \scaleup
    \plotsidesmallest{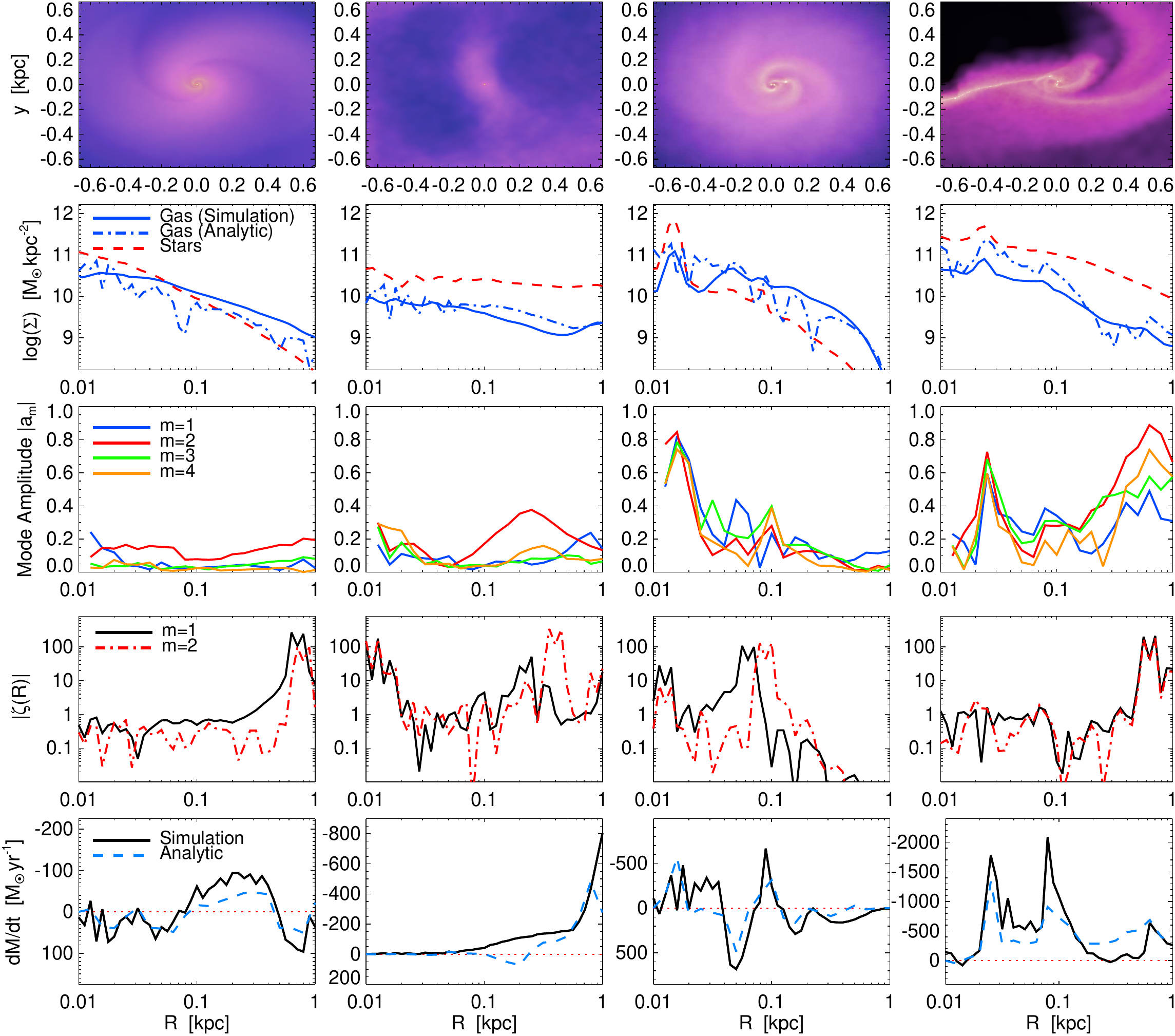}
    %\plotone{f1.pdf}
    \caption{Analogous results to Figure~\ref{fig:illustrate.merger} for  
    re-simulations of the central $\sim$kpc in 
    galaxy mergers, at representative times in the large-scale inflow history 
    (from left to right, these are runs {If9b5res, If1b1late, Ilowresq, Inf28b4} in \citealt{hopkins:zoom.sims}). 
    These simulations resolve gas inflows to within $\sim10\,$pc.  \label{fig:illustrate.100pc}}
\end{figure*}

\vspace{-0.2cm}
\subsection{The Simulations}
\label{sec:sims}

\begin{figure*}
    \centering
    \scaleup
    \plotsidesmallest{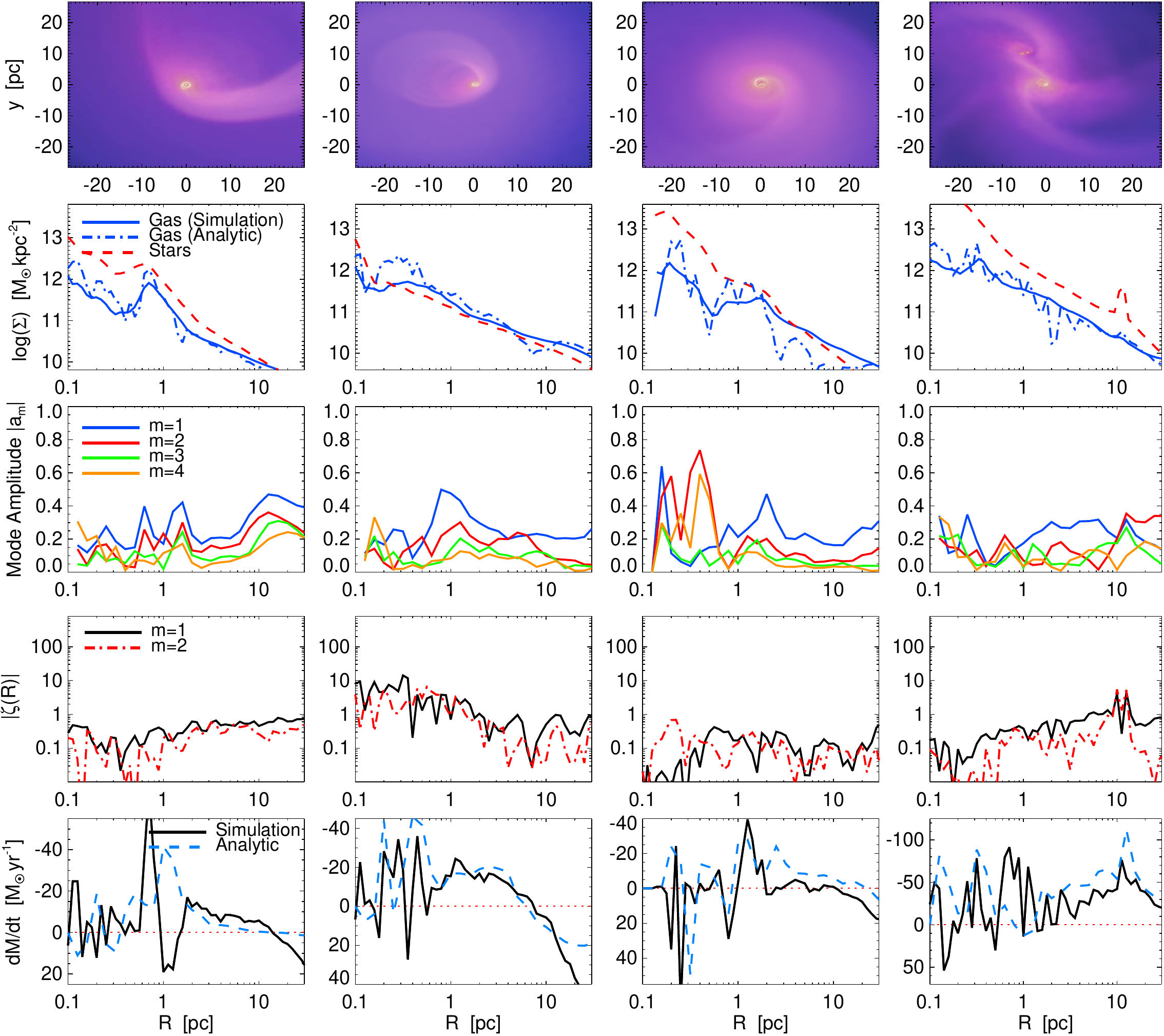}
    %\plotone{f1.pdf}
    \caption{Analogous results to Figure~\ref{fig:illustrate.merger} for
    re-simulations of the central $\sim10-50\,$pc in the intermediate-scale simulations shown in Figure~\ref{fig:illustrate.100pc}, at times near the peak of accretion and gas inflow (from left to right, these are runs {Nf8h1c1qs, Nf8h2b4, 
    Nf8h1c0hol, Nf8h1c1dens} in \citealt{hopkins:zoom.sims}).   These simulations have a resolution $\sim0.1\,$pc and extend deep into the potential of the central BH.
    %Representative simulations are shown, from 
    %weak to strong instability, chiefly from the 
    %ratio of BH to inflow mass (the large-scale bulge is sub-dominant 
    %to the BH inside these radii). 
    The (non-wound) $m=1$  eccentric disk and/or single-armed spiral mode is ubiquitous in the quasi-Keplerian potential.
%    when the inflow can yield a gas mass inside 
  %  the BH radius of influence that is a non-trivial fraction ($\sim0.1-1$) of the BH mass itself. 
    \label{fig:illustrate.10pc}}
\end{figure*}

\begin{figure}
    \centering
    \scaleup
    \plotone{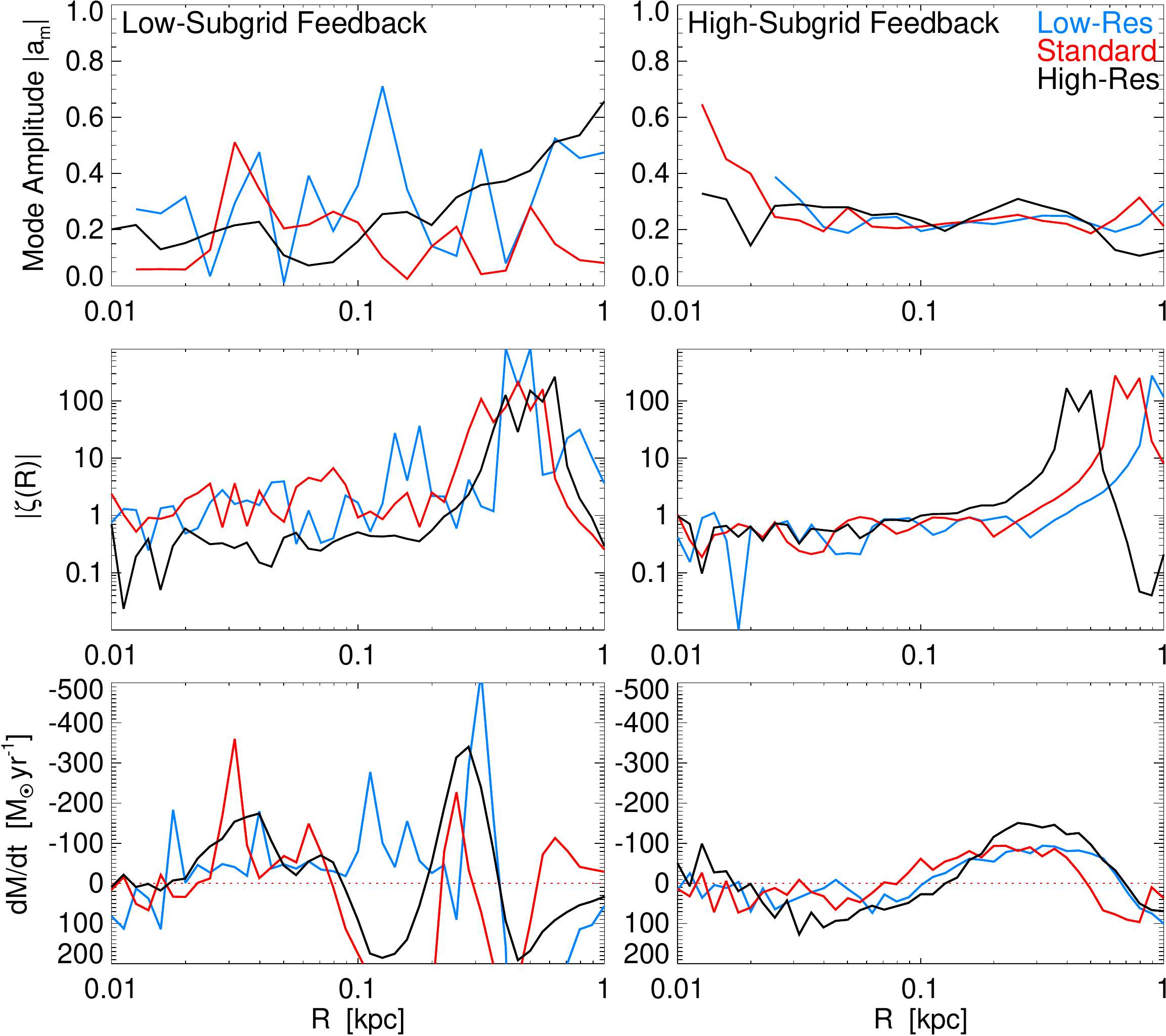}
    %\plotone{f1.pdf}
    \caption{Effects of simulation resolution on the modes we compare to our analytic model. 
    We plot the $m=2$ mode amplitude ({\em top}), $|\zeta(R)|$ for the $m=2$ mode 
    ({\em middle}), and inflow/outflow rate ({\em bottom}), 
    for intermediate-scale simulations as in Figure~\ref{fig:illustrate.100pc}.  
    For each, we show three otherwise identical simulations: one with low resolution 
    ($N_{\rm particles}=5\times10^{5}$; comparable to the resolution in the central regions of 
    our galaxy-scale runs before ``zoom-in''), 
    one with our standard resolution ($N=5\times10^{6}$; that of the other zoom-in runs 
    in Figure~\ref{fig:illustrate.100pc}), and one with very high resolution 
    ($N=5\times10^{7}$; high enough to obviate the need for an additional zoom-in 
    at $<10\,$pc scales). 
    We compare a model with very weak feedback ({\em left}; $q_{\rm eos}=0.02$ 
    giving an 
    effective ISM sound speed $c_{\rm eff}\sim10\,{\rm km\,s^{-1}}$) 
    to one with strong feedback ({\em right}; $q_{\rm eos}=0.25$, or $c_{\rm eff}\sim 50\,{\rm km\,s^{-1}}$). 
    The former shows more sensitivity to resolution because of sub-structure that forms in the 
    modes; the latter gives  a more smooth ISM so converges more quickly. At the order-of-magnitude 
    level, however, the mode amplitudes and inflow rates are similar. 
    \label{fig:illustrate.restest}}
\end{figure}

Thus far, we have made a number of simplifying assumptions in order to derive tractable analytic results.    In this section, we compare these analytic models to numerical simulations of gas inflow in galaxies.
We do not present any new simulations, but instead draw on results that have been described previously in a number of papers \citep[see e.g.][]{dimatteo:msigma, robertson:msigma.evolution,cox:kinematics,younger:minor.mergers,
hopkins:zoom.sims}.   The simulations we compare to include galaxy-galaxy mergers, isolated galaxy disks, and multi-scale ``re-simulations'' that follow gas from large scales in galaxies to sub-pc scales around a BH.  Table~\ref{tbl:sim.summary} summarizes the broad properties of these simulations and provides references to the original papers.

The simulations were performed with the parallel TreeSPH code {\small
GADGET-3} \citep{springel:gadget}, based on a formulation of smoothed particle hydrodynamics (SPH) that conserves energy and entropy simultaneously even when smoothing lengths evolve
adaptively \citep[see e.g.,][]{springel:entropy,hernquist:sph.cautions,oshea:sph.tests}.
The detailed numerical methodology is described in
\citet{springel:gadget}, \citet{springel:multiphase}, and
\citet{springel:models}.    The model galaxies consist of gas disks, stellar disks, stellar bulges, BH particles, and  dark matter halos, with \citet{hernquist:profile} 
halos, exponential disks, and (optional) bulges with mass fractions 
$B/T\sim0.1-1$.    Some of the simulations include sub-resolution prescriptions for BH accretion and feedback, but they are not important for the physics described here -- for our results, only the BH mass is critical, and then only on the smallest scales $\lesssim 10$ pc where it dominates the gravitational potential.  
{On all scales, extensive resolution tests have been performed 
(see references in Table~\ref{tbl:sim.summary}), with high-resolution case studies using $>100$ times 
as many particles.}

We consider a large number ($\sim50$) of simulations of isolated galactic disks, marginally to strongly self-gravitating  \citep[described in detail in][]{younger:minor.mergers}. 
We also consider a large number of mergers of these idealized disks, 
in which we vary the merger mass ratio (from 1:20 to 1:1), orbital parameters, and structural properties of the disks  \citep[for details, see][]{hopkins:disk.survival}. 
{These structural variations include changing the disks in mass, bulge-to-disk ratios, 
disk scale height, gas and stellar disk scale lengths and mass profiles, 
and scaling the disks following \citet{momauwhite:disks} to correspond to expected 
disk properties at various redshifts from $z=0-6$.}
Finally, we also include the ``re-simulations" of galactic nuclei described in detail in \citet{hopkins:zoom.sims}. In these calculations, we re-simulate the central $\sim$kpc of a representative sub-set of the mergers and isolated disks at several times in their evolution, at much higher resolution, for a total of $\sim 100$ simulations.  This technique allows us to study, using realistic initial conditions,  the formation of secondary instabilities when the gas/stars are self-gravitating (``bars within bars'') and the resulting gas inflow to small radii. These simulations are accompanied by a similar number of yet smaller-scale re-simulations that follow the gas inflow deep into the BH potential, with $\sim0.1\,$pc resolution  (the resolution studies 
mentioned above include both varying this spatial resolution and the scales at which the 
re-simulations are performed; neither of these changes our conclusions).

In both sets of ``re-simulations,'' we consider simulations with zero initial perturbations, 
with initial perturbations seeded randomly, and with initial perturbations inherited from 
the ``parent'' simulation. In Appendix A3 of \citet{hopkins:zoom.sims} we show that this does 
not qualitatively change our results, because the systems of interest are self-gravitating on 
small scales and so the local structure determines the equilibrium modes, not the initial 
conditions. Moreover our general analytic model appears to describe shocks due to modes from 
either initial condition equally well.

The simulations utilized here all include gas cooling and star formation, with gas forming stars at a rate motivated by the observed \citet{kennicutt98} relation. Specifically, we use a
star formation rate per unit volume $\dot \rho_{\ast} \propto
\rho^{3/2}$ with the normalization chosen so that an isolated Milky-way like
galaxy has a total star formation rate of about $1\,M_{\sun} \, {\rm yr^{-1}}$.  This formulation produces a projected Schmidt-Kenicutt relation (i.e., $\dot \Sigma_*$) similar to that measured by \citet{bouche:z2.kennicutt}.   We crudely account for feedback from stars with an effective equation of state that mimics turbulence in a multi-phase interstellar medium by adopting a sound speed that is much larger than the true thermal sound speed (e.g., \citealt{springel:multiphase}).   \citet{hopkins:zoom.sims} describe in detail the effects of different subgrid ISM sound speeds on angular momentum transport and inflow rates.   They also argue for effective turbulent speeds of $\sim 10-50 \ {\rm km\,s^{-1}}$ for densities $\sim 1-10^5$ cm$^{-3}$, respectively.    As we have shown in \S \ref{sec:response}, the torques and inflow rate induced in the gas by an asymmetric stellar potential are {\em independent} of the ISM sound speed in the orbit-crossing limit (e.g., eq.~\ref{eqn:mdot.full}).  This suggests that the use of a subgrid sound speed in the simulations does not have a significant effect on the physics of gas inflow.   However, the critical assumption in {\em both} the numerical simulations and the analytic calculations presented here is that stellar feedback maintains a reasonable fraction of the ISM in a diffuse phase, rather than being bound in star clusters.    In the simulations, gravitational instability can form some compact, dense clumps (analogous to molecular clouds or star clusters), but the enhanced subgrid sound speed ensures that most of the ISM does not suffer this fate. This is intended to mimic physics not included in our calculations, such as the observed low efficiency of star formation in dense molecular gas (e.g., \citealt{krumholz:sf.eff.in.clouds}) and the disruption of molecular clouds by radiation pressure, HII regions, etc. (e.g., \citealt{murray:molcloud.disrupt.by.rad.pressure}) -- the latter returns most of the gas from molecular clouds back into the diffuse ISM.

{There are of course considerable uncertainties in any model of the ISM and star formation in galaxies, 
especially on small scales near a BH; the adoption of a simple \citet{kennicutt98} relation is merely a convenient simplification. For this reason, \citet{hopkins:zoom.sims} consider a broad range 
of such prescriptions (see their \S~2 and Appendices). We briefly summarize some of these variations, but interested readers should see that paper for details. The slope of the star formation law was varied from $\dot{\rho}_{\ast}\propto \rho^{1-2}$, amounting to changes in the star formation efficiency on small scales by factors of $\gtrsim 100$. This spans the entire observed parameter space of star formation efficiencies on large galactic scales \citep[see e.g.][]{bigiel:2008.mol.kennicutt.on.sub.kpc.scales}, 
in dense molecular clouds \citep{krumholz:sf.eff.in.clouds}, and 
the central regions of AGN \citep{hicks:obs.torus.properties}; 
our ``typical'' cases are consistent with the latter observations. The primary difference the star formation efficiency makes is that it changes the gas-to-stars ratios at a given time; but since this quantity explicitly appears in our model, the effect is to move systems along our predicted scalings, not with respect to them. In our analytic estimates that follow, when the absolute value of star formation efficiency is important, we parameterize it as a free function, allowing for different detailed models of star formation. 

The simulations in Table~\ref{tbl:sim.summary} also include large systematic studies of the prescription for the sub-grid gas physics in the ISM.   Specifically, the effective equation of state of the gas was widely varied.  The simulations include cases where it is quite stiff on small scales \citep[near-adiabatic, similar to what is adopted in the studies of][]{mayer:bh.binary.sph.zoom.sim,
dotti:bh.binary.inspiral}, through to cases where the gas is allowed to cool to a cold isothermal floor (with $c_{s}\ll V_{c}$), and thus forms a clumpy, inhomogeneous medium 
\citep[on galactic scales, similar to what is assumed in][]{bournaud:disk.clumps.to.bulge,
teyssier:2010.clumpy.sb.in.mergers}. Intermediate cases are also sampled, allowing for e.g.\ an equation of state  similar to that presented in 
\citet{klessen:2007.imf.from.turbulence}, 
motivated by the models of \citet{spaans:2005.gmc.eos} for dense star-forming gas. In a large number of our nuclear-scale simulations, the gas dispersion is dominated by {\em resolved} turbulent gas motions, which maintain an effective $c_{s}$ even as the sub-grid $c_{s}$ becomes very small (the origin of this turbulence and its importance for e.g.\ AGN torii will be discussed in future work). As such, we can explicitly check that our formulation for gravitational torques still applies even in turbulent inhomogeneous gas distributions.}

Nevertheless, we consider our sub-grid treatment of the ISM to be the biggest uncertainty 
in the numerical models. Because we cannot resolve the extremely large densities of 
actual star-forming cores in these regions, star formation is almost certainly more diffuse/uniform throughout 
the disk, which may in turn have consequences for the ability of stars and gas to interact efficiently. 
And the ISM is, by definition, more smooth than the implicit sub-grid multiphase structure. 
An important goal for future work will be the investigation of more realistic ISM and star formation models 
on these scales. These are unlikely to make a large difference in the limit of stellar-dominated disks 
(our focus here), but may in the very gas-rich regime open up new channels for angular momentum 
exchange. 

In \S \ref{sec:sims:overview} we provide an overview of the simulation results and a comparison to the analytic results derived in \S \ref{sec:response} \& \ref{sec:deriv.gasden} for example simulations covering three different spatial scales:   galactic ($\sim$ kpc), intermediate-scale ($\sim 0.01-1$ kpc), and nuclear ($\sim 0.1-10$ pc).    We refer the reader to \citet{hopkins:zoom.sims} for more detailed information about the properties of the gas and stars in the simulations (e.g., $Q[r]$, $h/r$, etc).  Although the specific simulations discussed in \S \ref{sec:sims:overview} are quite instructive, it can also be misleading to focus on the results of an individual simulation.  The reason is that there is large variation in time and potentially large scatter introduced by modest differences in galaxy properties (due to, e.g., large-scale fragmentation of a spiral arm biasing the results of a given simulation).   For this reason we believe that it is important to consider the statistical properties of a large number of simulations.  In \S~\ref{sec:sims:inflow} we thus present a more comprehensive statistical comparison between the simulations and our analytic results.   

\vspace{-0.5cm}
\subsection{Overview of Simulation Results}
\label{sec:sims:overview}

Figure~\ref{fig:illustrate.merger} provides an overview of the dynamics during a major merger of equal-mass $\sim L_{\ast}$ galaxies with $f_{\rm gas}\approx0.4$.   The first row shows, from left to right, images of one of the two galaxies before first passage, after a near-final passage (when the impact parameter of the interaction is $\lesssim$ the size of the galaxy), and just before the final coalescence of the two systems; the final column shows an image of the remnant system just after coalescence.   The second row in Figure~\ref{fig:illustrate.merger}  shows gaseous and stellar mass surface density profiles at each time, while the third shows the amplitude of the stellar surface density perturbations for different azimuthal mode number $m$ as a function of radius 
($R=0$ is defined as the location of the BH); 
\begin{equation}
| a_{m}(R,t) | = \frac{| \int_{0}^{2\pi} \Sigma(R,\,\phi)\,\exp{(\imath\,m\,\phi)}\,{\rm d}\phi |}
{ \int_{0}^{2\pi} \Sigma(R,\,\phi)\,{\rm d}\phi }\  
\label{eqn:am}
\end{equation}
The phase angle $\phi_{0}$ of the mode maximum is determined using a similar expression, as are the    
potential perturbations, and other perturbed quantities of interest.   The second to last row in Figure~\ref{fig:illustrate.merger}  shows estimates of the orbit crossing parameter $\zeta$ for the $m = 1$ and $m = 2$ modes (the calculation of which is described below).   Finally, the bottom row in Figure~\ref{fig:illustrate.merger}  shows the instantaneous gas inflow rate as a function of radius $R$, determined directly as $\dot{M}=\Delta R^{-1}\,\int {\rm d}M_{\rm gas}\,v_{R}$ given the 
radial velocity $v_{R}$ of each gas particle of 
mass ${\rm d}M_{\rm gas}$ in a small radial annulus $\Delta R$. 
As before, we define $\dot{M}<0$ to be accretion/inflow, whereas $\dot{M}>0$ corresponds to outflow. 

In addition to the numerical results, Figure~\ref{fig:illustrate.merger} also shows several of the predictions of our analytic model.    First, given the measured inflow rate in the simulations $\dot{M}$, we use the balance between inflow and star formation implied by equation~\ref{eqn:sgas.mdot} to predict the gas surface density $\Sigma_{\rm gas}$ (2nd row).   This prediction is within a factor of $\sim 3$ of the measured gas density in the simulations, particularly at later times (right-most two columns) when our assumed balance between inflow and star formation is a better approximation.   In addition, given the measured $|a|$ at each radius, and the local gas properties ($\Sigma_{\rm gas}$, $\Omega$, and their derivatives), Figure~\ref{fig:illustrate.merger} also shows our prediction of the inflow rate $\dot{M}$ using equation~\ref{eqn:mdot.full} (bottom row).  Evaluating the sign of $\dot M$ in equation~\ref{eqn:mdot.full} (i.e., inflow or outflow), is non-trivial given its dependence on $\Omega_{p}$ and $\zeta$; one option would be to fit a global model to the mode structure as a function of time and radius, 
but since $\zeta$ depends on second derivatives this is very noisy.   Instead, we found it easier and more numerically robust to use the epicyclic approximation to estimate $\zeta$ and the sign of $\dot M$.
With this approximation, the mean local radial velocity (zero in the unperturbed state) obeys
\be
\frac{\partial v_{1,\,R}}{\partial R} \sim i\,m\,(\Omega-\Omega_{p})\,\zeta,
\label{zeta}
\ee
i.e.\ it is phase-shifted by $\pi/2$ and multiplied by $(\Omega-\Omega_{p})$ relative to $\zeta$; this implies that ${\rm sign}(\Omega-\Omega_{p})\,\cos{({\rm arg}[\Phi_{1}\,\zeta^{\ast}])}
=-\sin{({\rm arg}[\Phi_{1}]-{\rm arg}[\partial v_{1}/\partial R])}$. 
We use the true radial velocity of the stars in the simulations $v_{R}$ as our estimate of the perturbed velocity $v_{1,R}$ and determine its amplitude and phase as in equation~\ref{eqn:am}.   Together with the measured $\Phi_{1}$ in the simulations we can thus determine the sign of $\dot M$ induced by a given $m$.    For the results in Figure~\ref{fig:illustrate.merger}, we sum the contributions to $\dot M$ from $m=1$ to $m=5$, but find that this is almost always dominated by the largest mode or two (this will be an assumption of ours later).
Equation \ref{zeta} also implies that $|\zeta| \sim |\partial v_{1,\,R}/\partial R|/[m\,|\Omega-\Omega_{p}|]$; we estimate $\Omega_{p}$ from a global fit to the mode amplitude and axis $\phi_{0}$ as a function of time, assuming $\Omega_{p}$ is constant in time and radius for a given simulation.   The results for $|\zeta|$ are shown in the 2nd to last row in Figure~\ref{fig:illustrate.merger} for the major merger simulations.    

Figure~\ref{fig:illustrate.merger} shows that the predicted inflow rate as a function of radius is well reproduced by our analytic model at all times, providing support for the simplifying assumptions we made in the analytic derivation.   By contrast, the predictions of the no-shock, weak-torque limit (eq~\ref{eqn:mdot.weak.torques}) or the direct transport  by the mode in the gas (eq~\ref{eqn:mdot.kalnajs}) are smaller by a factor $>10$.    Although the results in Figure~\ref{fig:illustrate.merger} are for a major merger, we find similar consistency between the simulation results and our analytic predictions for both minor mergers and isolated, bar-unstable galactic disks.    Figure~\ref{fig:illustrate.merger} also shows that the orbit crossing parameter $\zeta$ is $\sim1$ at a range of radii, with $|\zeta|\sim10-100\gg1$ near the co-rotation resonance, implying strong orbit crossings and shocks.   As described in the text above equation~\ref{zeta}, our estimate of $\zeta$ is itself based on the epicyclic approximation; it is thus very approximate in mergers given that the galaxies are strongly disturbed and quantities such as $\Omega_{p}$ are not that well-defined.  Nonetheless, the results in Figure~\ref{fig:illustrate.merger} are consistent with the presence of strong shocks and dissipation in the simulations, which drive the large gas inflow rates.

Figures~\ref{fig:illustrate.100pc} and \ref{fig:illustrate.10pc} show 
analogous results to Figure~\ref{fig:illustrate.merger}, but 
for simulations of the central $\sim$kpc  (Fig. \ref{fig:illustrate.100pc}) and $\sim10$ pc (Fig. \ref{fig:illustrate.10pc}) of gas-rich mergers (from \citealt{hopkins:zoom.sims}).  
The simulations span different morphologies and degrees of clumpiness in the ISM gas, 
and include both cases that are strongly stellar-dominated and that have more comparable gas 
and stellar densities.   In each case, the agreement between the analytically predicted $\Sigma_{\rm gas}$ and $\dot M$ and the numerical results is again quite good.   
{As noted above, we do not necessarily expect our predictions to be applicable in gas-dominated 
systems; however, at least up to gas fractions $\sim0.5$ seen here, the inflow rates remain 
well-described by our formulation. }
And the Figures include both simulations with initial conditions set to have ``smooth'' structure 
(no initial modes) and those with a wide spectrum of initial modes inherited from 
the structure of their ``parent'' simulations. 
More than Figure \ref{fig:illustrate.merger}, Figures~\ref{fig:illustrate.100pc} and \ref{fig:illustrate.10pc} demonstrate that our analytic model even reproduces the sign of $\dot M$ (inflow vs. outflow) reasonably well.   Note also that the $m = 1$ mode becomes increasingly important relative to the $m = 2$ mode at small radii (Fig.~\ref{fig:illustrate.10pc}).   This is not only clear visually in the images, but also quantitatively in the mode amplitudes.\footnote{{The reasons for this are discussed at length in 
\citet{hopkins:zoom.sims} and \citet{hopkins:m31.disk}, but in general owe to the 
fact that $\kappa\approx \Omega$ in a quasi-Keplerian potential; "slow" $m = 1$ modes are 
therefore generic (though not unique) to small scales near a BH 
(see \citealt{tremaine:slow.keplerian.modes} for stellar-dominated systems, 
or \citealt{ostriker:eccentric.waves.via.forcing} for gaseous disks).}}
At the smaller radii shown in Figures~\ref{fig:illustrate.100pc} and \ref{fig:illustrate.10pc}, the modes are more often in the marginal orbit-crossing limit, with $|\zeta|\sim0.1-$a few, rather than $\zeta \gg 1$; this highlights the importance of our results in \S~\ref{sec:deriv.mom.inflow:marginal}, which show that shocks and rapid inflow can occur even where $|\zeta|\lesssim1$ (although not for $\zeta \ll 1$).

Figure~\ref{fig:illustrate.restest} briefly illustrates how the simulation properties 
depend on the resolution and uncertain microphysics 
of the numerical models. We show the mode amplitudes and $|\zeta(R)|$ (for simplicity restricting  
to just $m=2$), and inflow rates as a function of radius, 
for simulations of the central $\sim$kpc (the range shown in Fig.~\ref{fig:illustrate.100pc}). 
We consider otherwise identical initial conditions but with different resolution. 
We also consider two different extremes in the parameterization of the ISM gas equation of 
state, encapsulated in the parameter $q_{\rm eos}$ (see \citealt{hopkins:zoom.sims} for details). 
A low value of $q_{\rm eos}$ corresponds to an ISM with efficient cooling and without 
much turbulent or thermal pressure support (here $q_{\rm eos}=0.02$ corresponds to an effective sub-grid 
sound speed of just $\sim10\,{\rm km\,s^{-1}}$), 
and a high value corresponds to an ISM with strong pressure support 
(here $q_{\rm eos}=0.25$ corresponds to effective 
sound-speeds $\sim50\,{\rm km\,s^{-1}}$). For each, we compare simulations at three 
resolution levels: a low-resolution run, for which the mass and force resolution 
inside of $\sim1\,$kpc are comparable to what can be achieved directly (without any 
``zoom-in'' or ``re-simulation'') in the parent, galaxy-scale simulations; a 
standard-resolution run (the characteristic resolution of our re-simulations on this 
scale); and a very 
high-resolution run (the ``ultra-high'' resolution runs in \citealt{hopkins:zoom.sims}) 
with $\sim0.3$\,pc force resolution, small enough to resolve the same scales as our 
nuclear-scale simulations without the need for an additional re-simulation iteration. 
Even at low resolution, the major qualitative features are in place and the 
inflow rates are order-of-magnitude accurate. 
But in the case with weak assumed feedback, the ISM is subject to much greater 
resolved fragmentation and clumpy star formation (see Appendix~B in \citealt{hopkins:zoom.sims} 
for more discussion and examples). This leads to inflow rates 
and mode structures that are more highly variable in position and time, and 
the exact details (depending on how clumps collapse locally and turn into 
stars) are more sensitive to resolution. The case with stronger implicit 
feedback yields an ISM with less explicit substructure, 
which in turn allows rapid convergence with resolution. 
But in both low and high-feedback cases, 
the average behavior can still be reasonably approximated with the analytic prescriptions 
(accounting for the implicit, sub-grid ISM structure assumed in the numerical models). 
Because of the uncertain details of star formation and ISM physics in the simulations, 
we do not think any of these runs should be taken literally as an exact description of 
what really happens in AGN; however, they provide a means to test our 
analytic prescriptions in fully non-linear, chaotic systems. 

Note that on small scales, we have chosen times after the two BHs in a galaxy merger are 
assumed to merge. It is of course not clear if the final merger proceeds quickly or slowly 
on sub-pc scales. We do so for simplicity and generality (since isolated disks will not have a second 
black hole). But during the actual binary phase of a near equal-mass BH-BH merger, the mode 
structure and gravitational torques on small scales will be highly non-linear and feature a 
more complex geometry than what we show here. We can, however, still attempt to consider this in the 
context of our analytic framework, but with the secondary BH as the ``collisionless'' driver of 
non-circular gas motions and shocks, as opposed to modes in the stellar disk. 
For more discussion on the inspiral phase in these simulations and its implications for the 
presence of the $m=1$ modes discussed above, we refer to \citet{hopkins:zoom.sims}.

\vspace{-0.45cm}
\subsection{Inflow Rates}
\label{sec:sims:inflow}

We now present a statistical comparison between the inflow rates in our numerical simulations and the predictions of inflow produced by stellar torques (\S \ref{sec:response}).  In doing so, we shall show a number of Figures that contain the results of many of our simulations.  In these Figures, the critical point to focus on is less the results of any given simulation (which can be difficult to identify in the Figure), but rather the ensemble properties of all of the simulations (e.g., median, scatter, or systematic trend relative to analytic predictions).  In what follows, we take $F(\zeta) = 1$ for all but the $m = 1$ modes at $\lesssim 10$ pc, for which we take $F(\zeta) = 1/2$; these choices are motivated by the typical values of $\zeta$ shown in Figures \ref{fig:illustrate.merger}-\ref{fig:illustrate.10pc}.

\begin{figure}
    \centering
    \scaleup
    \plotonexa{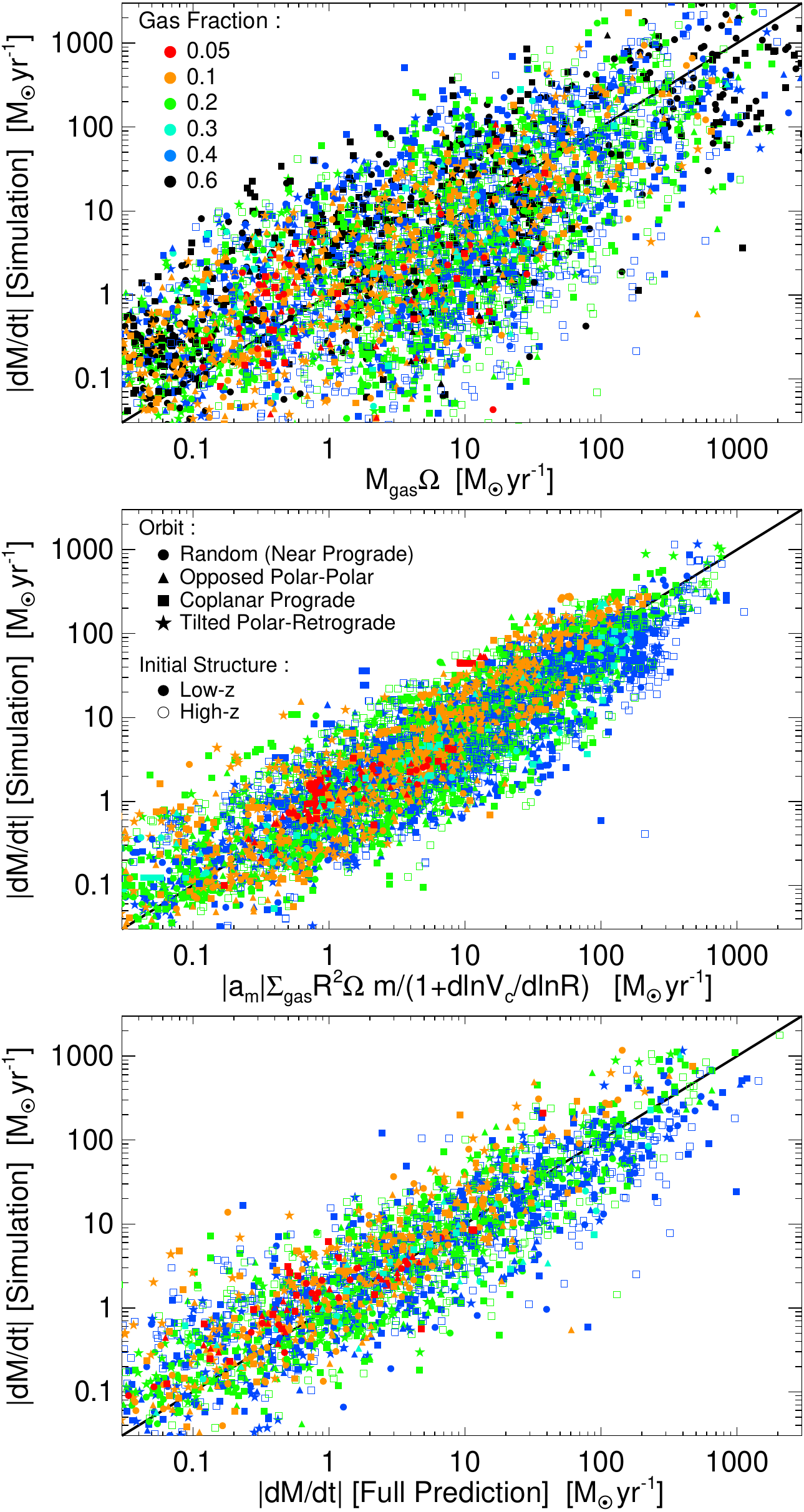}
    %\plotone{f1.pdf}
    \caption{Instantaneous gas inflow/outflow rates $|\dot{M}|$
    in major galaxy merger simulations versus several predictors of the inflow rate.   The simulations span a range of masses, gas fractions, sizes, and orbital parameters.   Some of these parameters are explicitly labeled (e.g., color denotes gas fraction, open (filled) points denote initial disks with structural parameters 
    similar to local $z\sim0$ (high-redshift) galaxies). 
    We calculate $|\dot{M}|$ through a radius $\approx 5\,\epsilon$ (where 
    $\epsilon$ is the simulation resolution limit; results are similar for 
    radii $R\sim0.1-1\,$kpc), at random times uniformly sampling 
    between first passage and $0.5\,$Gyr after peak activity. 
    {\em Top:} Simulation results compared to the dimensional scaling $M_{\rm gas}(<R)\,\Omega(R)$. 
    The scatter is very large, $\sim1.5-2\,$dex. 
    {\em Middle:} Simulation results compared to the simple estimate of inflow induced by strong orbit crossing in the gas (eq.~\ref{eqn:mdot.strongtorque.simple}). 
    Predicted and simulated inflow rates agree well, with a scatter of $\sim0.5$\,dex.
    {\em Bottom:} Simulation results compared to the full scaling derived for the strong-torque 
    regime (eq.~\ref{eqn:mdot.strong}). The scatter is reduced to $<0.3\,$dex. 
    \label{fig:inflow.rate.merger.vpred}}
\end{figure}
%\clearpage

\begin{figure*}
    \centering
    \scaleup
    \plotside{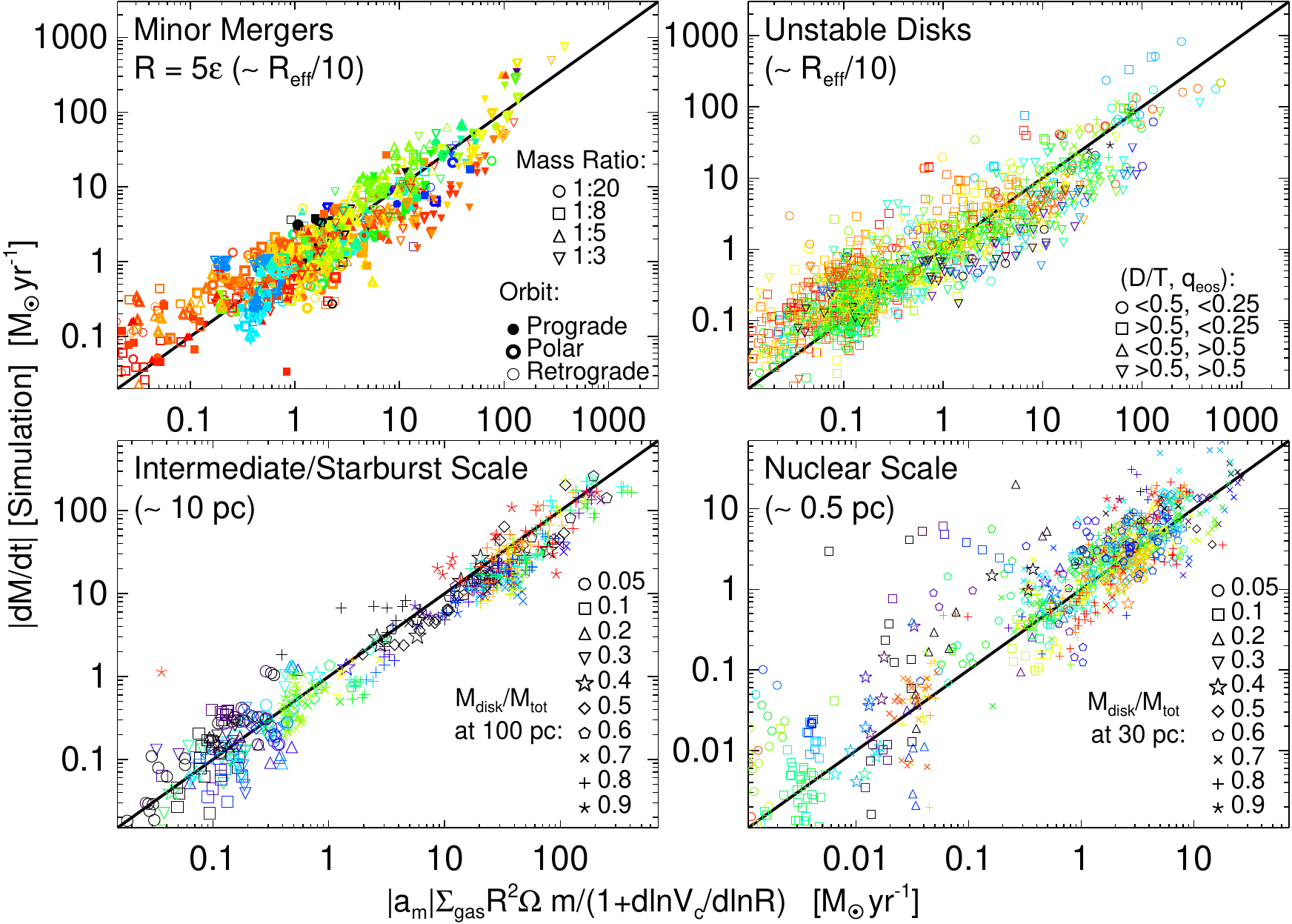}
    %\plotone{f1.pdf}
    \caption{Simulated vs. predicted inflow rates, as in 
    Figure~\ref{fig:inflow.rate.merger.vpred}, for minor mergers, isolated galactic disks, and re-simulations of galactic nuclei; the inflow rates are measured at  a radius of $5\,\epsilon$ (as Figure~\ref{fig:inflow.rate.merger.vpred}) 
    in each case, with the values of this radius in parentheses in each panel. 
    Colors denote the gas fraction, as in Fig.~\ref{fig:inflow.rate.merger.vpred}, while
    symbol types denote some of the key parameters varied in the ensemble of simulations.  In all panels, the predicted inflow rate (x-axis) is the simple strong orbit crossing estimate in equation~\ref{eqn:mdot.strongtorque.simple}.    {\em Top Left:} Galaxy-galaxy minor mergers. 
    {\em Top Right:} Isolated, bar-unstable galactic disks. 
    The parameters varied in the simulations
    include the initial disk-to-total mass ratio $D/T$ at the effective radius and the
    stellar feedback prescription ($q_{\rm eos}$, where $<0.25$ denotes modest feedback 
    and $>0.5$ denotes strong feedback; see \citealt{hopkins:zoom.sims}). 
%    ), the slope and normalization of the 
%   star formation law (``normal'' being the standard Kennicutt-Schmidt law), 
%    as well as galaxy masses, structural properties, and presence or absence of BHs.     
    {\em Bottom Left:} Intermediate-scale re-simulations of both mergers 
    and isolated unstable systems (e.g.,
    Fig.~\ref{fig:illustrate.100pc}).  The key parameter regulating the formation of secondary bars is the     disk-to-total mass ratio.
%     of sub-kpc radius, but again feedback prescriptions, mass, BH mass, and mass profile 
%    shapes are varied, along with the boundary conditions (matched to different 
 %   larger-scale simulations). 
    {\em Bottom Right:} Nuclear-scale re-simulations of the 
    intermediate-scale systems (e.g., Fig.~\ref{fig:illustrate.10pc}), 
    where the lopsided $m = 1$ disk forms and regulates the gas inflow.
    \label{fig:inflow.vs.sims}}
\end{figure*}

Figure~\ref{fig:inflow.rate.merger.vpred} compares the instantaneous inflow rate 
in major galaxy merger simulations to several analytic predictors of the 
inflow rate:   the simplest possible dimensional expectation, $M_{\rm gas}(<R)\,\Omega(R)$ (top panel);
the order of magnitude prediction of inflow produced by stellar torques 
(middle panel; equation~\ref{eqn:mdot.strongtorque.simple}); 
and the more detailed prediction of our analytic model 
(bottom panel; equation~\ref{eqn:mdot.full} from \S \ref{sec:deriv.mom.inflow:all}). 
In all three panels, we measure the inflow rate through 
a radius of $5\,\epsilon$ (where $\epsilon$ is the gravitational softening length of the simulation, 
roughly chosen so that $5\,\epsilon\sim R_{\rm eff}/10$), although we find similar results at other radii.
We calculate $\dot M$ for each simulation at $\sim 10^{8}\,$yr intervals, 
from just before first passage until $\approx0.5\,$Gyr after the peak of activity (when the galaxies fully coalesce), at which point significant inflow has ceased.  The instantaneous inflow 
rate is averaged over each time interval between points to average over 
both the chaotic motions of the gas and noise introduced by the finite 
simulation resolution.   The mode amplitude and other properties 
of the gas needed in equations~\ref{eqn:mdot.strongtorque.simple} and \ref{eqn:mdot.full} are 
calculated from the simulations themselves (at the same radius as the measured inflow rate) 
as in \S \ref{sec:sims:overview}.

Figure~\ref{fig:inflow.rate.merger.vpred} (top panel) shows that although 
there is a correlation between the true inflow rate and $M_{\rm gas}(<R) \, \Omega(R)$, 
the scatter is very large $\sim1-2\,$dex. 
Note that the large vertical scatter here (particularly that 
above the prediction) does not imply material falling in on
much less than the dynamical time; instead, it stems from a 
combination of time variability and scatter in the $x$-axis.  The bottom panels of Figure~\ref{fig:inflow.rate.merger.vpred} show that the analytic predictions 
in the strong torque regime, equations~\ref{eqn:mdot.strongtorque.simple} 
and \ref{eqn:mdot.full}, work remarkably well, predicting the 
instantaneous inflow rates with a scatter of less than $\sim 2-3$. 
Moreover, there are no systematic trends with simulation properties 
such as the gas fraction, galaxy size, orbital parameters, or ISM equation of state model.

Figure~\ref{fig:inflow.vs.sims} compares the inflow rate to equation~\ref{eqn:mdot.strongtorque.simple} (the order of magnitude strong torque $\dot M$) for our other sets of simulations:   minor mergers, isolated bar-unstable disks, intermediate (sub-kpc) scale re-simulations of galactic nuclei, and nuclear-scale ($<10\,$pc) re-simulations of disks around BHs. In each case we again measure the inflow rate at $\sim 5 \epsilon$ in the simulations and smooth the results in time.   The strong torque analytic model works well in all cases, despite a very large dynamic range in mass, spatial scale, and inflow rate, and despite the fact that different, independent non-axisymmetric modes control angular momentum loss at different radii. In addition, some of the simulations (particularly the mergers) are clearly in the strong orbit-crossing regime ($\zeta \gg 1$), while others (particularly the $<10\,$pc circum-BH disks) are in the marginal orbit-crossing regime ($\zeta \lesssim 1$) in which there are fewer (if any) formal orbit crossings.   The same inflow prediction nonetheless works well in both limits. {Gas fractions are explicitly labeled by color; although we expect our assumptions to break down somewhere around $\fgas\sim1$, it is immediately clear from Figure~\ref{fig:inflow.rate.merger.vpred} that up to $\fgas\sim0.5$, there is no systematic deviation from our scaling.\footnote{
We emphasize, however, that at gas fractions this high, the sub-grid ISM model of the simulations becomes 
particularly suspect (as small-scale clumping/fragmentation is not explicitly treated). It is safer to say 
that our results are robust to variations in $\fgas$ for $\fgas\ll1$. 
} Other parameters varied from simulation to simulation are generally not shown, but we have checked to see that there is no clear systematic difference between our predictions and the results of a given subset of the simulations. For example, the effective equation of state of the gas is varied widely at each scale as discussed in \S~\ref{sec:sims}, but there is no statistically significant offset of the inflow rates with respect to the predictions in each case (with the important caveat that we have no simulations with a true non-sub grid model).   Likewise, both in isolated disks and nuclear-scale simulations, the model of star formation has been varied in a large subset of simulations; the resulting points are indistinguishable. Making star formation less homogeneous by raising 
the star formation density threshold does not appear to change our conclusions, but this is again with the caveat 
that the simulations do no explicitly model individual GMCs.  
Changes to the ISM equation of state or star formation law 
do affect various global properties of the results -- the mass distribution, gas fractions, etc -- but these 
are explicitly included in our formulation of $\dot M$; the important point is that they introduce no significant deviations from our predicted scalings. Parameters such as the initial simulation redshift enter only indirectly, changing disk structural parameters, and Figure~\ref{fig:inflow.rate.merger.vpred} 
shows explicitly that they make little difference to our conclusions.}

One apparent outlier in Figure~\ref{fig:inflow.vs.sims} is the ``spray'' of several (blue/black) points at large true $\dot{M}$, but small predicted $\dot M$, in the nuclear-scale simulations (bottom right). 
This is a system in which the gas disk undergoes a large-scale fragmentation and 
a gas-rich clump happens to fall on a nearly radial orbit towards the BH.  {This is a special case. }
Because of the high density of points in Figure \ref{fig:inflow.vs.sims}, this small number of outliers makes it appear as if the scatter in this panel in much larger -- in fact, the correlation is just as significant as in each of the other panels (linear correlation coefficient $0.87$, or significance $>10\,\sigma$), and the intrinsic scatter is nearly identical (lognormal $0.37$\,dex, as compared to $0.33$\,dex for the points plotted in other panels). Recall as well that in many of the simulations shown here, the gas motions include a significant contribution from resolved `turbulent' motions and the gas mass distribution can be highly inhomogeneous/clumpy; we nonetheless still find good agreement with the predictions of the strong torque analytic model. 

\begin{figure}
    \centering
    \scaleup
    \plotone{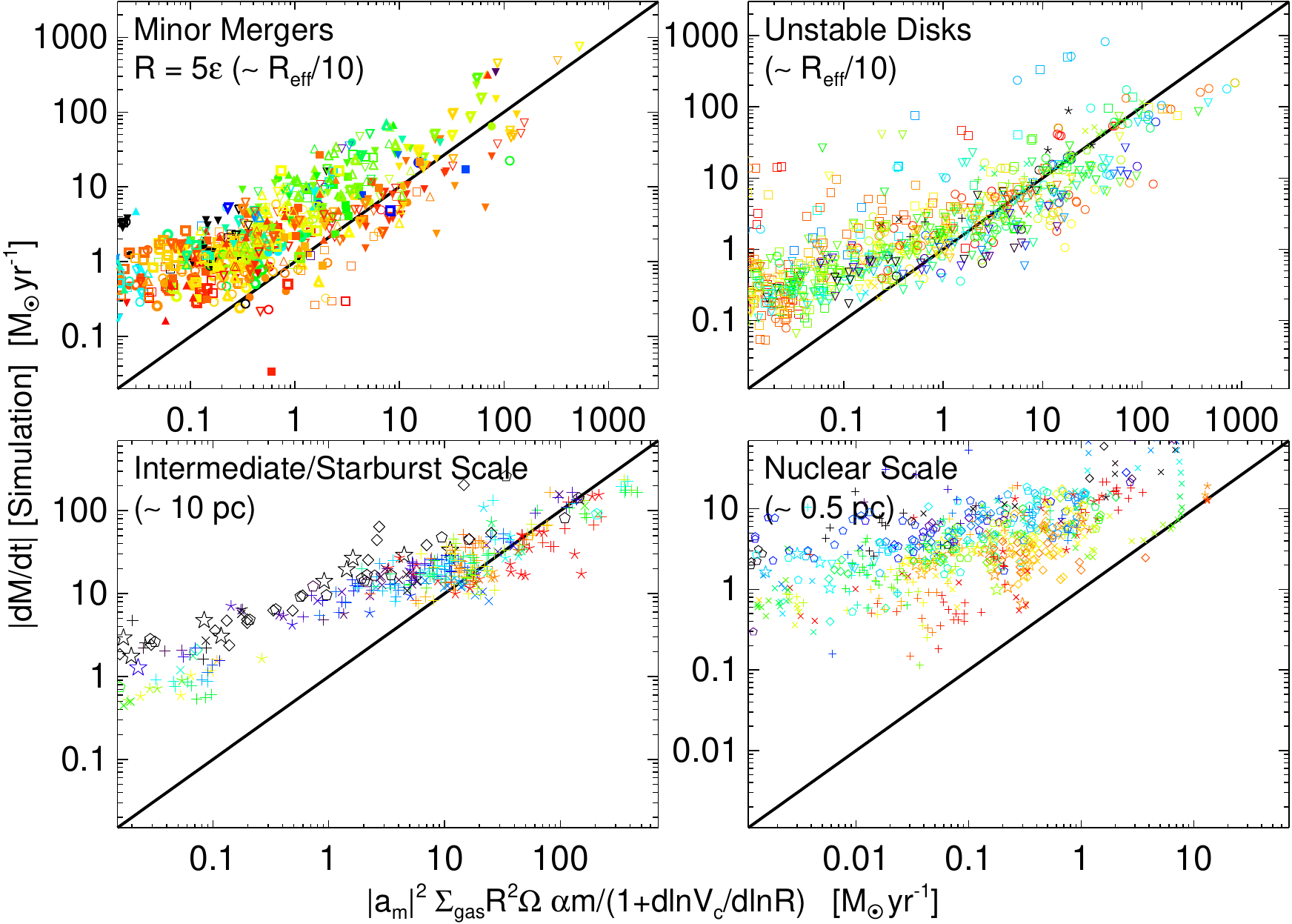}
    %\plotone{f1.pdf}
    \caption{Simulated vs. predicted inflow rates, as in Figure~\ref{fig:inflow.vs.sims}, but for the linear weak-torque, non-orbit crossing predictions (eq.~\ref{eqn:mdot.weak.torques} with  $\alpha(\Lambda)\sim1$).  
    This prediction works reasonably well at the highest $\dot{M}$, because the second-order formula in $|a|$  approaches that for strong torques (in Fig.~\ref{fig:inflow.vs.sims}) 
    when $|a|\rightarrow1$.   However, over the majority of the dynamic range in $\dot M$ the weak-torque prediction significantly under-predicts the true inflow rate.   Note that the axes here are the same as in Figure~\ref{fig:inflow.vs.sims}.  As a result many of the low simulated $\dot M$ points (y-axis) do not show up in this Figure because they have very small predicted $\dot M$ (x-axis) and are thus off the panel to the left.  \label{fig:inflow.vs.sims.weak}}
\end{figure}

Figure~\ref{fig:inflow.vs.sims.weak} shows the same simulation results as Figure~\ref{fig:inflow.vs.sims}, but compares them to the predicted inflow in the weak-torque regime, equations~\ref{eqn:tau.weak.torques} \& \ref{eqn:mdot.weak.torques}.   These scalings do fine at high $\dot M$, but fail to reproduce the numerical results at lower $\dot M$ and are clearly systematically discrepant. 
If we use the results of \citet{lynden-bell:1972.spiral.amplification} 
for direct angular momentum transport by the non-axisymmetric modes in the gas (also 
second-order in $|a|$; see eq.~\ref{eqn:mdot.kalnajs}), we obtain a similar, but more severe, discrepancy.  This is because the inflow rate induced by the direct transport by non-axisymmetric modes (eq.~\ref{eqn:mdot.kalnajs}) is smaller than the weak-torque inflow produced by the stars on the gas (eqs~\ref{eqn:tau.weak.torques} \& \ref{eqn:mdot.weak.torques}) by at least a factor of $\sim \Sigma_{\rm gas}/\Sigma_{\rm tot}$ (see \S  \ref{sec:deriv.mom.weak}).   Overall, Figure~\ref{fig:inflow.vs.sims.weak}  highlights that for the entire mass and radial range we have simulated, the dominant torque is well-modeled as originating from shocks induced in the gas by the stars.

\vspace{-0.5 cm}
\section{Bridging the Gap: from Galaxy to BH}
\label{sec:unified}

The literature contains a number of estimates of the accretion rate onto a central massive BH 
and/or through a given radial annulus, none of which properly account for the physics we have elucidated here.  Here we briefly review several of these estimates (\S \ref{sec:unified:other}) and then present a new analytic model for the inflow rate onto a central BH given the conditions at larger radii in its host galaxy (\S \ref{sec:unified:new}).

\vspace{-0.5 cm}
\subsection{Previous Accretion Models}
\label{sec:unified:other}

In semi-analytic and numerical models that lack the resolution to study gas transport from $\sim 0.1-1$ kpc to $\lesssim 0.1$ pc, simple subgrid prescriptions are necessarily used to estimate the accretion rate onto a central BH.   One common estimate of the BH accretion rate is the Bondi accretion rate for spherical accretion (e.g., \citealt{dimatteo:msigma}):
\begin{equation}
\label{eqn:mdot.bondi}
\dot{M}_{\rm Bondi} = 4\pi\,\alpha_{\rm B}\,\frac{G^{2}\,M_{\rm BH}^{2}\,
\rho_{\rm gas}}{c_{s}^{3}} 
\end{equation}
where 
$c_{s}$ and $\rho_{\rm gas}$ are the ambient sound speed and gas density, respectively, and $\alpha_{\rm B}$ is a constant that depends on the mass profile and gas equation of state; $\alpha_{\rm B}$ is sometimes taken to be as large as $\sim 100$ \citep{springel:models}.   Although perhaps appropriate for pressure-supported gas (e.g., in clusters of galaxies), the Bondi rate is not well-motivated for the cold, nearly Keplerian, gas that dominates the mass in disk galaxies.
%\footnote{{\bf Moreover, 
%even where appropriate, the Bondi rate as described above only applies to systems 
%with pure, non self-gravitating gas around a BH; without resolving the radii where the BH truly dominates the potential, the application of this estimator at different radii can give order-of-magnitude differences 
%\citep[see e.g.][]{colpi:2007.binary.in.mgrs,volonteri:2008.binary.bh.mdot}.}} 
Indeed, we shall see that equation~\ref{eqn:mdot.bondi} is not a good approximation to our numerical results.

For rotationally supported gas, the inflow rate is determined by the rate of angular momentum transport.   If angular momentum transport in galaxies were primarily produced by local stresses in a thin disk, the effective kinematic viscosity would be $\sim \alpha_V\,c_{s}\,h$, yielding an inflow rate
\begin{equation}
\label{eqn:mdot.viscous}
\dot{M}_{\rm viscous} = 3\pi\,\alpha_{\rm V}\,c_{s}\,\Sigma_{\rm gas}\, h
\end{equation}
where $h \sim c_s/\Omega$ is the disk's vertical thickness and $\alpha_{\rm V}$ is the dimensionless viscosity parameter (assumed spatially constant).

\begin{figure}
    \centering
    \scaleup
    \plotone{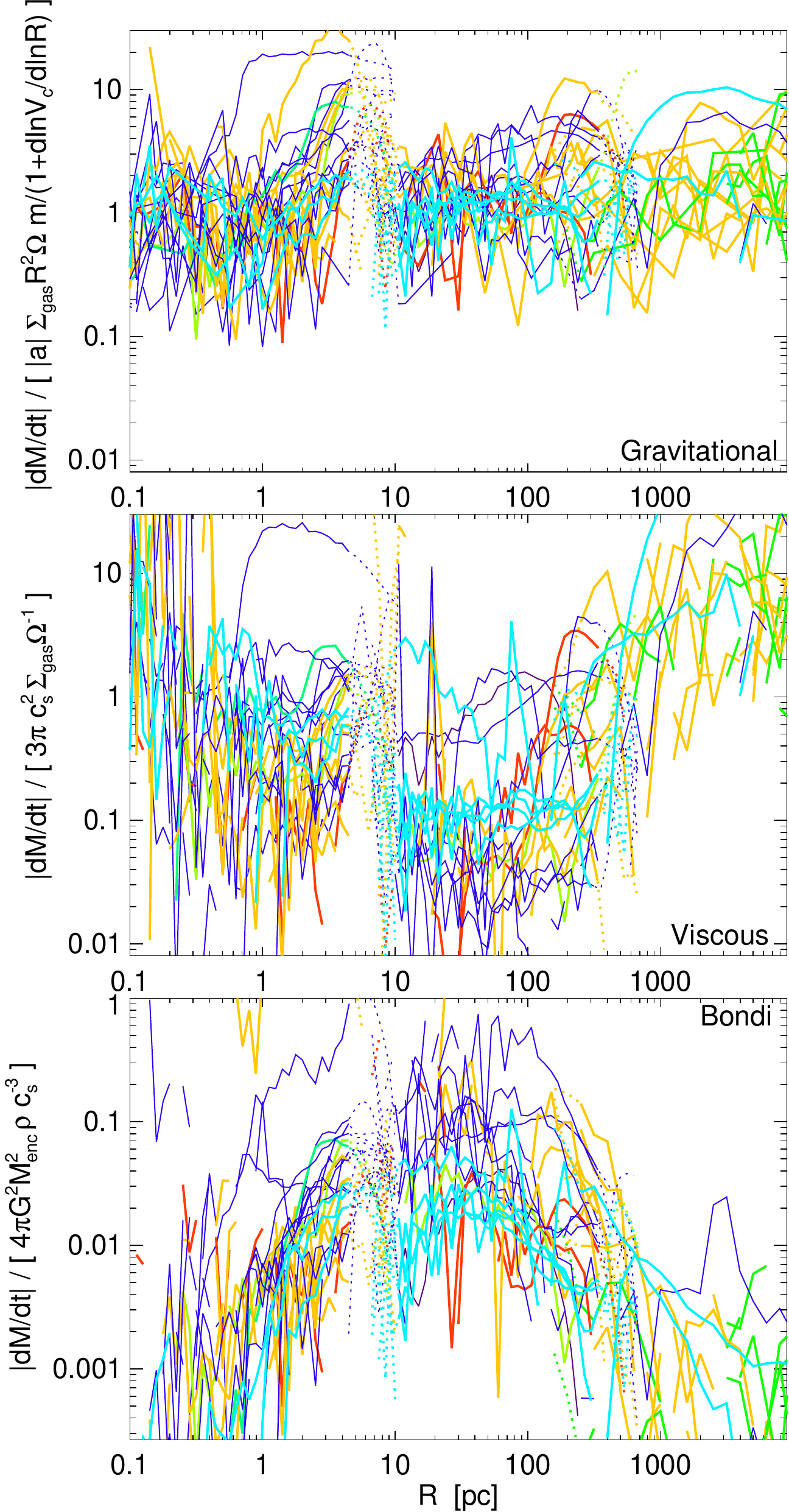}
    %\plotone{f1.pdf}
    \caption{
    Comparison of simulated and predicted gas inflow rates as a function of radius.  The simulations are shown during the strong inflow phases, as described in Figure~\ref{fig:inflow.rate.merger.vpred}.  
{\em Top:} The inflow rate in the
      simulations at each radius (${\rm d}M/{\rm d}t$) compared to the simple
      estimate for strong gravitational torques  (eq.~\ref{eqn:mdot.strongtorque.simple}).
      Each solid line denotes a  different simulation, with dotted lines showing the radii near the
      boundaries of our re-simulations, where the exact profile shape
      is suspect; colors correspond to initial gas fractions as in
      Figure~\ref{fig:inflow.rate.merger.vpred}.  
           {\em Middle:} Simulated inflow rate versus the scaling expected for
      accretion via local viscous stresses (eq.~\ref{eqn:mdot.viscous}).  
      {\em Bottom:} Simulated inflow rate versus a Bondi-Hoyle-like estimate  
      (eq.~\ref{eqn:mdot.bondi} with $M_{\rm BH}\rightarrow M_{\rm enc}(<R)$). 
     The Bondi-Hoyle and viscous accretion rate formulations 
      do not accurately capture the dominant physics in the numerical simulations, while the strong gravitational torque estimate does quite well. 
    \label{fig:mdot.model.comparison}}
\end{figure}

Both equations~\ref{eqn:mdot.bondi} and \ref{eqn:mdot.viscous}  depend quite 
differently on the ambient gas properties than the inflow rate produced by 
stellar gravitational torques and orbit crossing (eq.~\ref{eqn:mdot.strongtorque.simple} \&~\ref{eqn:mdot.full}).  
To highlight  this, Figure~\ref{fig:mdot.model.comparison} shows the ratio of the true 
inflow rate through a given annulus $R$ in our simulations to the 
viscous accretion rate (eq. \ref{eqn:mdot.viscous}) 
and the order of magnitude gravitational torque inflow rate 
from equation~\ref{eqn:mdot.strongtorque.simple}.  {Because we show the results as a function of 
scale, all simulations on various scales are shown together (but each is only shown at the 
well-resolved and self-consistently modeled radii appropriate to it). }
We also compare the simulation results to a generalization of the Bondi-Hoyle formula appropriate for larger radii in a galaxy, with $M_{\rm BH} \rightarrow M_{\rm enc}$, the total enclosed mass within a radius.  {This is strictly valid for inflow through an annulus $R$ only if $M_{\rm enc}$ changes relatively slowly with $R$ so that the spherical accretion sonic radius is at $> R$.  We include it to emphasize how poorly the simple Bondi-Hoyle scaling does at describing the radial variation of $\dot M$ in galactic simulations 
\citep[for discussion of how 
Bondi rates change when applied outside the Bondi radius, see][]{colpi:2007.binary.in.mgrs,volonteri:2008.binary.bh.mdot}.}  
The results  in Figure~\ref{fig:mdot.model.comparison} are for randomly chosen times near the peak of activity in each of the simulations from Figures~\ref{fig:inflow.rate.merger.vpred}-\ref{fig:inflow.vs.sims} 
\citep[the re-simulations from][]{hopkins:zoom.sims}. 
Note in particular that Figure~\ref{fig:mdot.model.comparison}  shows the results of three different sets of simulations, with the intermediate and nuclear-scale simulations using initial conditions drawn from the results of the larger-scale simulations; the boundaries between the different sets of simulations are shown with dotted lines.
Runs shown also include systematic numerical resolution studies discussed in \citet{hopkins:zoom.sims}.

Figure~\ref{fig:mdot.model.comparison} shows that the gravitational torque 
model works well at all radii, with $\lesssim0.3\,$dex scatter and without significant 
systematic deviations over the range of simulations we have carried out.    
By contrast, the viscous and Bondi accretion rate fare poorly by comparison.  Not only is there factor of $\sim 10$ scatter, but there are very large systematics as a function of radius (and in some cases, as a function of model parameters such as gas fraction and/or ISM equation of state parameterization $q_{\rm eos}$).

%\begin{figure}
%    \centering
 %   \scaleup
   % \plotone{mdotppr_mdotvsouter_vsmodel.pdf}
    %\plotone{f1.pdf}
  %  \caption{Comparison of the BH accretion rate in our simulations with the viscous ({\em left}; $\alpha_V = 0.1$) and Bondi ({\em right}) accretion rate  estimates evaluated at two large radii: 
    %$100\,$pc ({\em top}) and $1\,$kpc ({\em bottom}).  Solid diagonal line shows equality, dotted lines $\pm1$\,dex.  The viscous and Bondi accretion rate estimates both fare poorly, with very large scatter and a large systematic offset for the Bondi estimate.
  %In the simulations, the BH accretion rate is estimated using the inflow rate at $< 0.1$ pc in nuclear-scale re-simulations that have initial conditions taken from the results of galactic-scale simulations with the  appropriate conditions at $\sim 100$ pc or $\sim 1$ kpc.
 %\label{fig:mdotBH.model.comparison}}
%\end{figure}

\begin{figure}
    \centering
    \scaleup
    \plotone{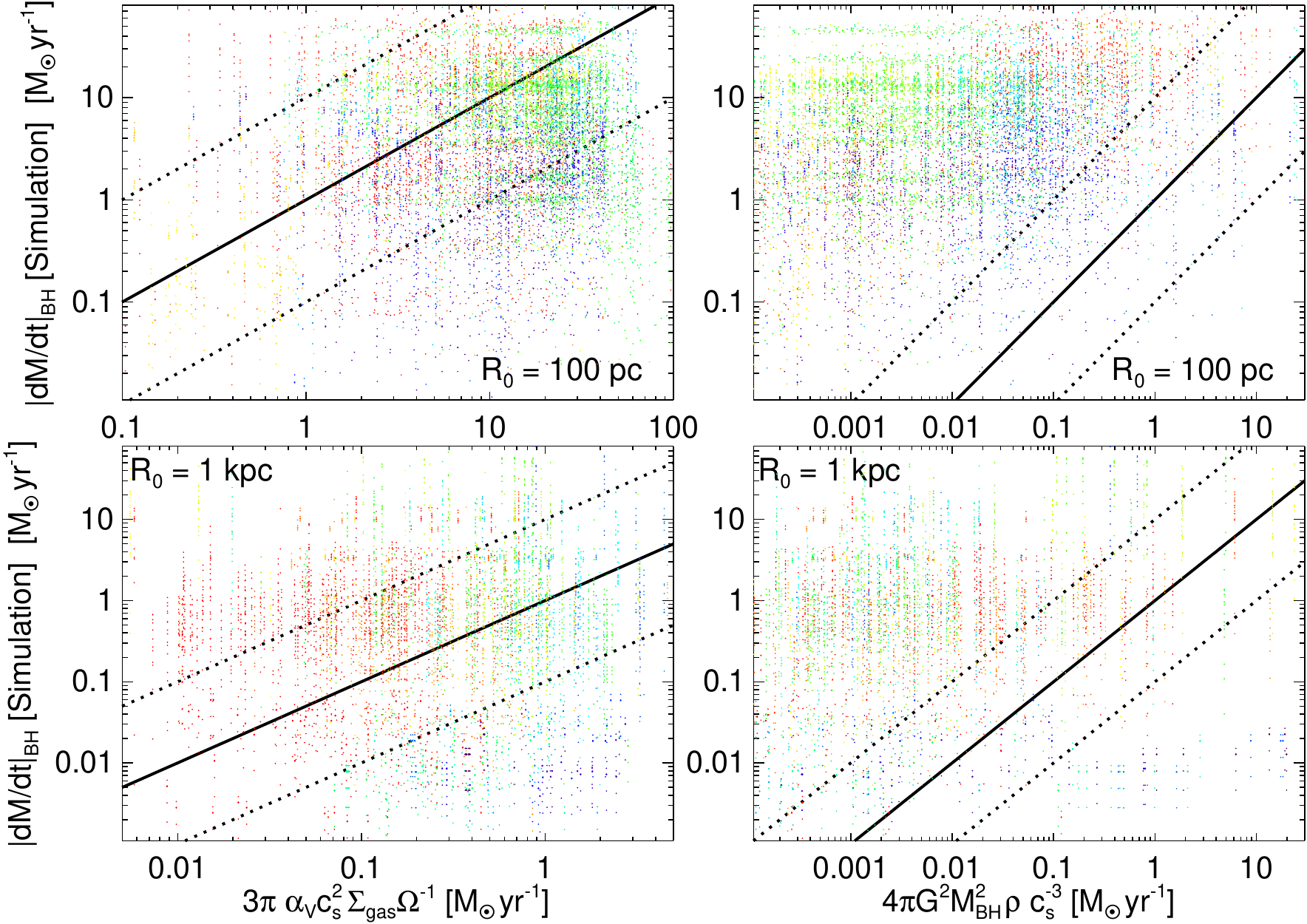}
    %\plotone{f1.pdf}
    \caption{Comparison of the BH inflow rate 
    in our simulations with the viscous ({\em left}; $\alpha_V = 0.1$) and Bondi ({\em right}) accretion rate  estimates evaluated at two large radii: 
    $100\,$pc ({\em top}) and $1\,$kpc ({\em bottom}).  Solid diagonal line shows equality, dotted lines $\pm1$\,dex.  
    Colors denote gas fraction (inside each radius) as in Figure~\ref{fig:inflow.rate.merger.vpred}. The viscous and Bondi accretion rate estimates both fare poorly, with very large scatter and a large systematic offset for the Bondi estimate.
    {In the simulations, the BH ``accretion rate'' is estimated using the inflow rate at $< 0.1$ pc in nuclear-scale re-simulations that have initial conditions taken from the results of intermediate-scale or galactic-scale simulations with the  appropriate conditions at $\sim 100$\,pc or $\sim 1$\,kpc, respectively.}
 \label{fig:mdotBH.model.comparison}}
\end{figure}

Figure~\ref{fig:mdotBH.model.comparison} demonstrates a similar point by comparing different predictors of the BH accretion rate given the conditions in the galaxy at large radii.  Specifically, we compare the viscous or Bondi-Hoyle accretion rate evaluated at a large radius $R_{0}$ (as would be done using such prescriptions as sub-grid models) to the true inflow rate near the 
accretion disk.  The latter is taken here to be the inflow rate at $<0.1\,$pc in our re-simulations that use initial conditions drawn from the galactic-scale conditions at $R_0$ (as in Fig.~\ref{fig:mdot.model.comparison}).\footnote{
{In a simple alpha-disk prescription, the inflow rate onto the outer edge of the disk 
corresponds to the actual accretion rate onto the BH. 
In practice, there may be a complex transition to this inner disk, for which 
some intermediate-scale model \citep[e.g.][]{power:2010.sph.bh.particle.eating} should be applied to 
map the efficiency of inflow through to the viscous disk. 
}
}
For the viscous accretion rate, we use $\alpha_{\rm V}=0.1$, although we consider this normalization to be arbitrary.  For the Bondi-Hoyle accretion rate, we use $\alpha_{\rm B}=1$, which assumes that the density and sound speed are relatively constant inside $R_0$.    

Figure~\ref{fig:mdotBH.model.comparison} shows that neither the Bondi nor viscous models predict the actual inflow rates at small radii.   For the Bondi rate, there is essentially no correlation between the true and predicted inflow rates.   For the viscous rate, $\alpha_V \sim 0.1$ gets the median $\dot M$ approximately correct, but there is factor $\gtrsim 10$ scatter and only a weak correlation between the true and predicted inflow rates.\footnote{It may appear surprising that the local viscous estimate is even accurate to a factor of $\sim 100$.   This is because the sound speed in our simulations is $\gg$ the true thermal speed of neutral molecular or atomic gas (which dominates the mass); using the latter for $c_s$ would lead to a much smaller $\dot M_{\rm viscous}$.  Our viscous estimate is thus more akin to an estimate of the inflow rate produced by turbulence generated by stellar feedback or strictly 
local, sub-sonic gravitational instabilities \citep[][]{gammie:2001.cooling.in.keplerian.disks,
lodato:2004.acc.disk.spiralwaves,durisen:2007.grav.instab.in.protoplantary.disks}.
}  The predicted inflow rate also depends systematically on the radius at which it is evaluated (as in Fig.~\ref{fig:mdot.model.comparison}). Variations in the sound speed $c_s$ drive much of the horizontal scatter in the predicted $\dot{M}_{\rm BH}$ in Figure~\ref{fig:mdotBH.model.comparison}, at a given true $\dot{M}_{\rm BH}$.   This is because the Bondi and viscous models both depend strongly on $c_{s}$ (eqs.~\ref{eqn:mdot.bondi} \& \ref{eqn:mdot.viscous}), whereas the true inflow rate is only a weak function of sound speed 
(at least for the stellar-dominated systems that are the particular focus here).

\vspace{-0.5cm}
\subsection{A Gravitational Torque Model}
\label{sec:unified:new}

In this section, we use the analytic results from \S~\ref{sec:response} to derive a physically motivated estimate of the inflow rate $\dot{M}$ at a given radius in a galaxy, which approximates the numerical simulations described in \S~\ref{sec:sims.main}. Our goal is to develop a model of the inflow rate 
that can be extrapolated over a large range of spatial scales, both to provide physical insight and to approximate the BH accretion rate on small scales from properties estimated on galactic or $\sim100\,$pc scales (e.g., in cosmological simulations or semi-analytic models).

We assume that the galaxy properties are known at a radius $R_{0}$, outside the BH radius of influence. 
We also assume that the stellar disk and bulge have approximately power-law 
surface density profiles, i.e., $\Sigma_{d,\,\ast} \propto R^{-\eta_{d}} $ and
$\Sigma_{b} \propto R^{-\eta_{b}}$. This is reasonably accurate in our simulations, 
so long as $R_{0}<R_{e}$ (the galaxy effective radius), and it allows us to make analytic progress.   
We will typically consider $\eta_b \sim \eta_d$ as well, i.e., that the bulge and 
disk have similar density profiles.

As argued in \S \ref{sec:deriv.gasden}, the gas in galaxies reaches a 
quasi steady-state in which its radial distribution is set by a 
competition between inflow to smaller radii and star formation (\citealt{thompson:rad.pressure}):
\be 
\label{eqn:steady.state.gas}
\frac{\partial}{\partial R}\,\dot{M}_{\rm inflow} \sim 2\pi\,R\,\dot{\Sigma}_{\ast}\ .
\ee
The qualitative behavior of solutions to equation (\ref{eqn:steady.state.gas}) depends on the ratio of the star formation time $t_*$ to the inflow time $t_{\rm inf}$.   For $t_{\rm inf} \ll t_*$, the solution is $\dot M_{\rm inflow} \simeq {\rm const} \gg \dot M_*$, while if $t_{\rm inf} \gg t_*$, $\dot M_{\rm inflow}$ declines precipitously at smaller radii as gas is consumed by star formation.   The observed star formation times in galaxies are $\sim 10-100$ dynamical times, which is comparable to the inflow time for $a \sim 0.01-0.1$.  This is similar to the mode amplitudes in our simulations and thus we expect $t_{\rm inf} \sim t_*$.    A reasonable analytic approximation to the numerical solutions of equation (\ref{eqn:steady.state.gas}) in this limit is given by taking $\dot M_{\rm inflow} \simeq \dot M_* \simeq \pi R^2 \dot \Sigma_*$.   More precisely, this approximate solution is valid when it itself implies $\dot M_{\rm inflow} \propto R^{p}$ with $p > 0$, so that $\dot M_{\rm inflow}$ declines at small radii (as it must in the correct solution).  This is not guaranteed for all $\eta_d$, $\eta_b$, $\eta_K$, etc., but is satisfied for the parameters we utilize below. The exact normalization will depend on details of the star formation law 
beyond just the power-law scaling (e.g.\, small-scale structure not treated explicitly in the simulations), 
so we consider it somewhat uncertain. 

In the potential of the galaxy, the dominant mode of angular momentum transport is an $m = 2$ bar mode.   To estimate $\dot M_{\rm inflow}$,  we use equation~\ref{eqn:mdot.full}, with
the WKB approximation for $|\Phi_{1}|$, 
\be 
|\Phi_{1}|\approx |a|_{\rm max}\,2\pi\,G\,\Sigma_{d}\,R\,|kR|^{-1}.
\ee
We will eventually take $k R \sim 1$ because the most important modes are global.   Since we are focusing on small scales in the galaxy ($\lesssim 0.1-1$ kpc), we would like to know the saturation mode amplitude $|a|_{\rm max}$ appropriate for self-excited modes (rather than the mode amplitude induced directly by an external perturber).  This is in general a complex function of non-linear processes such as heating by the mode, fragmentation, inflow, and momentum exchange with the halo.  \citet{hopkins:zoom.sims} show the saturation amplitude $|a|_{\rm max}$ in simulations on this scale 
as a function of various properties of the galaxy (their Figure~12); the results 
can be reasonably approximated by $|a|_{\rm max}\approx a_{0}\,f_{d}$ 
where $f_{d}=\pi\,G\,\Sigma_{d}/\Omega^{2}\,R \approx M_{d}(R)/M_{\rm enc}(R)$ and 
$a_{0}\sim 0.3$. 
Given this, equation~\ref{eqn:steady.state.gas} can now be approximated as
\begin{align}
\label{eqn:eqm.sigma.g}
\left(\frac{\Sigma_{g}}{\Sigma_{K}}\right)^{\eta_{K}-1} &= 
\alpha_{d}\,\left(\frac{\Sigma_{d}}{\Sigma_{t}}\right)^{2}\,\Omega\,t_{K} \\ 
\alpha_{d} &\equiv \frac{m\,a_{0}\,|2\nu_{\Sigma_{d,\,\ast}}+\nu_{\Sigma_{g}}-3\nu_{\Omega}|\,
(2+\nu_{\Sigma_{t}})^{2}\,F(\zeta) }{4\pi\,(2+\nu_{\Omega})\,|kR|} \\ 
%\alpha_{d} &\equiv \frac{m\,a_{0}\,
%(2+\nu_{\Sigma_{t}})^{2}\,F(\zeta) }{2\pi\,(1+s)\,|kR|} \\ 
\nu_{u} &\equiv \frac{\partial \ln u}{\partial \ln R} 
%a_{0}\,\tilde{c}(a_{0}) &= \frac{(3-s)^{1/2}}{\sqrt{2}\,(1+s)}\,
%\frac{|2\,(1+s)-m^{2}|}{2\,(1+s)+m^{2}} \label{eqn:a0}
\end{align}
where $\Sigma_{K}$ and $t_{K}$ are the parameters of the Schmidt-Kennicutt law (eq.~\ref{eqn:ks.law}),
$\Sigma_{t}=\Sigma_{d}+\Sigma_{b}$, and we take
$\Omega^{2}(R) \approx G\,M_{\rm enc}(<R)\,R^{-3}$.
Of course, the dominant term need not be just a bar, but the important point is that the 
dimensional scaling is the same for each  -- 
in the linear regime, modes are independent so 
each mode enters as $\sim\,m\,a_{m}\,F(\zeta_{m})$ and the relevant normalization 
reflects the sum over the spectrum of modes (in simulations, this typically does not differ 
from the single-mode normalization by more than a factor of a few).

Near where BH dominates the potential, an $m = 1$ mode 
often dominates the transport. Again, in real systems there can be a rich mode structure, 
but we focus on this case for convenience and tractability. 
For a power-law mass profile, $M(<R) = M_{\rm enc}(<R_{0})\,(R/R_{0})^{2-\eta}$, 
it is straightforward to solve for the radius $R_{\rm BH}$ at which the 
BH dominates the potential, i.e., where $M(<R_{\rm BH})=M_{\rm BH}$.  
%For a power-law disk mass profile, $M_{d}(<R) = M_{d}(<R_{0})\,(R/R_{0})^{2-\eta_{d}}$, 
%it is straightforward to use equation~\ref{eqn:eqm.sigma.g} to solve for the radius 
%$R_{\rm BH}$ at which the BH dominates the potential, i.e., where $M_{d}(<R_{\rm BH})=M_{\rm BH}$.  
%Using equation~\ref{eqn:eqm.sigma.g}, we determine $\Sigma_{g}$ and $\Sigma_{d}$ at this radius.   
The inflow rate in towards the BH from this radius then follows from equation~\ref{eqn:mdot.full}, 
but now using the coefficients evaluated for an $m=1$ mode in a 
roughly Keplerian potential.   Because the modes here are global, 
we assume that $\Phi_{1}(<R_{\rm BH}) = a\,G\,M_{d}(R_{\rm BH})/R_{\rm BH}$ 
in the BH potential (the local/WKB term is small as $R\rightarrow0$).    
The approximate analytic solution of equation (\ref{eqn:steady.state.gas}) is 
again $\dot M_{\rm inflow}(R) \simeq \dot M_*(R)$, which 
implies $\Sigma_{g}(R<R_{\rm BH}) \propto R^{-\eta_{\rm BH}}$, with 
$\eta_{\rm BH}=1/2\,(\eta_{K}-1)$. 
Specifically, in the limit of large gas supply, this leads to
\begin{align}
\label{eqn:sigmagas.bh}
\left( \frac{\Sigma_{g}}{\Sigma_{K}} \right)^{\eta_{K}-1} &= 
\alpha_{\rm BH}\,\sqrt{\frac{G\,M_{\rm BH}}{R_{\rm BH}^{3}}}\,t_{K}\,
\left(\frac{R}{R_{\rm BH}}\right)^{-1/2}\,f_{d}(R_{\rm BH})\\ 
\alpha_{\rm BH} &= a_{1}\,F(\zeta_{1})\,\frac{3\,\eta_{K}-4}{2\pi\,(\eta_{K}-1)}\,\frac{2+\nu_{\Sigma_{K}}}{2}
\end{align}
where $a_{1}\approx 0.2$ is defined by $|a|_{\rm max}=|\Phi_{1}(R_{\rm BH})|/\Phi \approx a_{1}\,f_{d}(R_{\rm BH})$ and the numerical value is based on the simulations of 
modes in the BH potential from \citet{hopkins:zoom.sims}.
%and $f_{d}(R_{\rm BH}) = M_{d}(<R)/M_{\rm enc}(<R)$ at $R_{\rm BH}$. 
Together with the solution for $R_{\rm BH}$ using equation~\ref{eqn:eqm.sigma.g}, this 
in turn gives the following BH inflow rate at a radius $R$:
\begin{align}
\label{eqn:mdot.all.1}
\dot{M} &= \tilde{\alpha}\,
\Sigma_{K}\,R_{0}^{2}\,\Omega(R_{0})\,
\left( \frac{R}{R_{\rm BH}}\right)^{\frac{3}{2}-\eta_{\rm BH}}\,
\mu_{\rm BH}^{\gamma}\,\\
\nonumber &\ \ \ \ \ 
%\times\,(2+R_{b})^{\frac{-3}{2\,(\eta_{K}-1)}}\,
%\times\,f_{d}^{\frac{3}{2\,(\eta_{K}-1)}}\,
\times\,f_{d}^{\frac{3\,\eta_{K}\,(1-\eta_{d})}{2\,(\eta_{K}-1)\,(2-\eta_{d})}}\,
[\Omega(R_{0})\,t_{K}]^{\frac{1}{\eta_{K}-1}}\\ 
\Omega(R_{0}) &\equiv \left(\frac{G\,M_{\rm enc}(R_{0})}{R_{0}^{3}}\right)^{1/2} \\ 
\mu_{\rm BH} &\equiv \frac{M_{\rm BH}}{M_{d}(R_{0})} \\ 
%R_{b} &\equiv \frac{M_{b}(R_{0})}{M_{d}(R_{0})} \\ 
\gamma &\equiv \frac{-4+(3-\eta_{d})\,\eta_{K}}{2\,(2-\eta_{d})\,(\eta_{K}-1)} \\
\tilde{\alpha} &\equiv 
(2+\nu_{\Sigma_{t}})\,a_{1}\,F(\zeta_{1})\,\alpha_{\rm BH}^{\frac{1}{\eta_{K}-1}} 
\end{align}

Equation~\ref{eqn:mdot.all.1}  is an estimate of the BH 
inflow rate in the limit of large gas supply 
(because we have used the $\dot M_{\rm inflow} \sim \dot M_*$ solution to 
eq.~\ref{eqn:steady.state.gas}).   If the gas supply at $R_{\rm BH}$ is less than a critical amount the inflow rate will be smaller than given by equation~\ref{eqn:mdot.all.1}.   The critical gas mass $M_{\rm gas,\,c}$ which defines the large gas supply limit is found by integrating $\Sigma_{\rm gas}$ in equation~\ref{eqn:sigmagas.bh}
from $R=0$ to $R=R_{\rm BH}$. 
Recall as well that the intermediate-scale physics described by
equation~\ref{eqn:eqm.sigma.g} itself sets the amount of gas that reaches $R_{\rm BH}$. The gas mass supplied to 
this radius, $M_{\rm gas,\,bh}$, can be estimated by integrating 
$\Sigma_{\rm gas}$ as implied by equation~\ref{eqn:eqm.sigma.g} 
over the same range. 
For $M_{\rm gas,\,bh} \gtrsim M_{\rm gas,\,c}$, 
equation~\ref{eqn:mdot.all.1} for the BH inflow rate is reasonably accurate 
while for $M_{\rm gas,\,bh} \lesssim M_{\rm gas,\,c}$ it overestimates the inflow rate.  
To quantify the suppression in $\dot M$ for $M_{\rm gas,\,bh} \lesssim M_{\rm gas,\,c}$, 
one can solve equation~\ref{eqn:steady.state.gas}  as a boundary value problem with 
a given gas mass $M_{\rm gas,\,bh}$ 
at $R_{\rm BH}$ (as, e.g., \citealt{thompson:rad.pressure} did 
using somewhat different assumptions about the inflow and star formation physics).   
For low gas masses the solution is an inflow rate that is 
approximately independent of radius and $\propto M_{\rm gas,\,bh}$, 
implying that the suppression of $\dot M$ is given by
\begin{align}
\dot{M}  &\rightarrow \dot{M}\,
\left[1 + \frac{M_{\rm gas,\,c}}{M_{\rm gas,\,bh}}\right]^{-1}
\label{mdot.sup} \\ 
\frac{M_{\rm gas,\,c}}{M_{\rm gas,\,bh}} &= 
\alpha_{g1}\,\left( \frac{\Sigma_{d}}{\Sigma_{t}} \right)^{-\frac{1}{\eta_{K}-1}} \\ 
\alpha_{g1} &= \frac{4\eta_{K}-5-\eta_{d}}{4\eta_{K}-5}\,\left( \frac{\alpha_{\rm BH}}{\alpha_{d}}
\right)^{\frac{1}{\eta_{K}-1}}
\end{align}

Following the same logic, a similar suppression factor in $\dot M$ should appear if the gas mass available at $R_0$ is too small. Specifically, the solution for the gas surface density profile between $R_{\rm BH}$ and $R_0$ in equation~\ref{eqn:eqm.sigma.g} is valid only in the limit of large mass supply.  It requires that the available gas mass at $R_0$, $M_{\rm gas}(R_0)$, exceed a critical value $M_{\rm gas,\,i}$ that can be found by by integrating $\Sigma_{\rm gas}$ in equation~\ref{eqn:eqm.sigma.g} from $R=0$ to $R=R_{0}$.   The resulting suppression factor is then 
\begin{align}
\dot{M}  &\rightarrow \dot{M}\,
\left[1 + \frac{M_{\rm gas,\,i}}{M_{\rm gas}(R_{0})}\right]^{-1}
\label{mdot.sup.2} \\ 
M_{\rm gas,\,i} &= \alpha_{g}\,\Sigma_{K}\,R_{0}^{2}\,f_{d}^{\frac{2}{\eta_{K}-1}}
\left( \Omega(R_{0})\,t_{K} \right)^{\frac{1}{\eta_{K}-1}} \\ 
\alpha_{g} &\equiv \frac{4\pi\,(\eta_{K}-1)}{4\,\eta_{K}-\eta_{d}-5}\,\alpha_{d}^{\frac{1}{\eta_{K}-1}}
\end{align}

%The critical gas 
%mass $M_{\rm gas,0}$ at which this transition happens can 
%be found by integrating equation~\ref{eqn:eqm.sigma.g} out to $R_0$, which gives 
%\begin{align}
%M_{\rm gas,\,0} &= \alpha_{g}\,\Sigma_{K}\,R_{0}^{2}\,f_{d}^{\frac{2}{\eta_{K}-1}}
%\left( \Omega(R_{0})\,t_{K} \right)^{\frac{1}{\eta_{K}-1}} \\ 
%f_{d} &\equiv \frac{M_{d}(R_{0})}{M_{\rm enc}(R_{0})} \\ 
%\Omega(R_{0}) &\equiv \left(\frac{G\,M_{\rm enc}(R_{0})}{R_{0}^{3}}\right)^{1/2} \\ 
%\alpha_{g} &\equiv \frac{4\pi\,(\eta_{K}-1)}{4\,\eta_{K}+\eta_{d}-3}\,\alpha_{d}^{\frac{1}{\eta_{K}-1}}
%\left[
%\left( \frac{2-\eta_{d}}{2\pi} \right)^{2}\,\alpha_{d}
%\right]^{\frac{1}{\eta_{K}-1}}
%\end{align}

We now combine terms and evaluate the numerical coefficients and exponents in equations~\ref{eqn:mdot.all.1} and \ref{mdot.sup}.    We adopt star formation parameters 
consistent with those measured by \citet{bouche:z2.kennicutt}:
$\Sigma_{K}=10^{8}\,\msun\,{\rm kpc^{-2}}$, $t_{K}=0.41\times10^{9}\,{\rm yr}^{-1}$, 
and $\eta_{K}=7/4$.  We find reasonable consistency with our 
simulation results and the profiles of observed cusp ellipticals by taking 
$\eta_{b}\approx\eta_{d}\approx1/2$ over this radial range 
\citep[see e.g.][]{lauer:bimodal.profiles,hopkins:cusps.ell}.  We assume that the $R>R_{\rm BH}$ mode is $m=2$, while the $R<R_{\rm BH}$ mode is $m=1$, and that both are global, so that 
$|kR|, \zeta \sim1$. We emphasize that all of these are gross simplifications and shouldn't be taken 
to be universally true -- however they provide a starting point for the order of magnitude normalization of 
our derived inflow rates. 

Finally, equation (\ref{eqn:mdot.all.1}) should be evaluated at the 
radius $R_{\rm acc}$ where accretion itself maintains stability to local gravitational perturbations, i.e., $Q \gg 1$.   Inside this radius, star formation ceases, and $\dot M \simeq {\rm constant}$. The value of $R_{\rm acc}$ in fact depends on the mechanism of angular momentum transport.  For local angular momentum transport, $R_{\rm acc} \sim 0.01$ pc, while for larger-scale torques, $R_{\rm acc} \sim 0.1$ pc \citep{goodman:qso.disk.selfgrav}.  Our calculations favor the latter physics, so we take $R_{\rm acc} \sim 0.01 \, R_{\rm BH}$, which corresponds to $R_{\rm acc} \sim 0.1$ pc for typical parameters 
(it makes little difference, to within the scatter and accuracy of our formula, if we take 
$R_{\rm acc}$ to be a fixed physical radius instead). 

With these assumptions we find that our analytic gravitational torque inflow model predicts a BH inflow rate of 
\begin{align}
\label{eqn:mdot.predictor}
\frac{\dot{M}_{\rm BH,\,grav}}{\msun\,{\rm yr^{-1}}} &\approx \alpha(\eta_{K}) \,
%\dot{M}_{\rm BH,\, grav} &\approx
%\alpha(\eta_{K})\,M_{\sun}\,{\rm yr^{-1}}\,
f_{d}^{7/6}\,(1+1.71\,f_{d}^{-4/3})^{-1}\,\\
%\dot{M}_{\rm BH,\, grav} &\approx
%1.12\,\alpha(\eta_{K})\,M_{\sun}\,{\rm yr^{-1}}\,
%f_{d}^{11/6}\,(1+1.76\,f_{d}^{-2})^{-1}\,\\
\nonumber &\ \left( \frac{M_{\rm BH}}{10^{8}\,\msun} \right)^{1/6}\,
\left( \frac{M_{d}(R_{0})}{10^{9}\,\msun} \right)^{1}\,
\left( \frac{R_{0}}{100\,{\rm pc}} \right)^{-3/2}\,\\
\nonumber &\ \nonumber \left( \frac{R_{\rm acc}}{10^{-2}\,R_{\rm BH}} \right)^{5/6}\,
{\left[1+\frac{f_{0}}{f_{\rm gas}(R_{0})}\right]^{-1}} \nonumber 
\end{align}
where $\alpha(\eta_K)$ is a dimensionless constant discussed below.  In a simpler form, equation~\ref{eqn:mdot.predictor} becomes
\begin{align}
\label{eqn:mdot.predictor.simple}
%\nonumber 
\frac{\dot{M}_{\rm BH,\,grav}}{\msun\,{\rm yr^{-1}}} 
\approx
\alpha(\eta_K)\,f_{d}^{5/2}\,M_{\rm BH,\,8}^{1/6}\,M_{d,\,9}\,R_{0,\,100}^{-3/2}\,(1+f_{0}/f_{\rm gas})^{-1} 
%\exp{\left( -\frac{f_{0}}{f_{\rm gas}}\right)}
%e^{-f_{0}/f_{\rm gas}} \\ 
%e^{-f_{0}/f_{\rm gas}} \\ 
%\dot{M}_{\rm BH} &= {\rm MIN}\left[ 
%\dot{M}_{\rm BH,\,grav},\, \dot{M}_{\rm Edd}(M_{\rm BH})
%\right] 
\end{align}
with
\begin{align}
f_{0} &\approx 0.31\,f_{d}^{2}\,\left( \frac{M_{d}(R_0)}{10^{9}\,\msun} \right)^{-1/3} \\ 
f_{\rm gas}(R_{0}) &\equiv M_{\rm gas}(R_{0}) / M_{d}(R_{0})
\label{eqn:mdot.predictor.fgas}
%M_{\rm gas,\,0} &\approx 2.6\times10^{8}\,\msun\,
%f_{d}^{2}\,\left(\frac{M_{d}(R_{0})}{10^{9}\,\msun} \right)^{2/3}\,
%\left( \frac{\alpha_{g}}{0.0312} \right)
\end{align}
Recall that $f_d$ in equations~\ref{eqn:mdot.predictor} and \ref{eqn:mdot.predictor.simple} is the total (stellar and gas) disk mass fraction evaluated at $R_0$.    Note also that these predictions for $\dot M_{\rm BH,\,grav}$ can, under a variety of circumstances, exceed the Eddington accretion rate.   How much above Eddington the BH can in fact accrete depends on the uncertain physics of radiation-pressure dominated, super-Eddington accretion; much of the mass may in fact be unbound at radii smaller than we consider here.  This is less likely to be a concern for $\dot M_{\rm BH,\,grav} \lesssim \dot M_{\rm Edd}$.

Despite the complex expressions leading to the final result for $\dot M_{\rm BH, grav}$, the power-law exponents in equations~\ref{eqn:mdot.predictor} and \ref{eqn:mdot.predictor.simple} are reasonably insensitive to the values of $\eta_{d}$ and $\eta_{K}$ 
over the plausible range ($0\lesssim\eta_{d}\lesssim1$, 
$1.4\lesssim \eta_{K}\lesssim1.75$). 
The normalization, however, is fairly sensitive, especially to $\eta_{K}$, ranging from $\approx1\,\msun\,{\rm yr}$ to $\approx 18\,\msun\,{\rm yr}$ for $\eta_{K}=7/4$ to $3/2$,
respectively (for $\eta_{d}=1/2$). This is one reason why we explicitly include the constant $\alpha(\eta_{K})\sim1-10$ in equations~\ref{eqn:mdot.predictor} and \ref{eqn:mdot.predictor.simple}.  The dependence on $\eta_K$ is due to the fact that a larger $\eta_K$ implies a higher star formation rate in dense gas and thus a lower BH inflow rate. {Star formation on the small scales of interest here is uncertain -- 
we choose to parameterize that in $\alpha$ so that the 
 model has the advantage that it can be rescaled in a straightforward manner  for many different models of star formation (via an appropriate choice of $\eta_K$ or $t_K$). We have also made the over-simplified assumption 
 of just one mode at each scale, when the real behavior reflects a sum over a mode spectrum; this 
 does not change the dimensional scalings, but does enter the normalization $\alpha$.
 The exact values we adopt here 
are therefore unlikely to be universally valid, but are convenient, observationally motivated, and 
match the choice in most of our simulations.} 

\begin{figure}
    \centering
    \scaleup
    \plotone{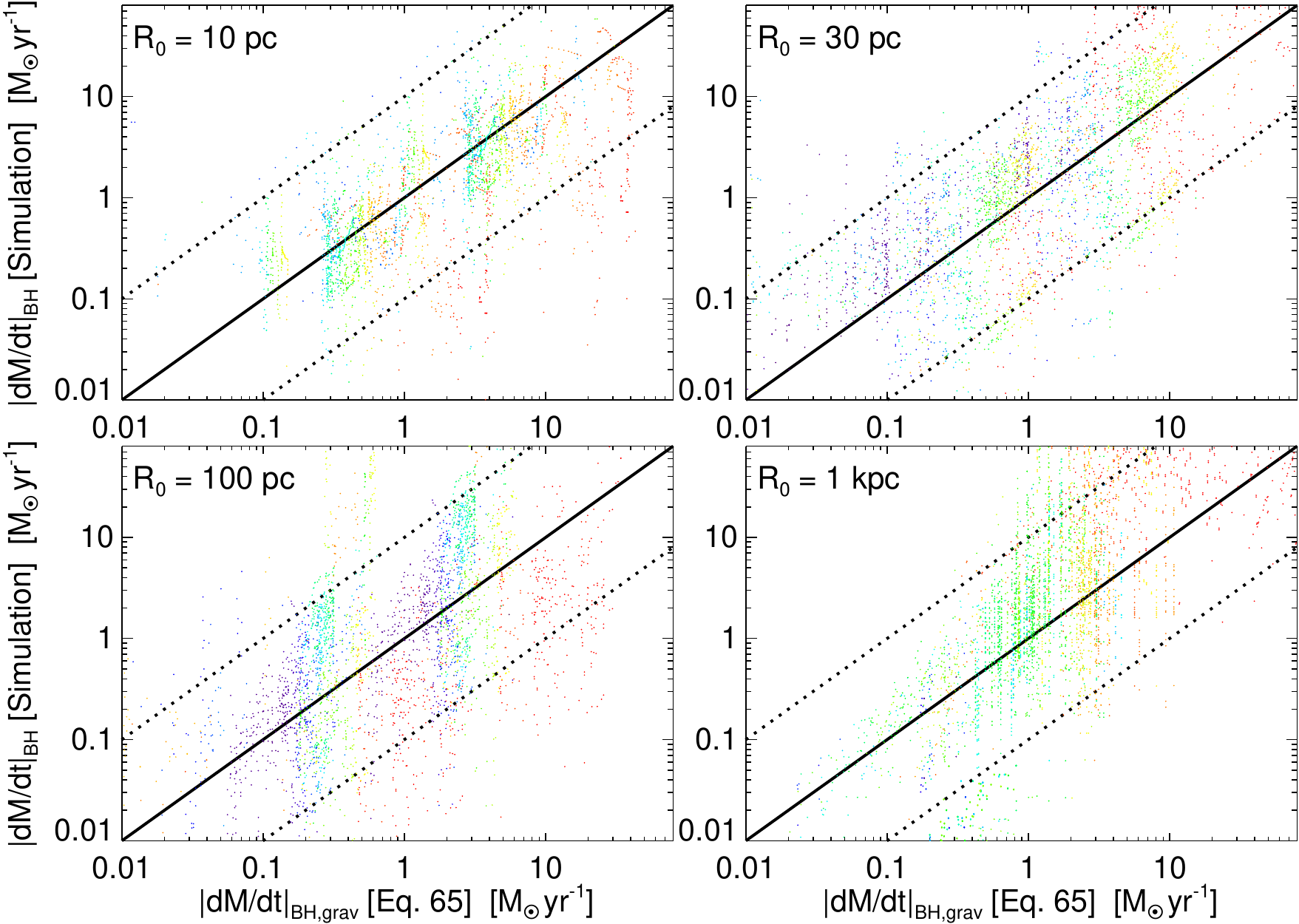}
    %\plotone{f1.pdf}
    \caption{Comparison of the BH inflow rate (into $<0.1\,$pc) in our simulations with the predictions of our analytic model of inflow driven by stellar gravitational torques (eq.~\ref{eqn:mdot.predictor.simple} with $\alpha(\eta_K) = 5$).
    In each panel, the properties used to infer $\dot{M}_{\rm BH}$  are measured at a different radius $R_{0}$, from $30\,$pc to $1\,$kpc (each from simulations modeling the appropriate scales). 
    Colors denote gas fraction (inside the given radius) as in Figure~\ref{fig:inflow.rate.merger.vpred}. 
    The analytic prediction works reasonably well and the agreement is much better than for the  Bondi or viscous predictions at all radii (see Fig.~\ref{fig:mdotBH.model.comparison}).   There remains, however, order-of-magnitude scatter -- mostly owing to time variability in the BH inflow rate that should be included in models. 
    \label{fig:mdot.grav.test}}
\end{figure}

Figure~\ref{fig:mdot.grav.test} compares equation~\ref{eqn:mdot.predictor.simple}, with $\alpha(\eta_K) \approx 5$ and with the relevant quantities
evaluated at different $R_{0}$ in the simulations, to the actual inflow rates into $<0.1\,$pc. 
There is, of course, very large scatter -- even with full knowledge of the mass profile and mode amplitudes, the inflow rates are chaotic and time-variable. 
And unsurprisingly, the accuracy of equation~\ref{eqn:mdot.predictor.simple} decreases for larger values of $R_{0}$.    Moreover, because the profile shapes and mode structure do not 
exactly match what we have assumed at all radii, 
the appropriate normalization of $\dot{M}_{\rm BH,\,grav}$ depends weakly 
on $R_{0}$.   In spite of its simplicity, however, equation~\ref{eqn:mdot.predictor.simple} captures the 
broad trends in the BH inflow rate, and is a much more accurate approximation than either the viscous accretion rate or Bondi-Hoyle accretion 
rate (compare  Figures~\ref{fig:mdot.grav.test} and \ref{fig:mdotBH.model.comparison}).  

As a sanity check on the predictions of equation~\ref{eqn:mdot.predictor.simple}, we briefly use it to estimate the inflow rate for two well studied systems:   the Milky Way and NGC 1068.
For the Milky Way, which has $M_{\rm BH} \simeq 4 \times 10^6 M_\odot$ \citep{ghez:mw.bh.mass} and $f_{\rm gas} \simeq 0.4$, $f_d \simeq 0.2$ and $M_{\rm gas}(R_0) \simeq 3 \times 10^7 M_\odot$ at $R_0 \simeq 100$ pc \citep{serabyn.morris:mw.nuclear.structure}, 
equation~\ref{eqn:mdot.predictor.simple} predicts $\dot M_{\rm BH,\,grav} \sim  3\times10^{-3}\, M_{\odot} \,{\rm yr^{-1}}$.   Although much larger than the current inflow rate onto Sgr A*, this is not unreasonable for a {time averaged inflow rate} given that BHs of $\sim 10^{7} M_\odot$ are growing significantly in mass in the local universe \citep{heckman:local.mbh}.   
For NGC 1068, a nearby Type 2 Seyfert with $M_{\rm BH} \simeq 1.5 \times 10^7 \, M_\odot$ 
\citep{greenhill:1997.maser.twist} and $L \sim 10^{45} \, {\rm ergs \, s^{-1}} \sim L_{\rm Edd}$ 
\citep{pier:1994.1068.spectrum}, 
the gas mass within $R_0 \simeq 100$ pc is $\sim 3 \times 10^7 M_\odot$  while the total dynamical mass is 
$\sim 3 \times 10^8 M_\odot$ \citep{helfer:1995.1068.molecular}.    
The total disk mass within this radius is uncertain 
but \citet{davies:sfr.properties.in.torus} find evidence for a stellar disk with $\sim 10^8 M_\odot$ within $\sim 100$ pc (or less).   Adopting  $f_{\rm gas} \sim 1/3$ and $f_d \sim 0.4$, equation~\ref{eqn:mdot.predictor.simple} predicts $\dot M_{\rm BH,\,grav} \sim 0.03\, M_{\odot} \,{\rm yr^{-1}}$ comparable to the Eddington accretion rate $\sim 0.1 \, M_{\odot} \,{\rm yr^{-1}}$ for NGC 1068.   There is, of course, no reason to expect that our estimate should accurately predict the inflow rate in a given system at a given time (as indeed it does not for the Milky Way!).   It is, however, reassuring that our estimates of $\dot M$ for the Milky Way and NGC 1068 are observationally plausible in a time-averaged sense. 

It is worth highlighting how and why our inflow rate prediction (eq.~\ref{eqn:mdot.predictor.simple}) depends on several key physical parameters:
\begin{itemize}
{\item $f_{d}$:} The inflow rate decreases 
rapidly with decreasing disk fraction (i.e., increasing bulge to disk ratio).
This is expected for inflow driven by global gravitational instability in the disk.
{\item $M_{\rm BH}$}:   The absolute inflow rate is not a strong function of $M_{\rm BH}$. 
In units of the Eddington accretion rate, the inflow rate actually decreases with increasing BH mass ($\sim M_{\rm BH}^{-5/6}$ for $\eta_d = 1/2$). This is because on small scales, the BH plays the same role as the bulge, suppressing global modes within $R_{\rm BH}$.
{\item $M_{d}$:} $\dot M_{\rm BH,\,grav}$ is linear in $M_d$, the {\em total} (gas+stellar) disk mass. 
{\item $f_{\rm gas}$:} Provided that sufficient gas is available 
($f_{\rm gas}\gtrsim f_{0}$), the inflow rates are nearly independent of the 
gas supply, because the gas densities at small scales are set by an equilibrium between 
inflow and star formation, which self-adjusts to a particular inflow rate.
Of course, at very high $\fgas$, our assumption of stellar-dominated disks breaks 
down, and it is not clear how robust the extrapolated model will be.
{\item $\dot{M}_{\ast}$:} The star formation rate enters into the expression for $\dot M_{\rm BH,\,grav}$ via the values of $t_{K}$, $\Sigma_{K}$, and $\eta_K$ from the Schmidt-Kennicut law (see eq.~\ref{eqn:mdot.all.1} for the precise dependencies).  A lower star formation efficiency (corresponding to larger $t_{K}$ and/or smaller $\eta_{K}$) at small radii in gas-rich galactic nuclei, 
as several authors have suggested \citep{thompson:rad.pressure,
    begelman:direct.bh.collapse.w.turbulence,larson:column.density.bh.vs.sf}, 
    would significantly increase the final BH accretion rate, 
    making it easier to grow a central massive BH. 
{\item $c_{s}$:} In our model, the inflow rate is independent of the 
sound speed $c_{s}$ and other thermal properties of the gas.  This is because the torques and inflow rate are determined by supersonic non-axisymmetric motions induced by global gravitational instabilities.
Again, caution is needed when systems have very large $\fgas$ and fragmentation can compete 
with global torques. 

\end{itemize}

Evaluating equation~\ref{eqn:mdot.predictor.simple} requires the input parameters $M_{\rm BH}$, $M_{\rm gas}(R_{0})$, $M_{d}(R_{0})$, $M_{b}(R_{0})$ and a choice of $R_{0}$ (ideally the 
smallest resolved radii, since Fig.~\ref{fig:mdot.grav.test} shows that the analytic approximations here are best at smaller radii $\lesssim 100$ pc -- but they are still far superior to the Bondi and viscous models even at $\sim$ kpc).   In analytic and semi-analytic models, these are all clearly-defined,  and implementing equation~\ref{eqn:mdot.predictor.simple} is straightforward. 
In numerical simulations, it is less clear how to define the bulge and disk separately  
in an on-the-fly manner.   One possibility that may be sufficient at the level of accuracy of our model is to define the disk mass fraction via the total angular momentum of the material within $R_0$ of the BH.

As noted above, there is considerable variability on a range of timescales in our simulations of AGN fueling -- see \citet{hopkins:zoom.sims} for a more extensive discussion.  
Very roughly, we find approximately equal power in fluctuations on each dynamical time resolved in the simulation.  Recently, \citet{levine:sim.mdot.pwrspectrum}  show this explicitly for timescales from $\gtrsim10^{7}$ to $\sim100\,$yr (see their Figures~3-5) 
using AMR simulations that cover a range of radii similar to our simulations.   Given this variability, it is unlikely that {\em any} BH accretion rate model could predict inflow rates on small scales given the properties of a galaxy on large scales with much better than order-of-magnitude accuracy (as in Fig.~\ref{fig:mdot.grav.test}).   Moreover, this variability can have important consequences for the effects of AGN feedback or the ionization of the intergalactic medium, to cite just two examples.    We thus believe that it is important to include variability when modeling BH growth and AGN feedback.\footnote{For example, following \citet{levine:sim.mdot.pwrspectrum}, in semi-analytic models or galactic-scale simulations one could approximate the variability on timescales smaller than some 
resolved timescale $t_{0}\sim \Omega_{0}(R_{0})$, and down to a minimum variability
timescale $dt$, using a Fourier series
\be
\dot{M}= \langle\dot{M}\rangle\,\prod_{i=0}^{N}
\exp{ \left\{ 
\frac{\sigma_{0}\,\ln{10}}{\sqrt{N/2}}\,f_{i}\,
\cos{\left[ \frac{\Omega_{0}\,t}{(\Omega_{0}\,dt)^{i/N}} + 2\pi\,\delta_{i}
\right]}
\right\}}
\ee
where $\sigma_{0}\sim0.5-0.6$ is the amplitude of variations (motivated by the simulations) and the $f_{i}$ and $\delta_{i}$ are a Gaussian random 
numbers drawn with dispersion of unity
and uniform random number from $0-1$, respectively; $N\gtrsim10$ appears sufficient for
a reasonable approximation of the variability in the simulations.}

\vspace{-0.5cm}
\subsection{Implications of our Gravitational Torque Model}
\label{sec:differences}

Having derived the new inflow rate predictor in equation \ref{eqn:mdot.predictor.simple} and shown that its functional form is significantly different from the Bondi or viscous accretion rate estimates, we briefly speculate on conditions under which we do or do not expect this to make a significant difference relative to previous results.   Because the problem of black hole fueling with feedback is highly nonlinear, it is entirely possible that the results of applying equation \ref{eqn:mdot.predictor.simple} in numerical simulations could be more subtle than we anticipate.  Thus it will be critical to explore this numerically as well.

We anticipate that our new fueling model may only have limited effects in the following circumstances:
\begin{itemize}

\item{Eddington-Limited Accretion:} In e.g.\  simulations of galaxy mergers, the AGN passes through a significant phase in which very large inflows from galactic scales lead to a large gas reservoir in the central $\sim 100$ pc.   The implied accretion rates for a range of models are comparable to, or in excess, the Eddington rate \citep{dimatteo:msigma,debuhr:momentum.feedback}.  This suggests that the Eddington limit is the rate-limiting factor in many cases, not the rate of mass supply.

\item{Feedback-Regulated Accretion:} Once BHs reach a certain mass, models that include BH feedback often lead to self-regulated BH growth.  In this limit, both simulations and analytic arguments suggest that the accretion rate is determined by the balance between feedback and inflow, independent of the precise functional form of the accretion rate model \citep{debuhr:2010.mom.fb.bhgrowth}.   Because the BH gains most of its mass during this phase, we suspect that the new accretion rate model proposed here will not significantly change previous inferences about the conditions required to reproduce the $M_{\rm BH}-\sigma$ relation and other BH-galaxy correlations.

\end{itemize}

There are, however, other contexts in which we suspect that the more physical treatment of BH fueling presented here will have a significant effect on the results of analytic models and galactic-scale simulations:
\begin{itemize} 

\item{AGN Lightcurves:}   Most BHs, most of the time, accrete below the Eddington rate.   In this limit, the predicted AGN accretion rate is more sensitive to the details of the accretion rate model.
As a result, our new accretion rate predictions may significantly affect the AGN lightcurves and the fraction of time a BH spends at a given accretion rate.  As emphasized by \citet{hopkins:lifetimes.interp}, this will in turn affect the interpretation of the AGN luminosity function.

\item{Early BH Growth:} The Bondi accretion rate is $\propto M_{\rm BH}^{2}$, which suppresses accretion onto low mass BH seeds at high redshift.  By contrast, the model presented here depends only weakly on BH mass --  low BH mass in fact allows global instabilities to form more readily and thus should promote rapid seed growth. 

\item{BH Growth Versus Galaxy Morphology:}   Our accretion model includes an explicit, strong dependence on the nuclear morphology of the galaxy.  BH growth via secular processes will 
be relatively more efficient in disk-dominated galaxies, but 
suppressed in bulge-dominated galaxies, potentially requiring 
more violent events (mergers) to grow BHs. 
Because the morphology as a function of radius is also important, the  
bulge structural properties are also relevant for the resulting BH accretion rate: e.g., low mass but sufficiently 
compact bulges can suppress accretion on small scales, despite allowing inflow to the central 
$\sim$kpc.    This could introduce differences between the predicted accretion rates in different types of bulges (e.g.\ classical or pseudo-bulges).   The dependence of the accretion rate on galaxy morphology may end up being one of the most important differences between the predictions of our model and previous results.

\end{itemize}

\vspace{-0.5cm}
\section{Discussion and Conclusions}
\label{sec:discussion}

The dominant mechanism generating significant gas inflows from $\sim10\,$kpc to $\sim100\,$pc in galaxies has been extensively studied using numerical simulations over the past two decades:  a perturbation to the galaxy (e.g., a merger or disk instability) generates a non-axisymmetric structure. If the perturbation is sufficiently strong in the collisionless component of the system, it drives a corresponding perturbation in the gaseous component. This, in turn, forces the gas to shock, which dissipates energy and relative motion, allowing for net angular momentum loss and inflow.  We have recently demonstrated that qualitatively similar physics can operate from $\sim 0.1-10$ pc in galactic nuclei if they also have a dominant collisionless (stellar) component, although the relevant non-axisymmetric mode is preferentially $m = 1$ instead of $m = 2$ as on larger scales \citep{hopkins:zoom.sims}.   The critical physics of stellar-induced shocks in gas has thus been seen in simulations of galaxy-galaxy mergers, unstable galactic disks, and secondary and tertiary gravitational instabilities in galactic nuclei \citep{noguchi:merger.induced.bars.gas.forcing,barneshernquist96,barnes:review,
berentzen:gas.bar.interaction,hopkins:disk.survival,hopkins:zoom.sims}. 
In this paper, we have developed an analytic framework to model this physics and to describe how gravitational perturbations redistribute angular momentum, and drive shocks and gas inflow, in galaxies. 
We have also tested these models against a large suite of numerical simulations.

Dissipation and the co-existence of collisional (gas) and collisionless (stars) disky components 
are critical for generating large gaseous inflow rates. Dissipationless components can re-distribute angular momentum via resonances \citep{lynden-bell:1972.spiral.amplification,goldreichtremaine:spiral.resonances,
hernquistweinberg92,athanassoula:bar.slowdown,
weinberg:bar.res.requirements} but these are much less efficient than gravitationally-induced shocks, with inflow rates smaller by a factor of $\gtrsim 10-100$.   And without a significant fraction of the mass in a collisionless component that drives the gas into shocks, the gas orbits would self-adjust to eliminate any long-lived strong shocks. 

The canonical criterion for gas orbits to shock in the presence of a non-axisymmetric potential perturbation is that 
$|\zeta|=|d R_{1}/dR_0|>1$, where $R_1$ is the perturbation to a circular orbit initially at radius $R_0$ due to the non-axisymmetric potential (eq.~\ref{eqn:strong.torques.crit.complex}).   This criterion is, however, only a sufficient condition for shocks, not a necessary one.  In the case of shocks near spiral arms or bars in a self-gravitating disk, 
the traditional condition on $|\zeta|$ is often adequate, since $|\zeta|\gg1$ for moderate-amplitude perturbations.   But this is not always the case, particularly for disks in a quasi-Keplerian potential near a BH.  We have demonstrated that even when $|\zeta| \lesssim 1$, a stellar perturbation can produce shocks in a gaseous disk with a finite sound speed (eq.~\ref{eqn:shock.criterion}-\ref{eqn:shock.criterion.alt}).   Physically, the key point is that when the non-axisymmetric potential is stellar dominated, gas ``collisions'' can happen not just at strict orbit crossings, but at a finite separation between streamlines.   The key requirements are that the amplitude of the non-axisymmetric stellar perturbation exceed the critical value given in eq.~\ref{eqn:shock.criterion.alt}, and that a reasonable fraction of the mass is in a disky collisionless component.  We have not quantified in detail how large the mass fraction in the collisionless component must be, but simple force considerations suggest $\gtrsim 50\%$.  

Given the presence of shocks in the gas, we have derived the resulting angular momentum loss/gain and the gas inflow/outflow rates in two distinct limits:   very strong shocks ($|\zeta| \gg 1$) and marginal orbit crossings ($|\zeta| \lesssim 1$) (eq.~\ref{eqn:mdot.full}).   This calculation includes a proper treatment of the phase function for angular momentum exchange between the  gas and stars, i.e., when the net torque produces gas inflow and when it produces outflow.  For typical conditions in galaxies, we find flow towards an inner Lindlad resonance inside of co-rotation and towards an outer Lindblad resonance outside of co-rotation. However, the behavior can be more complex depending on the details of the system. 

One of our key results is that the gas inflow produced by gravitationally-induced shocks is robust and rapid, with 
\be
\dot{M}\sim \frac{|a|\,M_{\rm gas}}{t_{\rm dyn}} \sim |a|\,M_{\rm gas}\Omega
\label{mdot_disc}
\ee
where $|a|$ is the fractional non-axisymmetric perturbation to the potential (eq.~\ref{eqn:mdot.strongtorque.simple} \& \ref{eqn:mdot.full}).  Note several key facts about this scaling.  First,  $\dot M$ is {\em linear} in $|a|$.
In contrast, when shocks do not occur, angular momentum transport by non-axisymmetric perturbations is at least second-order in $|a|$ (\S \ref{sec:deriv.mom.weak}).   Moreover, this 
second-order transport is also suppressed by powers of $kR$ and $m$ where $k$ ($m$) is the radial (azimuthal) wavenumber of the non-axisymmetric mode (see, e.g., \citealt{kalnajs:1971}'s exact solution for the logarithmic spiral; eq~\ref{eqn:mdot.kalnajs}).    Given these differences, we find inflow rates that are a factor of $\sim 10-100$ larger than would traditionally be predicted by linear theory (for a given set of gas properties).   

A second key property of equation~\ref{mdot_disc} is that {the inflow rate is independent of the sound speed of the gas $c_s$}, i.e., of the gas thermodynamics.  Physically, this is because the angular momentum transport is governed by the rate at which gas orbits are forced to shock via the non-axisymmetric stellar potential.   Because the relevant gas motions are by definition supersonic and induced by a collisionless component, the precise gas thermodynamics is not important.   This property of inflow produced by gravitationally-induced shocks is fundamentally different from angular momentum transport by local viscous stresses ("$\alpha$") or by linear spiral waves, both of which depend on $c_s$.\footnote{More precisely, transport by linear spiral waves only depends on $c_s$ indirectly, if the system maintains $Q \sim 1$; see eq.~\ref{eqn:mdot.kalnajs}.}   The lack of an explicit dependence on the gas thermodynamics, which is particularly uncertain in a multi-phase interstellar medium with stellar feedback, points to the robustness of our predicted inflow rates in this regime.   One potential subtlety is that our analytic work assumes that a significant fraction of the ISM consists of relatively diffuse gas, rather than gas bound up in, e.g., star clusters; this is also true to some extent in our numerical work (see \S \ref{sec:sims} for a discussion of this point).   This assumption is motivated by the observed low star formation efficiency in dense gas and the small fraction of a molecular cloud's mass that is converted into stars before the cloud is disrupted (e.g., \citealt{krumholz:sf.eff.in.clouds}).  But further numerical work including a more detailed treatment of stellar feedback and a multi-phase ISM would shed considerable light on whether a highly inhomogeneous ISM qualitatively modifies the physics of gas inflow in galaxies.  
It is well-known, for example, that mode development depends significantly on gas thermodynamics in 
pure gas disks \citep[e.g.][]{gammie:2001.cooling.in.keplerian.disks,
nayakshin:sfr.in.clumps.vs.coolingrate,
cossins:2009.grav.instab.vs.coolingrate}.
We suspect that the physics described here will be applicable to any collisionless plus collisional disk system, provided that the post-shock cooling time of the gas is short compared to the local dynamical time 
and the collisionless component is gravitationally dominant 
(well-satisfied in galaxies at the radii of interest).

Another way to appreciate the importance of equation~\ref{mdot_disc} is to compare the resulting inflow time of gas in galaxies ($\sim |a|^{-1} \, t_{\rm dyn}$) to the star formation timescale.  The latter is observed to be $\sim 10-100 \ t_{\rm dyn}$ (e.g., \citealt{bouche:z2.kennicutt}).   Since the non-axisymmetric mode amplitudes produced by mergers or instabilities of self-gravitating disks can readily reach $|a| \sim 0.01-1$ (Fig. 12 of \citealt{hopkins:zoom.sims}), equation~\ref{mdot_disc} implies that the gas inflow time can be comparable to or shorter than the timescale for the gas to turn into stars.   This is the fundamental reason that we find BH inflow rates $\sim 1-10 \, M_\odot$ yr$^{-1}$ in our simulations (sufficient to fuel luminous quasars) without overproducing star formation in galactic nuclei.  
Similar inflow rates have also been found in independent simulations by 
\citet{levine:sim.mdot.pwrspectrum}. 
By contrast, a number of previous calculations (e.g., 
\citealt{goodman:qso.disk.selfgrav,sirko:qso.seds.from.selfgrav.disks,thompson:rad.pressure,
nayakshin:forced.stochastic.accretion.model}) 
assumed either a local $\alpha$ model or linear transport by spiral waves, neither of which accurately describe our analytic or numerical results.  Because of this incorrect treatment of angular momentum transport, these previous calculations also significantly overestimated the inflow time of gas in galaxies, and thus overestimated the fraction of the inflowing gas that turns into stars. 
In our simulations, we find that star formation and inflow reach an approximate equilibrium, in which the inflow rate is comparable to the local star formation rate (except at sufficiently small radii, where the inflow rate becomes independent of radius).   This balance determines the quasi-steady gas surface density as a function of radius (\S~\ref{sec:deriv.gasden}), and is in good agreement with the numerical results from $\sim 10^{-3}-10$ kpc (e.g., Figs \ref{fig:illustrate.100pc} \& \ref{fig:illustrate.10pc}).
Of course, star formation in these models is still ``sub-grid'' and pegged to an empirical law, so 
avoiding runaway local collapse may still require processes such as stellar feedback.

For the purposes of analytic or semi-analytic calculations and galactic or cosmological numerical simulations, it would be useful to have an estimate of the BH inflow rate based only on galaxy properties at much larger radii (generally $>100\,$pc).  Calculations in the literature often assume simple models such as Eddington-limited accretion for a fixed timescale, spherical Bondi accretion \citep{springel:models,dimatteo:msigma}, 
local viscous transport (e.g., \citealt{sirko:qso.seds.from.selfgrav.disks,debuhr:momentum.feedback}), or linear transport by spiral waves (e.g., \citealt{thompson:rad.pressure}). None of these models, however, accounts for the key physics of gravitationally-induced shocks in the gas.   We have explicitly demonstrated that both spherical accretion and local viscous transport 
are relatively poor approximations to the inflow seen in our numerical simulations (Figs.~\ref{fig:mdot.model.comparison} \& \ref{fig:mdotBH.model.comparison}).

Using our analytic models of gravitationally-induced gas inflow we have developed a more accurate predictor of the BH accretion rate given the large-scale properties of a galaxy
(eqs~\ref{eqn:mdot.predictor}-\ref{eqn:mdot.predictor.fgas}). This model is highly simplified, and clearly far from exact, but it includes well-motivated dependencies for some of the key parameters of the problem.  For example, the inflow rate is strongly suppressed for larger bulge-to-disk ratios, as is generally the case for global gravitational instabilities.   In addition, the inflow rate is essentially independent of the gas sound speed so long as cooling is efficient and most of the gas resides in a rotationally-supported disk.  By again comparing to our suite of numerical simulations, we have demonstrated that this new BH inflow predictor represents a major improvement relative to previous models (Fig. \ref{fig:mdot.grav.test}); see \S \ref{sec:differences} for a brief discussion of some of the implications of this result.   However, even this new BH predictor has a factor of $\sim 10$ scatter when compared to the simulations.  This does not appear to be due to any systematic dependence on galaxy structural properties, gas fraction, etc. that is not already captured in our scalings.  Rather, the scatter is a consequence of physical variability in the inflow rate for a given set of conditions in the galaxy at large radii.    Indeed, because gravitational instabilities operate on all scales, and because the systems of interest are chaotic, we find that there is variability on essentially all timescales (see also \citealt{levine:sim.mdot.pwrspectrum}).  We believe that it is important to include this variability in future models of BH growth and AGN feedback.

A number of important regimes of gas inflow require significant further study.   We have assumed throughout this paper that the non-axisymmetric mode amplitudes are large enough to produce shocks at nearly all radii.   This is indeed borne out by our (and others') simulations of galaxy-galaxy mergers and gas-rich galactic nuclei, but it is by no means clear that this will always be the case.   For low mode amplitudes, we do not expect shocks and so the angular momentum transport mechanisms studied in this paper will not be present.  Understanding which processes dominate inflow at lower inflow rates will be of  considerable interest for models of low-luminosity AGN and weak enhancements  in nuclear star formation. 

The regime in which galaxies are gas-dominated is also of considerable 
interest. In this limit, it is unlikely that strong orbit crossings and/or coherent long-lived shocks will occur, 
if the mode remains linear. 
Generically, we expect this to decrease the efficiency of gas inflow.   
This has in fact been seen and studied in detail in the context of galaxy-galaxy mergers (e.g., \citealt{hopkins:disk.survival.cosmo}).   
{Of course, if strongly non-linear modes can be sustained, then our assumptions are probably not applicable and there can in principle be a wide range of inflow rates -- a pure radial inflow at the free-fall velocity is a 
valid non-linear configuration!} 
And in very gas-rich systems, local fragmentation may be much more efficient, generating an 
inhomogeneous ISM that may provide mechanisms for angular momentum transport that we have not 
considered in this paper. 
Understanding this physics on smaller scales in galactic nuclei is particularly important for high-redshift galaxy formation and the formation of the first BHs, where it has been posited that they might form in direct collapse of primordial gas clouds
\citep[][and references therein]{begelman:direct.bh.collapse.w.turbulence}.   In our simulations of gas dominated systems to date \citep{hopkins:zoom.sims}, we find that the gas inflow from $\sim 10$ pc to $\sim 0.1$ pc is significantly more efficient after some fraction of the gas has turned into stars.   {This is qualitatively consistent with the results from galaxy merger simulations, but needs to be studied in more detail 
\citep[for comparison, see e.g.][]{agertz:disk.fragmentation.model,teyssier:2010.clumpy.sb.in.mergers}. }

\acknowledgments 
We thank Todd Thompson for useful conversations, 
and the anonymous referee for helpful suggestions.
Support for PFH and EQ was provided by the Miller  Institute for Basic Research in Science, University of California  Berkeley. EQ was also supported in part by NASA grant NNG06GI68G and  the David and Lucile Packard Foundation.
\\

\bibliography{/Users/phopkins/Documents/lars_galaxies/papers/ms}

\end{document}